\documentclass[pre,aps,10pt,showpacs,amsmath,amssymb,showkeys,floatfix,nofootinbib,longbibliography]{revtex4-1}
\usepackage[english]{babel}
\usepackage{txfonts}
\usepackage{graphicx}
\usepackage{nicefrac}
\usepackage[caption=false]{subfig}
\newcommand{\aap}{Astron. Astrophys.}  
\newcommand{\mnras}{Mon.\ Not.\ R.\ Astr.\ Soc.}  
\newcommand{\gafd}{Geophys.\ Astrophys.\ Fluid\ Dyn.}  
\newcommand{\apjl}{Astrophys. J. Lett.}  
\newcommand{\pof}{Phys. Fluids}  
\newcommand{\jfm}{J. Fluid Mech.}  

\renewcommand{\vec}[1]{\mbox{\boldmath$#1$}}

\def\la{\mathrel{\mathchoice {\vcenter{\offinterlineskip\halign{\hfil
$\displaystyle##$\hfil\cr<\cr\sim\cr}}}
{\vcenter{\offinterlineskip\halign{\hfil$\textstyle##$\hfil\cr  
<\cr\sim\cr}}}
{\vcenter{\offinterlineskip\halign{\hfil$\scriptstyle##$\hfil\cr
<\cr\sim\cr}}}
{\vcenter{\offinterlineskip\halign{\hfil$\scriptscriptstyle##$\hfil\cr
<\cr\sim\cr}}}}}

\begin{document}

\preprint{APS/123-QED}

\title{Parametric instability in periodically perturbed dynamos}

\author{Andr\'e Giesecke}
\email{a.giesecke@hzdr.de}

\author{Frank Stefani}
\affiliation{
Institute of Fluid Dynamics\\
Helmholtz-Zentrum Dresden-Rossendorf\\
Bautzner Landstrasse 400, D-01328 Dresden, Germany
}
\author{Johann Herault}
\affiliation{
Aix Marseille Univ, CNRS, Centrale Marseille, Institut de Recherche
sur les Phenom{\`e}nes Hors Equilibre (IRPHE), UMR 7342, 
49 rue F. Joliot-Curie, 13013 Marseille, France}

\date{\today}

\begin{abstract}
{
We examine kinematic dynamo action driven by an axisymmetric
large-scale flow that is superimposed with an azimuthally propagating 
non-axisymmetric perturbation with a frequency $\omega$.  Although we
apply a rather simple large-scale velocity field, our simulations
exhibit a complex behavior with oscillating and azimuthally drifting
eigenmodes as well as stationary regimes.  Within these
non-oscillating regimes we find parametric resonances characterized by
a considerable enhancement of dynamo action and by a locking of the
phase of the magnetic field to the pattern of the perturbation.  We
find an approximate fulfillment of the relationship between the
resonant frequency $\omega_{\rm{res}}$ of the excitation and the
eigenfrequency $\omega_0$ of the undisturbed system given by
$\omega_{\rm{res}}=2\omega_0$, which is known from paradigmatic
rotating mechanical systems and our prior study [Giesecke et al.,
Phys. Rev. E, {\bf{86}}, 066303 (2012)].  We find further, broader,
regimes with weaker enhancement of the growth rates but without
phase locking.  However, this amplification regime arises only in case
of a basic (i.e., unperturbed) state consisting of several different
eigenmodes with rather close growth rates.  Qualitatively, these
observations can be explained in terms of a simple low-dimensional
model for the magnetic field amplitude that is derived using Floquet
theory.
}

{
The observed phenomena may be of fundamental importance in planetary
dynamo models with the base flow being disturbed by periodic external
forces like precession or tides and for the realization of dynamo
action under laboratory conditions where imposed perturbations
with the appropriate frequency might facilitate the occurrence of
dynamo action.
}
{}{}{}
\end{abstract}

\pacs{47.65.--d, 91.25.Cw, 52.65.Kj}
\keywords{dynamo, parametric resonance, periodic perturbation}
\maketitle

\section{Introduction}

Usually it is assumed that magnetic fields of galaxies, stars, and
planets are generated by the magnetohydrodynamic dynamo effect, which
describes the transfer of kinetic energy from a flow of an
electrically conducting fluid into magnetic energy.  Regarding stellar
or planetary bodies, magnetic fields are supposed to be powered by
convection-driven flows \cite{2010GGG....11.6016K}.  However, there
are also alternative concepts for planetary dynamos with a flow driven
by mechanical stirring, like precession (which has been considered as
an alternative forcing for the geodynamo \cite{1968Sci...160..259M}
and the ancient lunar dynamo
\cite{2011Natur.479..212D,2013JFM...737..412N}), or tidal forcing
(which has been proposed to be responsible for the ancient Martian
dynamo \cite{2008JGRE..113.6003A} and, as well, for the early
lunar dynamo \cite{2011Natur.479..215L}).

Although mechanical stirring, whether by precession or by tides, is
capable to drive a dynamo by its own
\cite{2005PhFl...17c4104T,2014ApJ...789L..25C}, it is more likely that
in the case of realistic planetary models a combination of different types
of forcing is responsible for the total flow, with tides and/or
precession acting as a periodic perturbation to a base flow
(e.g., convectively driven motions \cite{2012MNRAS.422.1975O}).
Despite the small amount of energy provided instantaneously by the
perturbation flow, its impact can be large since the perturbation may
act as a kind of catalyst that allows a transfer of energy from the
available rotational energy, which is huge but does not contribute
to magnetic induction, into a more usable form of flow
\cite{2015AnRFM..47..163L}.  In particular, a periodic perturbation of
some base flow may cause a parametric resonance resulting, e.g., in
the excitation of free nonaxisymmetric inertial waves like have
been found in simulations and experiments of precession driven flows
\cite{1999JFM...382..283K,2003JFM...492..363L,FLM:7951619,2015NJPh...17k3044G,herault2016}.
Such time-dependent nonaxisymmetric flow perturbations in turn may
significantly enhance the ability of the system to drive a dynamo via
yet another resonance attributed to the magnetic field generation
process by coupling low-order rotational modes of the fluid flow
\cite{2007GApFD.101..507A}.

A paradigmatic system related to the dynamo effect is the disk dynamo,
in which a current is guided from the edge of a rotating conductive
disk to the disk axis in such a way that the induced magnetic field
amplifies the original (seed) field.  A positive effect on magnetic
self-excitation arises when periodically varying the rotation rate of
the disk with an appropriate frequency \cite{2010PhLA..374..584P}.
Similar to periodically perturbed mechanical problems, this system can
be described by a Mathieu equation, which represents a special case of
the Hill equation
\begin{equation}
\frac{d^2y}{dt^2}=a(t)y
\end{equation}
with $a(t+T)=a(t)$ being a $T-$periodic function.  A parametric
resonance occurs if the excitation frequency $\omega$ is twice the
eigenfrequency $\omega_0$ of the unperturbed system, i.e.,
$\omega_{\rm{res}}=2\omega_0$.  In this case the field generation of
the disk dynamo is based on a simple axially symmetric variation of
the velocity of a solid body.  More relevant for astrophysical
objects are fluid flow driven dynamos where the induced electrical
currents have more degrees of freedom.  In that case a perturbation
can enter the induction equation via the induction term $\vec{u}\times
\vec{B}$, which in turn may involve a periodic contribution through a
perturbed velocity field
$\vec{u}=\vec{u}_0(\vec{r})+\epsilon\tilde{\vec{u}}(\vec{r},t)$ with a
base flow $\vec{u}_0(\vec{r})$ and a time- and space-periodic function
$\tilde{\vec{u}}(\vec{r}, t)$.  Flow models with periodical
perturbations have been used, for example, to explain the strong
nonaxisymmetric activity observed in close binary systems
\cite{2002MNRAS.334..925S} or the superposition of axisymmetric and
nonaxisymmetric contributions of magnetic fields in spiral galaxies
\cite{1990MNRAS.244..714C} where the spiralling arms provide a
periodic perturbation in terms of rotating density waves.

An improvement of fluid flow generated dynamo action is also of great
importance for the conduction of successful dynamo experiments.  This
can, first, be achieved by optimizing the pattern of the flow like 
had been done for the Riga and the Karlsruhe dynamo. However, in other
cases, where this is not readily possible such as in the dynamo
experiments in Cadarache \cite{2007PhRvL..98d4502M} and Madison
\cite{PhysRevLett.96.055002} or for the planned precession dynamo at
Helmholtz-Zentrum Dresden-Rossendorf \cite{stefani_dresdyn_2012}, other
types of improvement of dynamo action must be explored.

The present study is mainly motivated by the idea of facilitating
dynamo action of a prescribed large-scale flow by imposing small
periodic disturbances.  For this purpose, we investigate an idealized
system in a finite cylinder with dynamo action driven by a large-scale
prescribed axisymmetric flow that is periodically perturbed by a
nonaxisymmetric distortion propagating around the symmetry axis of a
cylinder.  We assume a given periodic perturbation of the velocity
field which is caused and maintained by some unspecified mechanism.
We examine the corresponding impact on growth rate and frequencies of
the magnetic field in dependence on the drift frequency and/or
amplitude of the perturbation.

The study is a continuation and generalization of previous work
related to the impact of azimuthally drifting equatorial vortices on
dynamo action in the von-K{\'a}rm{\'a}n-sodium (VKS) dynamo
\cite{2012PhRvE..86f6303G}.  Here, we present a more general and
systematic approach to the phenomenon of parametric resonance in
combination with dynamo action. Basically, we show how time-periodic
perturbations connect different dynamo modes from an unperturbed
model, thereby leading to an amplification of the process of magnetic
field generation. Finally, we conclude, how such an operation may be
realized in natural and/or experimental dynamos.

The paper is divided into two parts.  We start with numerical
simulations of the kinematic induction equation in cylindrical
geometry, and we present results for the growth rates in dependent on
the amplitude and the frequency of a paradigmatic perturbation
pattern. The geometric structure of the perturbation is prescribed by
an azimuthal wave number $\widetilde{m} = 2$ and a frequency $\omega$
of the propagation of the perturbation pattern around the axis of
symmetry. The space-time periodic behavior is qualitatively similar to
distortions caused by azimuthally drifting equatorial vortices that
have been found in water experiments with a geometry and a forcing
similar to the VKS dynamo \cite{2007PhRvL..99e4101D} or the velocity
perturbations caused by tidal interactions in a two-body system.

In the second part we develop a simple low dimensional model for the
amplitude of the magnetic field and show that a periodic perturbation,
even when it is small, is indeed capable of significantly enhancing
magnetic field generation and may trigger the transfer from a stable
state to an unstable state with an exponentially growing magnetic field.


\section{Numerical model}

\subsection{Velocity field}

We perform three-dimensional simulations of kinematic dynamo action
driven by a prescribed velocity field.  The basic flow field in our
simulations is a cylindrical adaptation of the so-called {\it{S2T2}}
flow, which consists of two poloidal and two toroidal large-scale flow
cells \cite{1989RSPSA.425..407D}.  This axisymmetric flow field
resembles the mean flow driven by two opposing and counter-rotating
impellers and has been applied with slightly different definitions for
the radial behavior as a mean flow in various kinematic studies of the
VKS dynamo
\cite{2004phfl,2005physics..11149S,2005PhFl...17k7104R,2010PhRvL.104d4503G,2012NJPh...14e3005G,2012PhRvE..86f6303G}.
It is well known that this kind of flow drives a dynamo at rather low
critical magnetic Reynolds numbers (${\rm{Rm}}^{\rm{crit}} \sim\!30
\dots 40$ with pseudo-vacuum boundary conditions and
${\rm{Rm}}^{\rm{crit}}\sim\!60 \dots 70$ with insulating boundary
conditions) with a magnetic eigenmode characterized by an azimuthal
wave number $m=1$ (equatorial dipole) \cite{2010GApFD.104..505G}.

\begin{figure}[h!]
\vspace*{-1.8cm}
\includegraphics[width=0.475\textwidth]{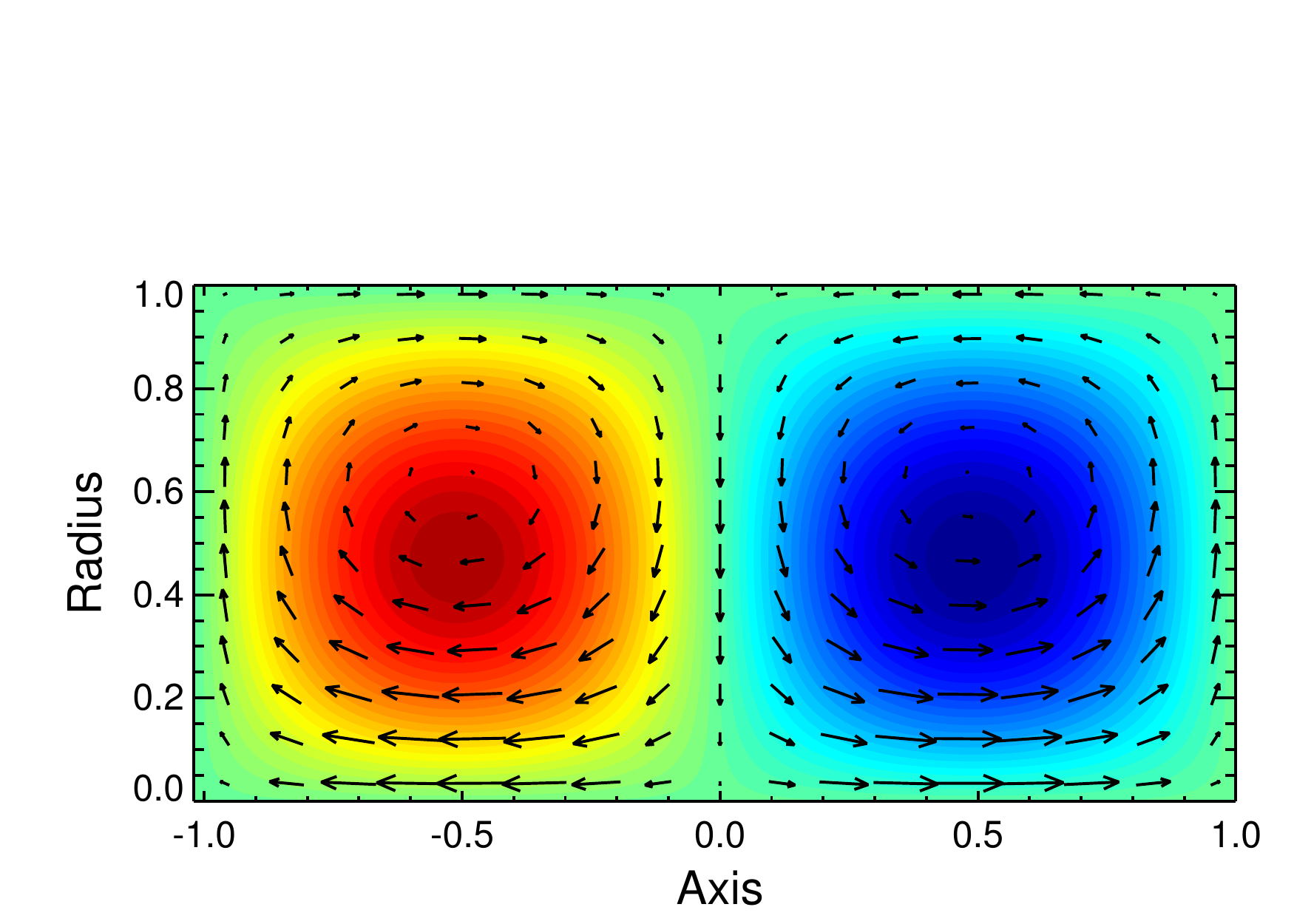}
\caption{Contour plot of the axisymmetric velocity field. The colors
  denote the azimuthal velocity component and the arrows represent the
  poloidal component of the axisymmetric velocity field.
  \label{fig::velfield_sim}}
\end{figure}

In the present study we consider a cylinder with radius $R=1$ and
height $H=2$ so that $r\in [0;1]$ and $z\in [-1;+1]$.  The total flow
$\vec{U}_{\rm{tot}}$ in the simulations is given by the sum of an
axisymmetric poloidal and toroidal contribution
$\vec{U}_{\rm{p}}+\vec{U}_{\rm{t}}$ and a nonaxisymmetric
contribution $\epsilon\vec{U}_{\widetilde{m}}$ with an azimuthal wave
number ${\widetilde{m}}$ and an amplitude factor $\epsilon$:
\begin{equation}
\vec{U}_{\rm{tot}}=\vec{U}_{\rm{p}}+
\vec{U}_{\rm{t}}+\epsilon\vec{U}_{\widetilde{m}}.
\end{equation}
The axisymmetric flow field is derived from an
axisymmetric time-independent scalar potential $\Psi (r,z)$ given by
\begin{equation}
\Psi(r,z)=J_1(\kappa r)\sin\left(\frac{2\pi z}{H}\right)\label{eq::skalar_pot}
\end{equation}
with $J_1$ the cylindrical Bessel function of order 1, $\kappa=3.8317$
the first zero of $J_1$ in order to enforce $u_r=0$ on the radial
boundary and $H$ the height of the cylinder. The actual axisymmetric flow is
composed of a poloidal part given by
\begin{equation}
\vec{U}_{\rm{p}}(r,z)=
\nabla\times \Psi (r,z) \vec{e}_{\varphi}\label{eq::flow_def}
\end{equation}
and a toroidal component given by 
\begin{equation}
\vec{U}_{\rm{t}}(r,z)=-\sqrt{\kappa^2+\left(\frac{2\pi}{H}\right)^2}
\Psi (r,z)\vec{e}_{\varphi}. 
\label{eq::flow_tor}
\end{equation}
The prefactor for the toroidal component ensures that the flow
fulfills the Beltrami property, i.e., $\nabla\times\vec{U}=k\vec{U}$
with $k=-\sqrt{\kappa^2+(2\pi/H)^2}$  which maximizes the kinetic
helicity $h=(\nabla\times\vec{U})\cdot \vec{U}$.
Figure~\ref{fig::velfield_sim} presents a contour plot of the
axisymmetric flow field where the colors denote the toroidal flow
component and the arrows represent the poloidal flow component.
\begin{figure}[h!]
\vspace*{-0.3cm}
{\centering{\includegraphics[width=0.49\textwidth]{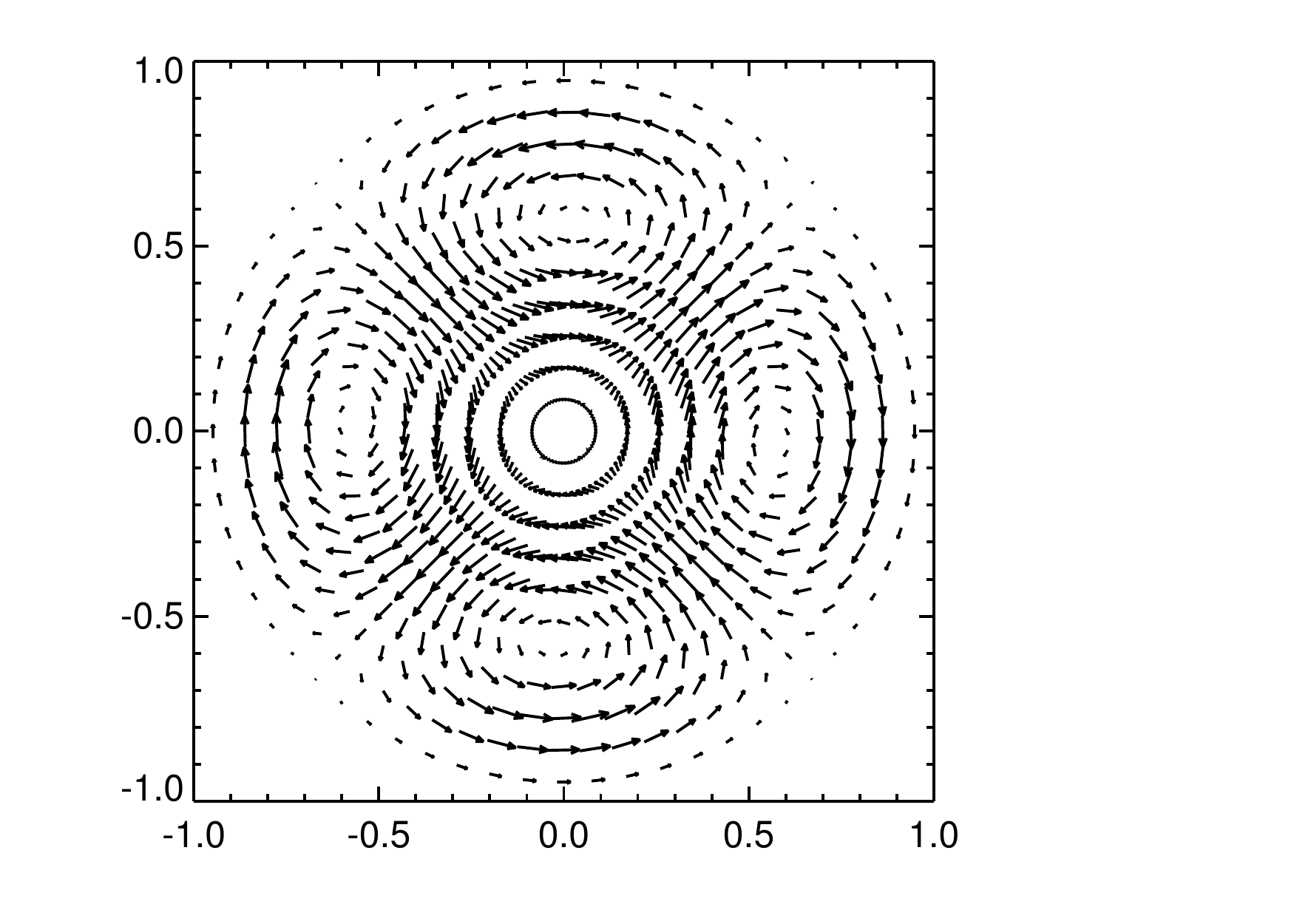}} 
\caption{Cut of the velocity perturbation in the equatorial plane. The
axial dependence of the perturbation ensures that $\vec{U}_m$ vanishes
at the top and bottom boundaries. The perturbation flow is parallel to
the equatorial plane and has no component along the
axis.\label{fig::velfield_sim_pert}}} 
\end{figure}

The nonaxisymmetric perturbation is derived from a time-dependent
scalar potential $\widetilde\Psi(r,\varphi,z,t)$ which is defined by
\begin{equation}
\widetilde\Psi (r,\varphi,z,t)  =  r \cos (2\pi r -1) 
\cos \left(\frac{2\pi z}{H}\right) 
\cos(\widetilde{m}\varphi+\omega t)\label{eq::potpert}
\end{equation}
with $\widetilde{m}$ the azimuthal wave number of the perturbation and
$\omega$ the frequency. The flow perturbation is then given by 
\begin{equation}
\vec{U}_{{\widetilde{m}}}(r,\varphi,z,t)  =  
\nabla\times \widetilde\Psi (r,\varphi,z,t)\vec{e}_{z}.\label{eq::flow_pert}
\end{equation}
Note that the perturbation flow $\vec{U}_{\widetilde{m}}$ has no
component along the axial direction.  The particular distribution
given by~(\ref{eq::flow_pert}) vanishes at the top and the bottom of
the cylinder with the maximum at the midplane.  The pattern of the
nonaxisymmetric perturbation for the case $\widetilde{m}=2$ is shown
in the equatorial plane in Fig.~\ref{fig::velfield_sim_pert}.  The
orientation of the perturbation flow differs from the model applied in
Ref.~\cite{2012PhRvE..86f6303G} where equatorial vortices were modelled
with vanishing radial component and a local symmetry axis
perpendicular to the symmetry axis of the cylinder like have been
observed in water experiments with von-K{\'a}rm{\'a}n-like flow
driving \cite{2007PhRvL..99e4101D}.

The timescale used in the simulations is the advective timescale
$\tau=R/U_{\rm{max}}$ defined by the radius of the cylinder ($R=1$ in
our study) and the maximum speed 
\begin{equation}
\nonumber
U_{\rm{max}}=\max\left(\!\sqrt{U_{\rm{p}}^2(r,z)+U_{\rm{t}}^2(r,z)}\right).
\end{equation}
with $U_{\rm{p}}$ and $U_{\rm{t}}$ taken from Eqs.~(\ref{eq::flow_def})
and~(\ref{eq::flow_tor}). 
 
The frequency $\omega$ of the azimuthal propagation of
the perturbation is given in units of $u_{\varphi}^{\max}/R_{\max}$
with the maximum of the azimuthal velocity component
$u_{\varphi}^{\max}$ of the unperturbed velocity field defined by
Eqs.~(\ref{eq::skalar_pot}) and~(\ref{eq::flow_tor}), and $R_{\max}$ the corresponding radius at
which the azimuthal velocity takes its maximum value.

\subsection{Results}

The numerical solutions that are presented in the following are
obtained by time stepping the magnetic induction
equation
\begin{equation}
\frac{\partial}{\partial t} \vec{B}= \nabla\times
\left(\vec{U}\times\vec{B}\right) + \eta \Delta\vec{B},
\label{eq::induction}
\end{equation}
where $\vec{B}$ is the magnetic flux density and $\eta$ denotes the
constant magnetic diffusivity. 
We apply a finite volume method with a constraint transport scheme
that ensures the exact treatment of the solenoidal property of
$\vec{B}$ (for details see \cite{2008giesecke_maghyd} and
\cite{2010GApFD.104..249G}).   
For the sake of numerical performance, we use pseudo-vacuum boundary
conditions $\vec{B}\times\vec{n}=0$, which in comparison with more
realistic insulating boundary conditions do not dramatically impact
the solutions except a shift of the growth rates to larger values thus
reducing the critical magnetic Reynolds number required for the onset
of dynamo action.  The magnetic Reynolds number characterizes the flow
amplitude and is defined as
\begin{equation}
{\rm{Rm}}=\frac{U_{\max} R}{\eta}
\end{equation}
with $U_{\max}$ the maximum of the axisymmetric velocity field and $R$
the radius of the cylinder. The Reynolds number is
determined by the axisymmetric basic flow and hardly changes when
adding the perturbation flow with an amplitude $\epsilon < 1$. 

All our  models have run for at least several diffusion times (given by
$\tau=R^2/\eta$), which is generally by far sufficient to identify the
leading eigenmode and to accurately calculate its growth rates.

\subsubsection{Basic state} 

\begin{figure}[t!]
\includegraphics[width=0.49\linewidth]{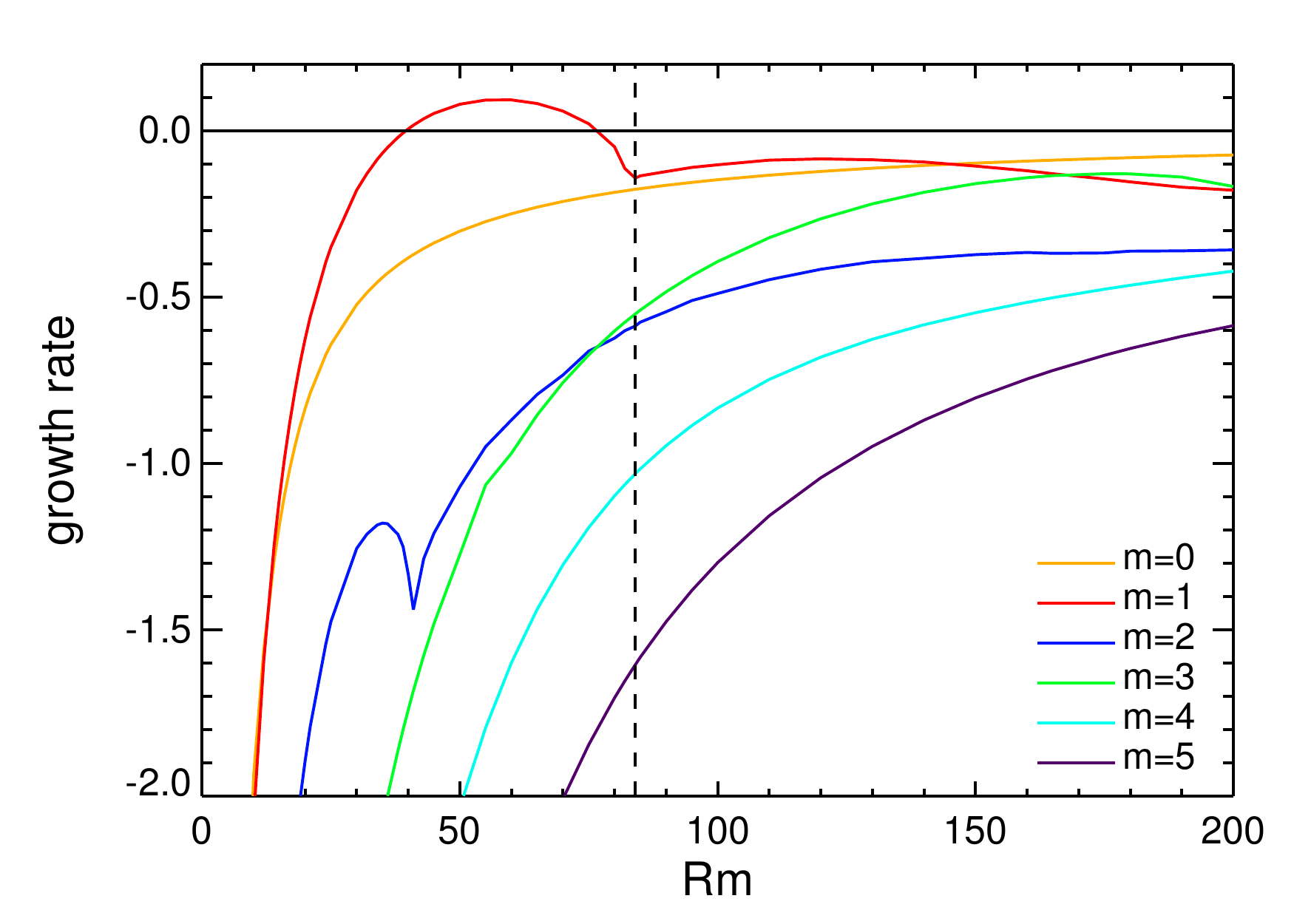}
\\[0.4cm]
\includegraphics[width=0.49\linewidth]{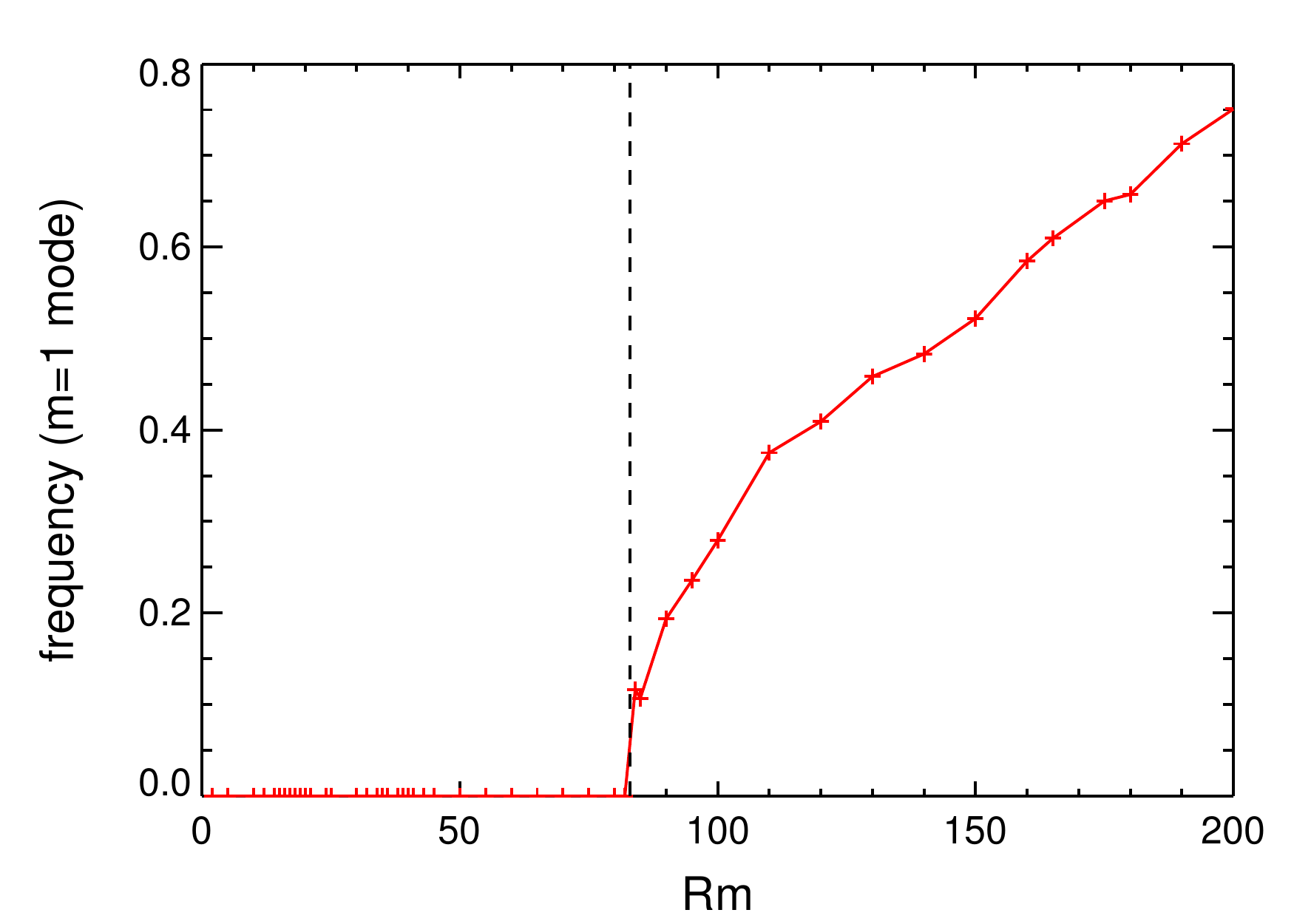}
\caption{Growth rates (top) and frequencies (bottom) versus
  ${\rm{Rm}}$ for the axisymmetric flow field given
  by~(\ref{eq::flow_def}) and (\ref{eq::flow_tor}). 
}
\label{fig::gr_vs_rm}
\end{figure}

\begin{figure}[b!]
\centering
\includegraphics[width=0.245\linewidth]{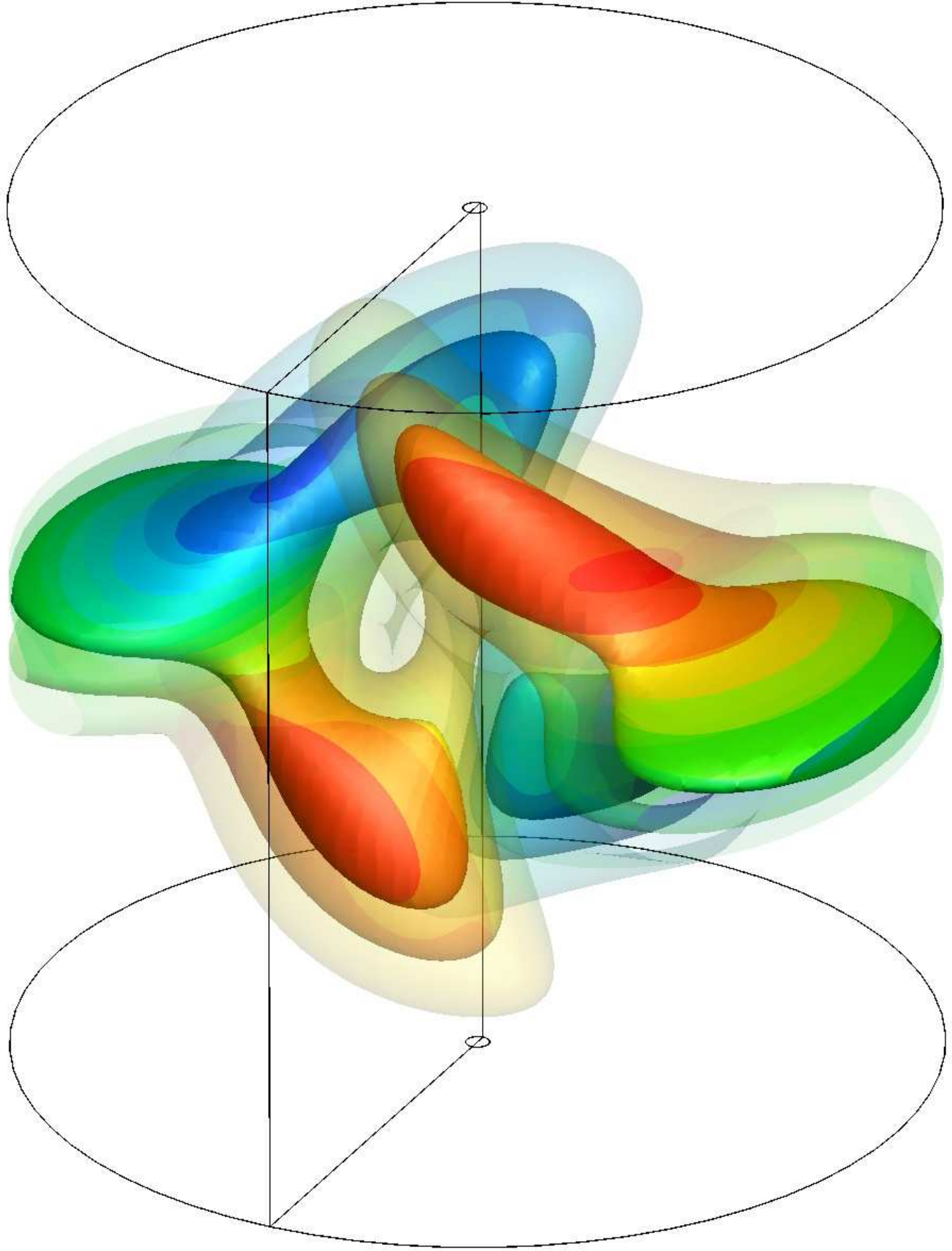}
\includegraphics[width=0.245\linewidth]{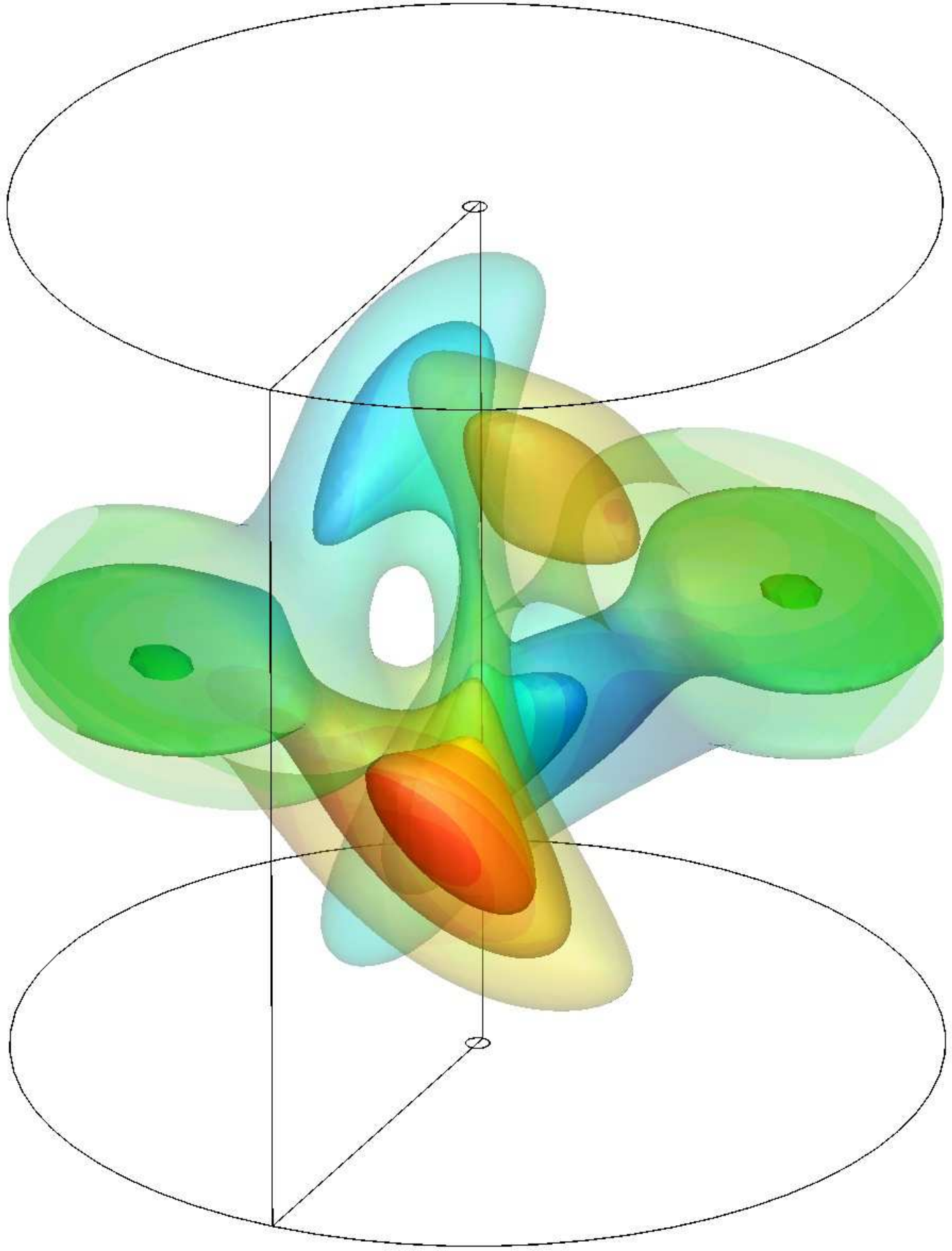}
\caption{Magnetic field energy density $B^2$ for ${\rm{Rm}}=60$
  (left) and  ${\rm{Rm}}=120$ (right). The isosurfaces are colored
  with the azimuthal field.\label{fig::unperturbed_eigenfunction}}
\end{figure}

\begin{figure}
\includegraphics[width=0.95\textwidth]{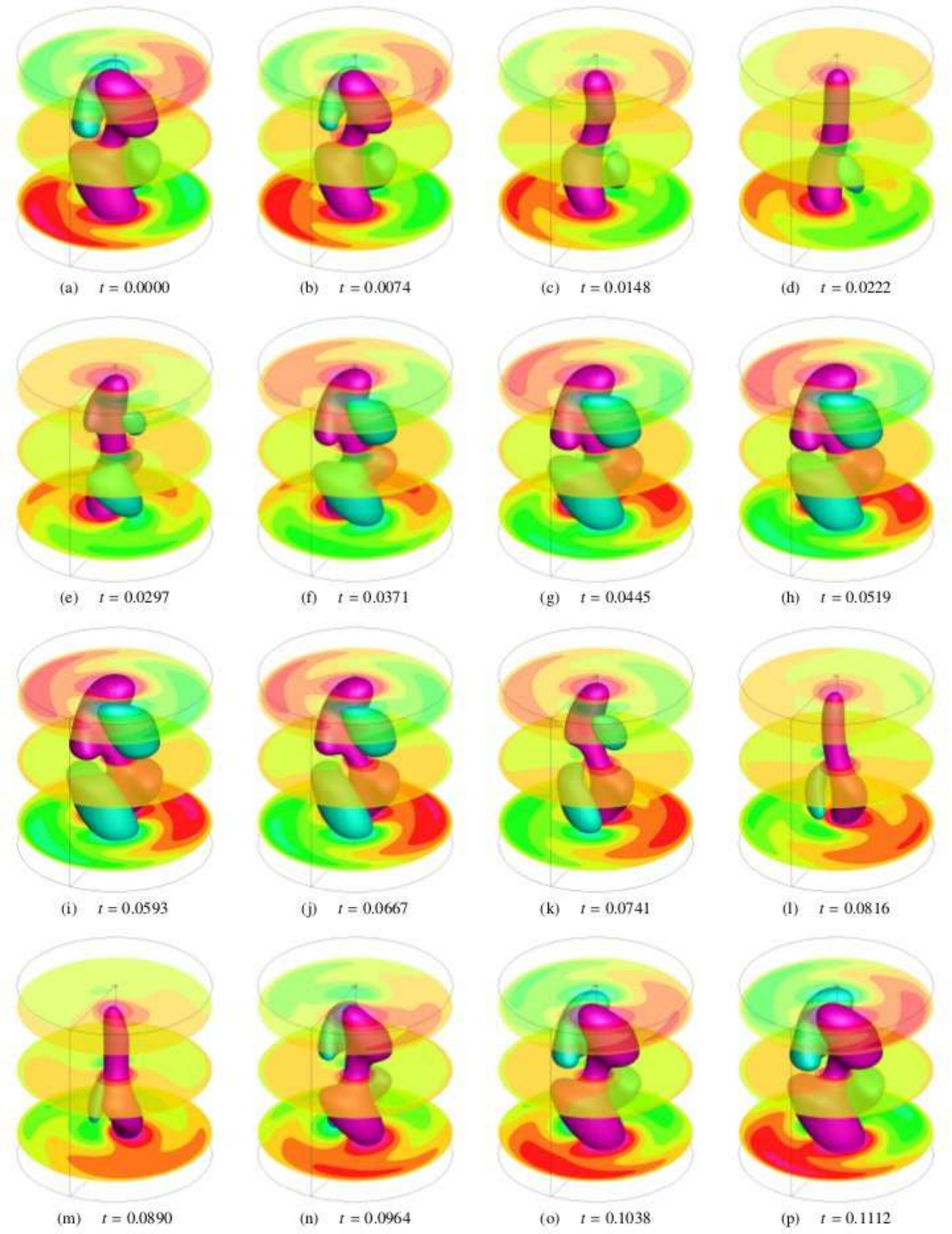}
\caption{
Evolution of the structure of the axial magnetic field component $B_z$
during one oscillation period. The isosurfaces show time snapshots of
$B_z$ at $50\%$ of the time-averaged absolute value. The exponential
decay has been removed. ${\rm{Rm}}=120$, unperturbed run
($\epsilon=0$).  
\label{fig::timeseries_unpert}
}
\end{figure}

The upper panel of Fig.~\ref{fig::gr_vs_rm} shows the magnetic field
growth rate against the magnetic Reynolds number for the basic flow
without perturbation.  Due to the axisymmetry of the flow field all
azimuthal modes are decoupled and follow separate curves.  Regarding
the $m=1$ mode (red curve in Fig.~\ref{fig::gr_vs_rm}), we can
roughly divide the solutions into two regimes. Below ${\rm{Rm}}\approx
83$ the $m=1$ mode is clearly dominant and does not oscillate. Dynamo
solutions with a positive growth rate are found in the range
$39\alt{\rm{Rm}}\alt 78$.  Above ${\rm{Rm}}\approx 83$ the mode with
$m=1$ changes its character into an oscillating mode (see bottom panel
in Fig.~\ref{fig::gr_vs_rm}) which always describes a decaying
solution that does not cross the dynamo threshold.  Despite the
different temporal behavior, the structure of both eigenmodes is quite
similar (see Fig.~\ref{fig::unperturbed_eigenfunction}).  A
timeseries showing the evolution of the magnetic field at
${\rm{Rm}}=120$ is shown in Fig.~\ref{fig::timeseries_unpert}.

Indeed, we find time-dependent solutions with the characteristic
evolution shown in Fig.~\ref{fig::timeseries_unpert} in all our
simulations with a magnetic Reynolds number above
${\rm{Rm}}=83$. These results are typical for a dynamo beyond an
"exceptional point", which is a point in the spectrum of a linear
operator where two eigenfunctions coalesce which, until this point,
had different growth rates and zero frequencies (see,
e.g., Refs. \cite{katobook,2004CzJPh..54.1039B}).  In the present case the
observed behavior can be explained only by means of two eigenfunctions
with different radial structures having complex-conjugate eigenvalues
with exactly (and not only approximately) the same growth rates and
opposite frequencies.  The appearance of such modes with
complex-conjugate eigenvalues is quite common in dynamo
theory. Dynamos exhibiting such exceptional points are, for example,
found in simple spherically symmetric $\alpha^2$ dynamo models with
radially varying $\alpha$ profiles, which served also as a simple
model to understand reversals of the geomagnetic field
\cite{2005PhRvL..94r4506S,2006E&PSL.243..828S}. The typical
exceptional point pattern for the formation of such modes modes is
nicely visible in the red curves of Fig. \ref{fig::gr_vs_rm}.

\subsubsection{Nonaxisymmetric perturbation}

In the following we present results from simulations with the
perturbation added to the basic flow at ${\rm{Rm}}=30$ where the
growth rates of the firts unperturbed eigenmodes are well separated and at
${\rm{Rm}}=120$ where the growth rates of the unperturbed eigenmodes
are rather close.  In both cases the unperturbed solutions do not
exhibit dynamo action. However, for ${\rm{Rm}}=120$ the unperturbed
solutions exhibits an oscillation of the amplitude with a frequency of
$|\omega_0|\approx 0.41$.


\newcommand{\siz}{0.25}
\newcommand{\jmp}{-0.5cm}
\newcommand{\nla}{-0.05cm}
\captionsetup[subfigure]{hangindent=20pt,margin=0.0cm,singlelinecheck=false,
                         format=plain,indention=0.3cm,justification=justified,
                         captionskip=-0.2cm,position=bottom}

\begin{figure}
\includegraphics[width=0.95\textwidth]{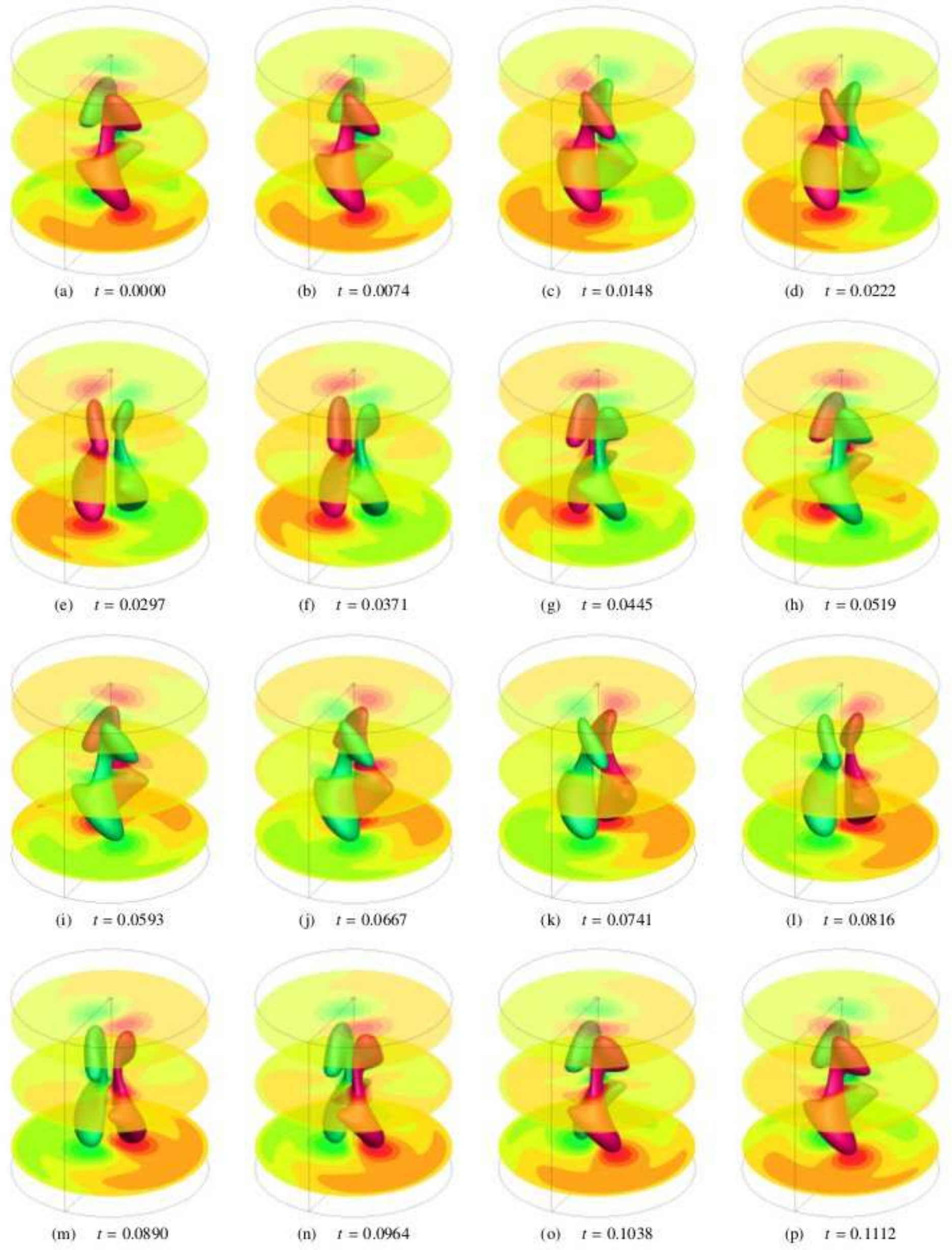}
\caption{
Evolution of the structure of the axial magnetic field component $B_z$
with velocity perturbation in the resonant case during one rotation
period.  The isosurfaces show time snapshots of $B_z$ at $50\%$ of the
time-averaged absolute value. The exponential growth has been
removed. ${\rm{Rm}}=120, \epsilon=0.3, \omega=1$.
\label{fig::timeseries_pert} 
}
\end{figure}

The addition of the perturbation slightly changes the geometric
structure of the eigenmodes (see Fig.~\ref{fig::timeseries_pert})
and has a clear impact on growth rate and/or frequency
(Figs.~\ref{fig::gr_vs_vortex1} and \ref{fig::gr_vs_vortex2}).  In
all cases, the growth rates tend to their unperturbed values for large
perturbation frequencies $|\omega|$ and we find sharp, narrow maxima
symmetrically distributed around the origin. At ${\rm{Rm}}=30$ we find
one maximum at $\omega=0$ [Figs.~\ref{sf::6a} and~\ref{sf::6b}]
whereas at ${\rm{Re}}=120$ we find two maxima around $\omega \approx
\pm 1$ with the exact location of the maxima slightly depending on the
amplitude of the perturbation $\epsilon$ [Figs~\ref{sf::6c}
and~\ref{sf::6d}].  In the vicinity of these maxima the growth rates
show a parabolic behavior and in the following we call this regime the
resonant regime [marked by the shaded regions for $\epsilon=0.3$ in
Figs.~\ref{sf::6b} and~\ref{sf::6d}].

\cleardoublepage

\renewcommand{\siz}{0.45}
\captionsetup[subfigure]{farskip=0.0cm,nearskip=0.1cm,margin=0.0cm,
                         singlelinecheck=false,format=plain,indention=0.0cm,
                         justification=justified,captionskip=0.0cm,position=bottom}
\begin{figure}[t!]
\subfloat[$\quad{\rm{Rm}}=30, \epsilon=0.3$.]
{\label{sf::6a}\includegraphics[width=\siz\linewidth]{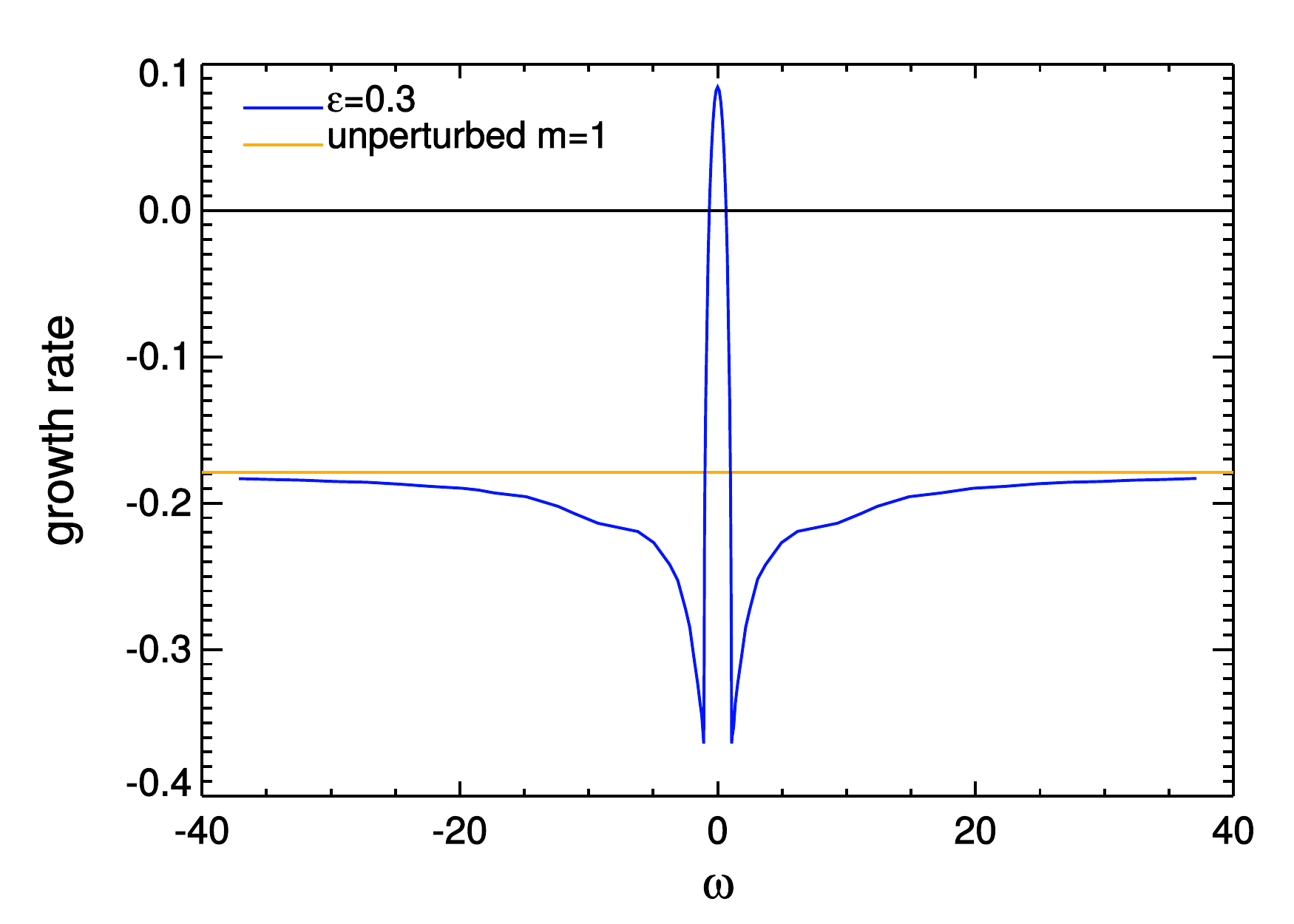}}
\subfloat[Zoom on the center of (a). The shaded region denotes the
regime with parametric resonance] 
{\label{sf::6b}\includegraphics[width=\siz\linewidth]{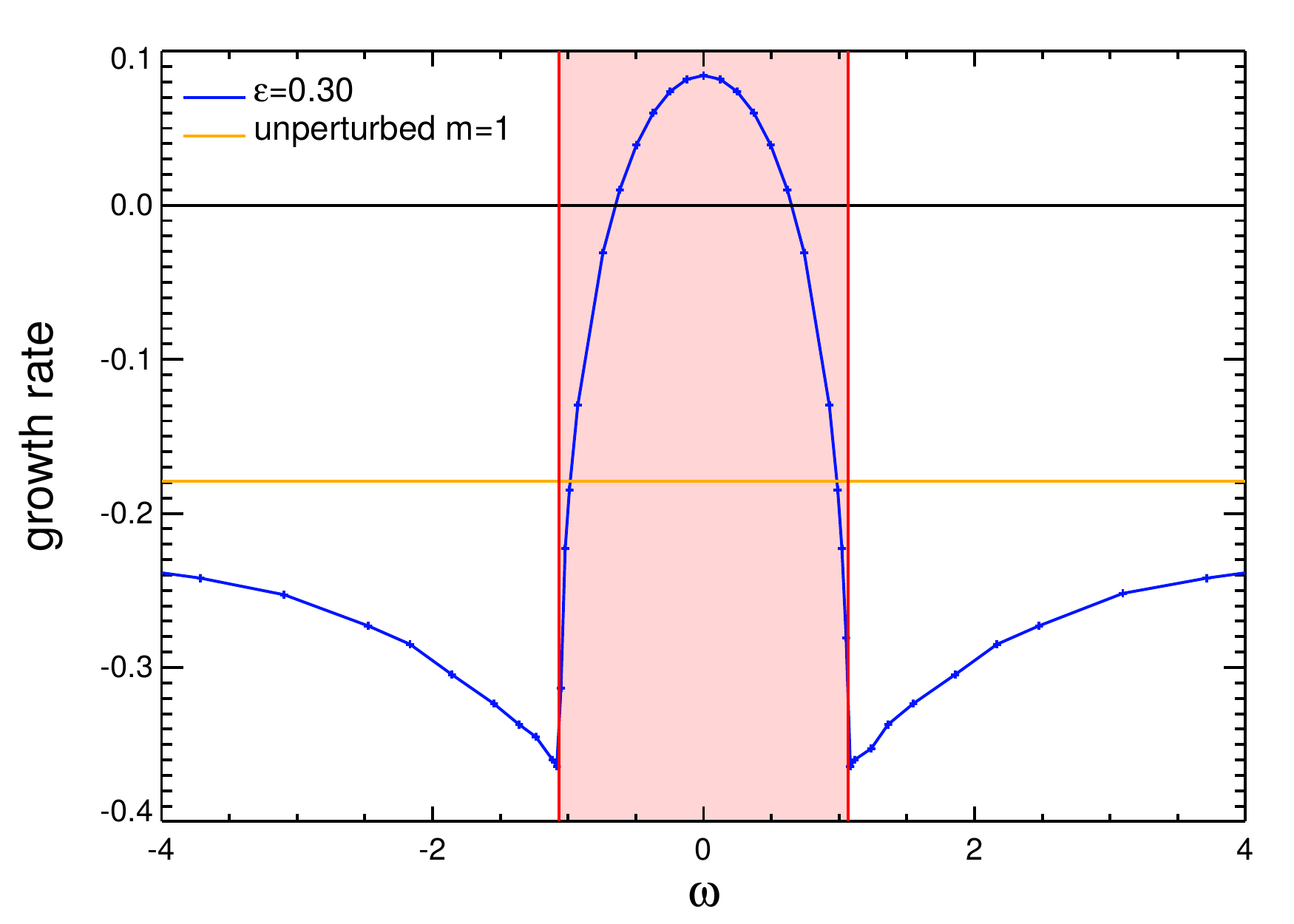}} 
\\[-0.2cm]
\subfloat[${\rm{Rm}}=120$. The shaded red region denotes the
regime with parametric amplification for $\epsilon=0.3$.]
{\label{sf::6c}\includegraphics[width=\siz\linewidth]{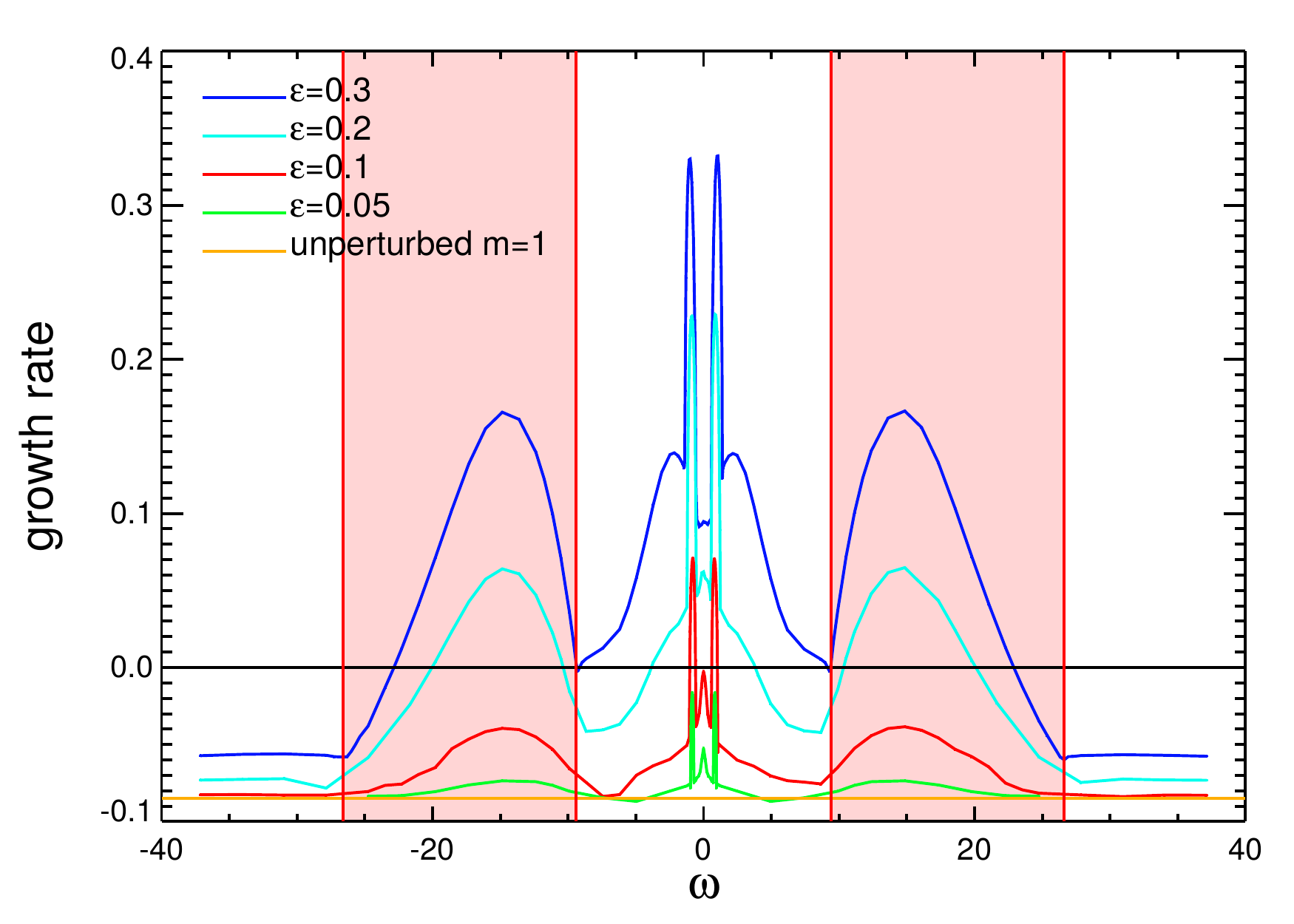}}
\subfloat[Zoom on the center of (c). The shaded red region denotes the
regime with parametric resonance for $\epsilon=0.3$.] 
{\label{sf::6d}\includegraphics[width=\siz\linewidth]{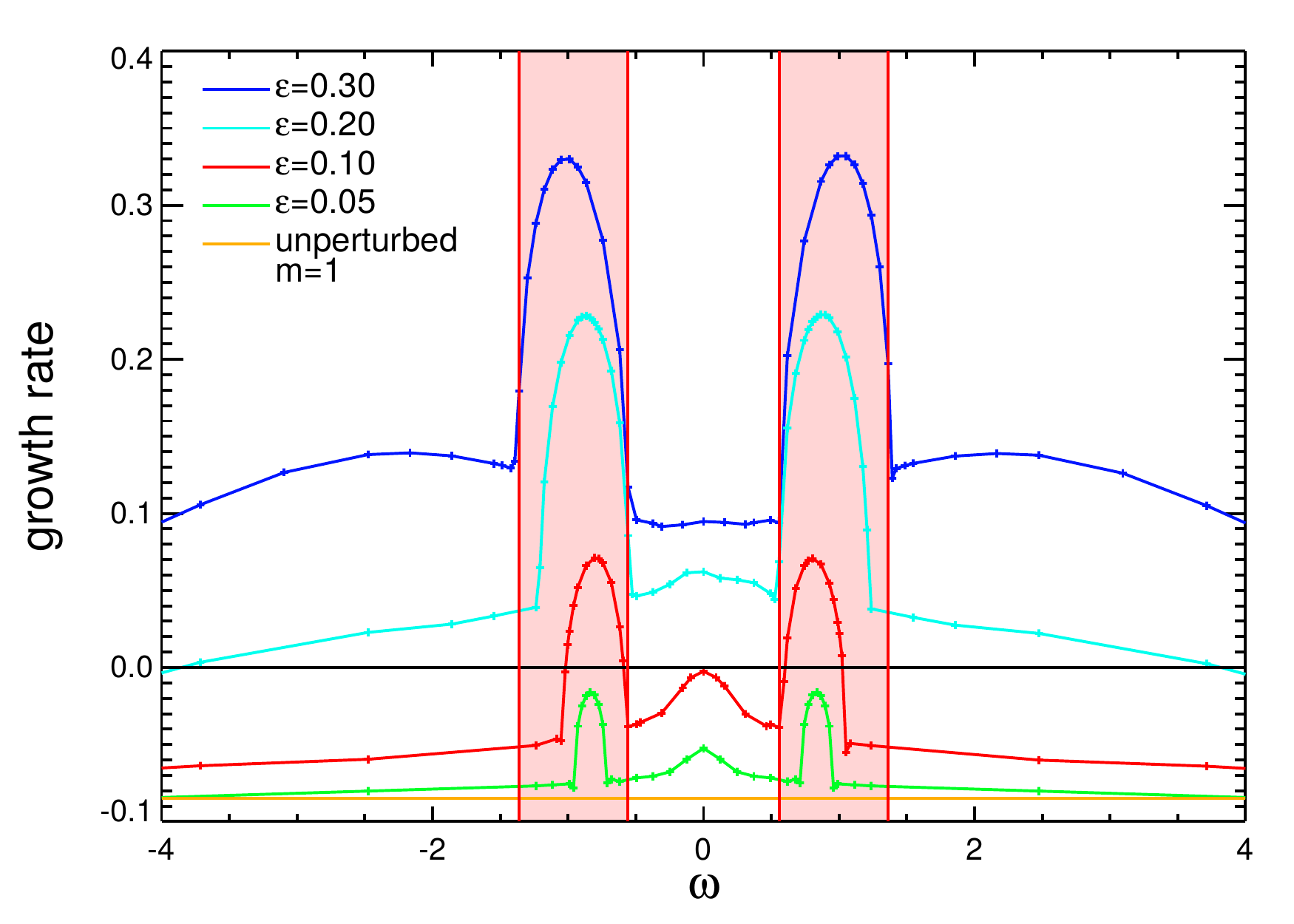}}
\caption{Growth rates against the perturbation frequency for
${\rm{Rm}}=30$ [(a), (b)] and ${\rm{Rm}}=120$ [(c), (d)] for various
perturbation amplitudes $\epsilon$.  The right-hand side provides a
detailed view of the behavior close to the origin.  
\label{fig::gr_vs_vortex1}}
\end{figure}
\begin{figure}[t!]
\subfloat[${\rm{Rm}}=30, \epsilon=0.3$]
{\label{sf::7a}\includegraphics[width=\siz\linewidth]{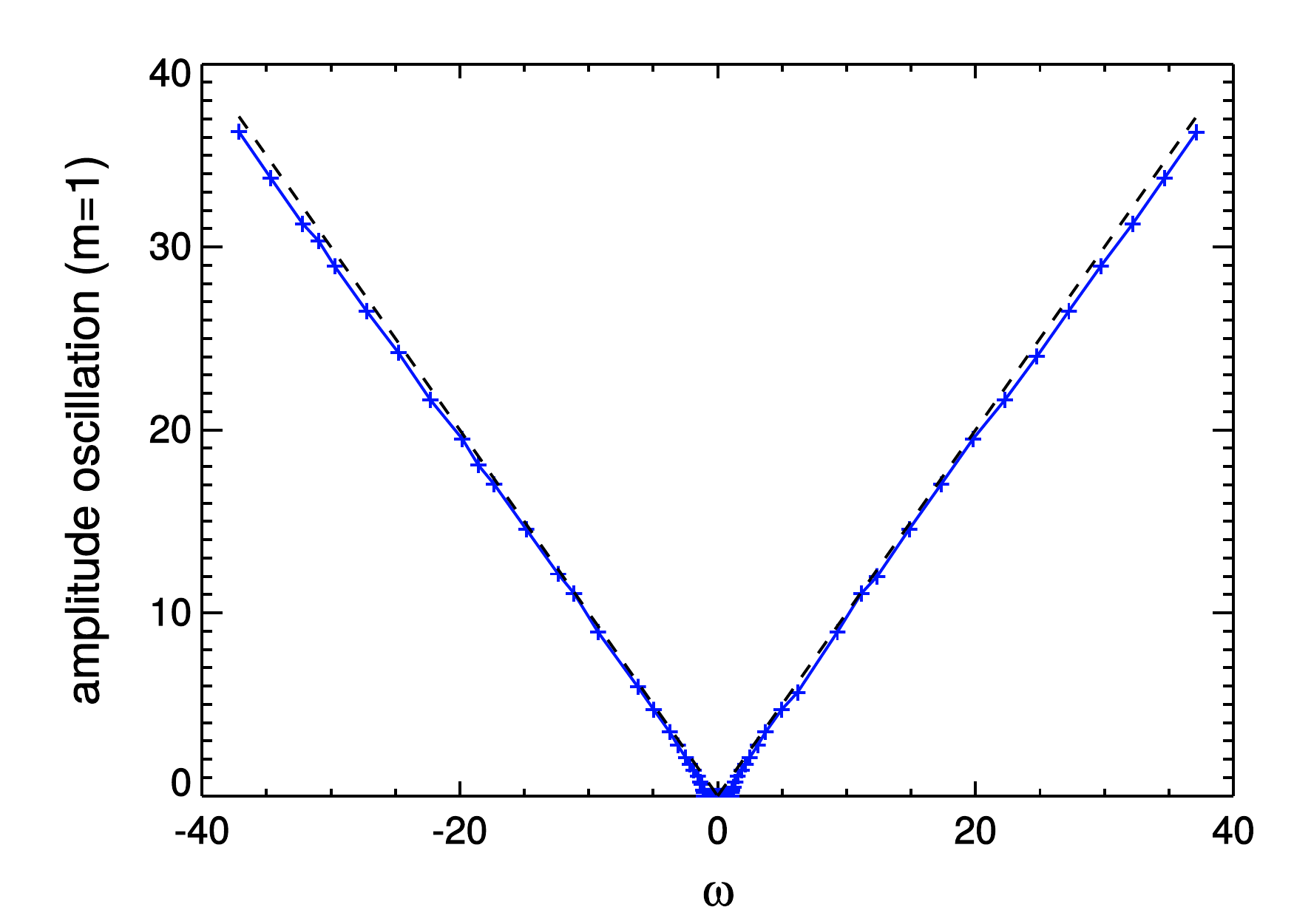}}
\subfloat[Zoom on the center of (a). The red shaded region denotes the regime with parametric resonance.]
{\label{sf::7b}\includegraphics[width=\siz\linewidth]{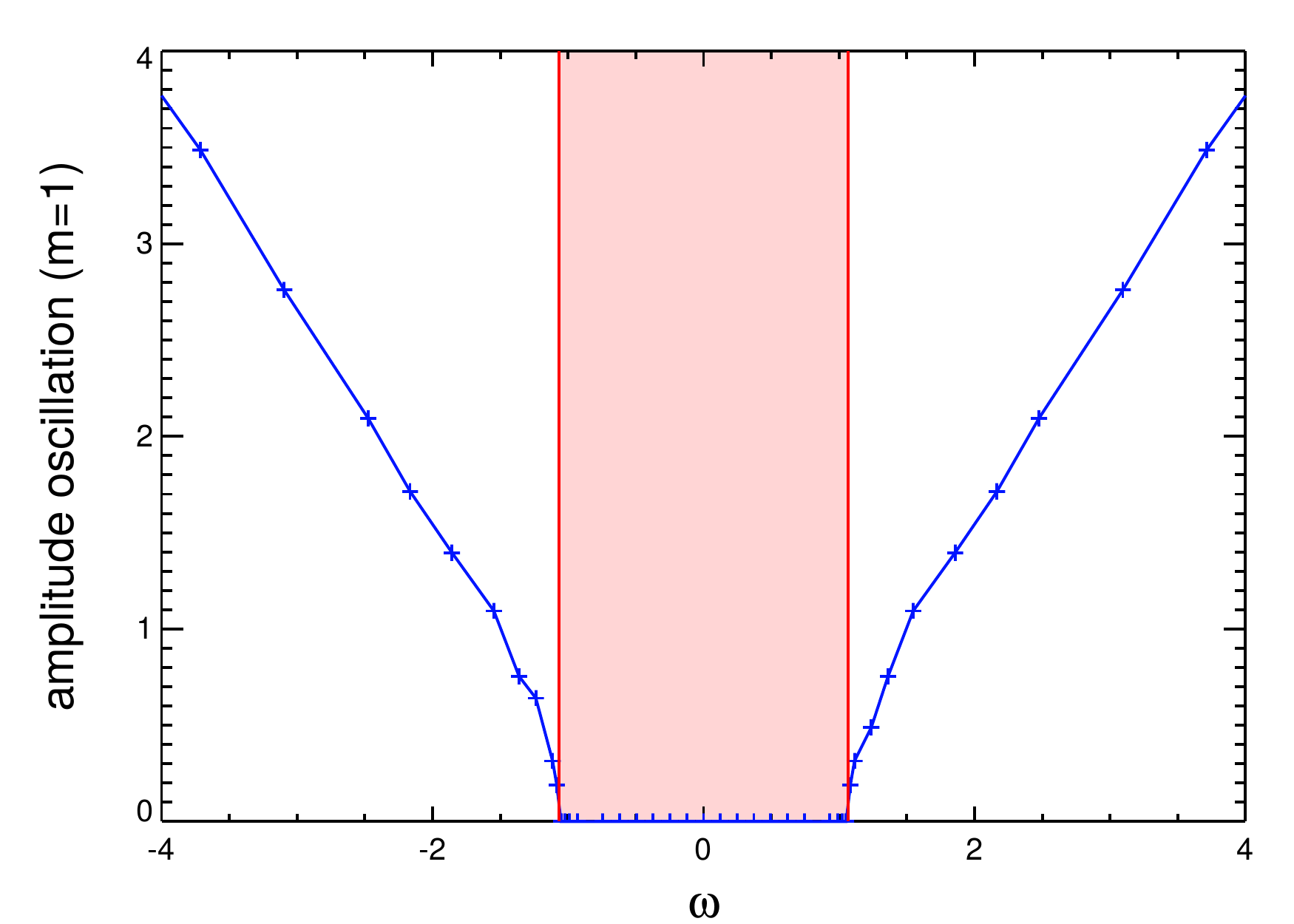}}
\\[-0.2cm]
\subfloat[${\rm{Rm}}=120, \epsilon=0.3$. The red shaded region denotes the regime
with parametric amplification that goes along 
with a small shift of the amplitude oscillation.]
{\label{sf::7c}\includegraphics[width=\siz\linewidth]{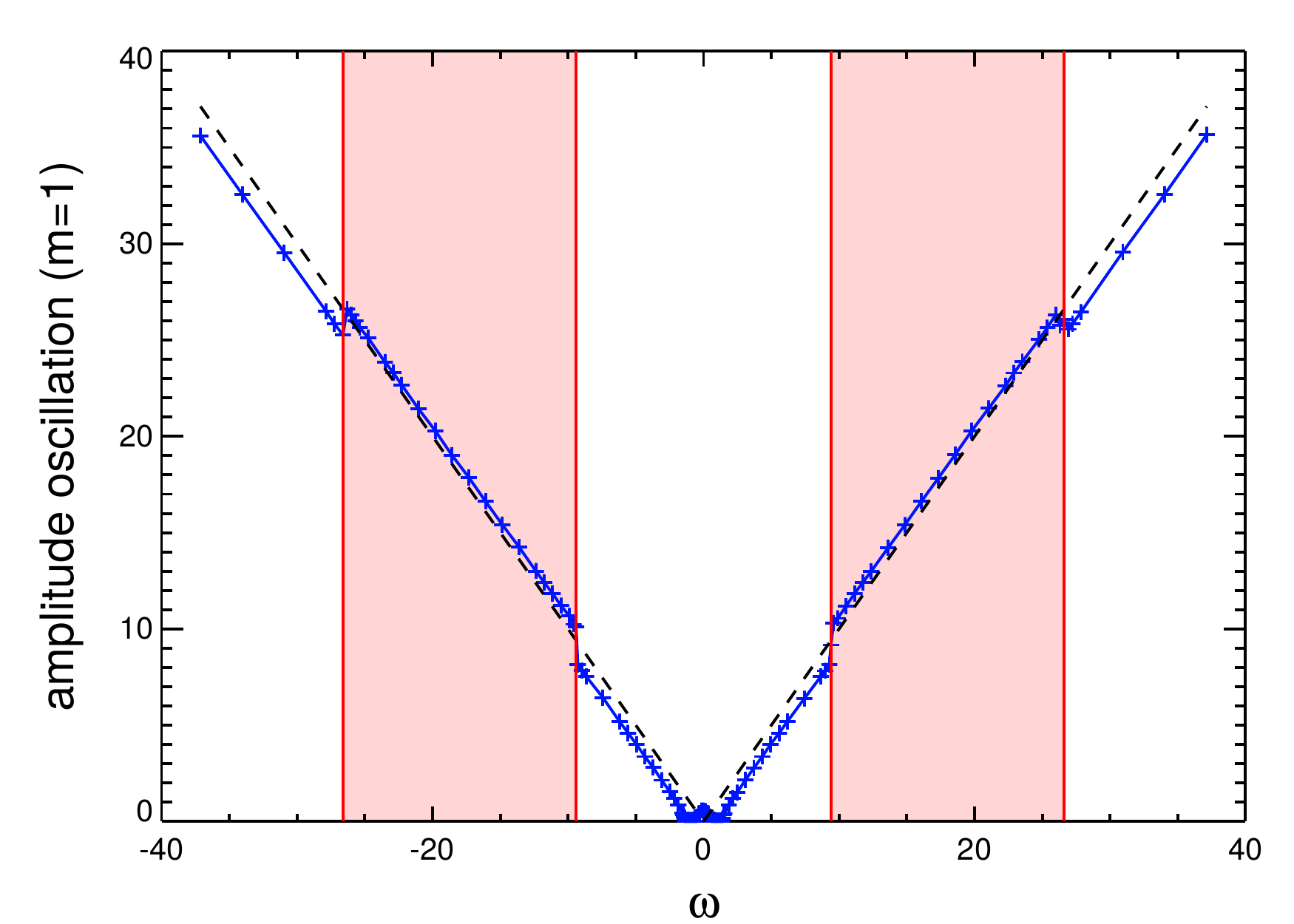}}
\subfloat[Zoom on the center of (c). The red shaded region denotes the
regime with parametric resonance at $\epsilon=0.3$.]
{\label{sf::7d}\includegraphics[width=\siz\linewidth]{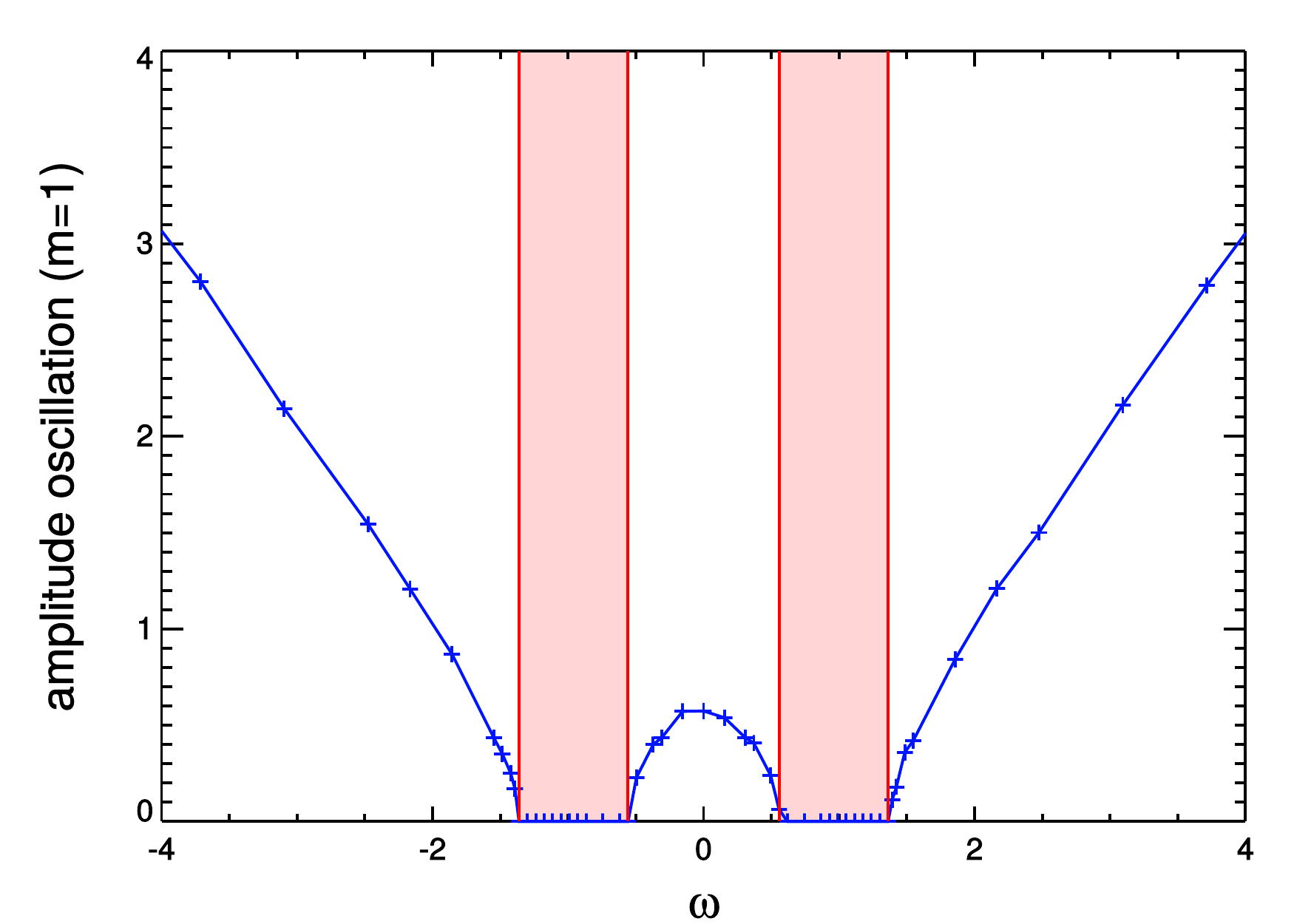}}
\caption{ 
Frequencies of the amplitude oscillations versus perturbation drift
frequency for ${\rm{Rm}}=30$ [(a), (b)] and ${\rm{Rm}}=120$ [(c), (d)].  
The right-hand side provides a detailed view on the behavior
close to the origin.  
\label{fig::gr_vs_vortex2}}
\end{figure}

Regarding the frequencies (Fig.~\ref{fig::gr_vs_vortex2}) we see an
amplitude oscillation except in the resonant regime [marked by the
shaded regions in the right panels of Figs.~\ref{sf::7b}
and~\ref{sf::7d}] where the amplitude oscillations vanish. In that
regime the field pattern {\it{exactly}} follows the perturbation
pattern, i.e., the magnetic field exhibits an azimuthal drift $\omega_{\rm{d}}$
that is determined by the frequency of the perturbation (phase
locking, $\omega_{\rm{d}} = \omega/2$ for the $\widetilde{m}=2$
perturbation; see time series in Fig.~\ref{fig::timeseries_pert}).
For large $|\omega|$
the azimuthal phases of magnetic field and velocity perturbation are
not connected and we see an amplitude oscillation of the magnetic
frield with an oscillation frequency $\omega_{\rm{a}}$ roughly proportional
to the excitation frequency $\omega_{\rm{a}}\sim \omega$.

In the following, we refer to the above described resonances as
parametric resonances that approximately fulfill the relation for the
location of the resonance maximum $\omega_{\rm{res}}=2\omega_0$ with
$\omega_0$ the eigenfrequency of the unperturbed problem which would
be $\omega_0 \approx 0.41$ for the case shown in Figs.~\ref{sf::7c}
and~\ref{sf::7d}.  The deviation from this relation increases for
increasing $\epsilon$, most probably because of the growing impact of
the perturbation on the base state.  A dependence of the location of
the maxima on the parameter $\epsilon$ is also observed in our model
introduced in Sec.~\ref{sec::ldm} (see Sec.~\ref{subsec::m3} and
Fig.~\ref{fig3d::gr_vs_omg_vs_eps}).

For weak amplitudes of the perturbation we see a weakly pronounced
third maximum between the main maxima [e.g., see red curve for
$\epsilon=0.1$ in Figs~\ref{sf::6c} and~\ref{sf::6d}].  Similar to
the solution shown in Fig.~\ref{fig::mpm3} below this interim
maximum is not connected to a regime with phase locking and vanishes
for increasing $\epsilon$.

At ${\rm{Rm}}=30$ we find no further features beside the parametric
resonances in the behavior of the growth rates. This can be attributed
to the large differences of the unperturbed growth rates (see
Fig.~\ref{fig::gr_vs_rm}, top panel) of the distinct azimuthal
eigenmodes which reduces the interaction between the individual
modes. The behavior changes at ${\rm{Rm}}=120$ where, in addition to
the sharp resonance peak, a number of further extremes emerge at
larger $|\omega|$ with a rather broad shape.  These maxima do not
occur in conjunction with phase locking, and the enhancement of dynamo
action is less intense than within the regime of parametric resonance.
In the following, we refer to this phenomenon as a {\it{parametric
amplification}} to distinguish from the mere parametric resonance.  In
the amplification regime, the frequency of the amplitude oscillation
varies roughly $\propto \omega$ with a small jump within the outer
maxima ([ee Fig.~\ref{sf::7c}].  This jump indicates that in this
regime a different eigenmode dominates compared to the mode around the
origin (with the parametric resonance).

The nearly perfect symmetry with respect to the origin differs from
the behavior found in a previous study \cite{2012PhRvE..86f6303G}
where an imposed breaking of the equatorial symmetry of the flow
causes a faster azimuthal drift of the leading dynamo mode
proportional to the degree of symmetry breaking and a corresponding
asymmetric behavior of the resonance maximum with respect to the
origin.

\begin{figure}[h!]
{\includegraphics[width=0.49\textwidth]{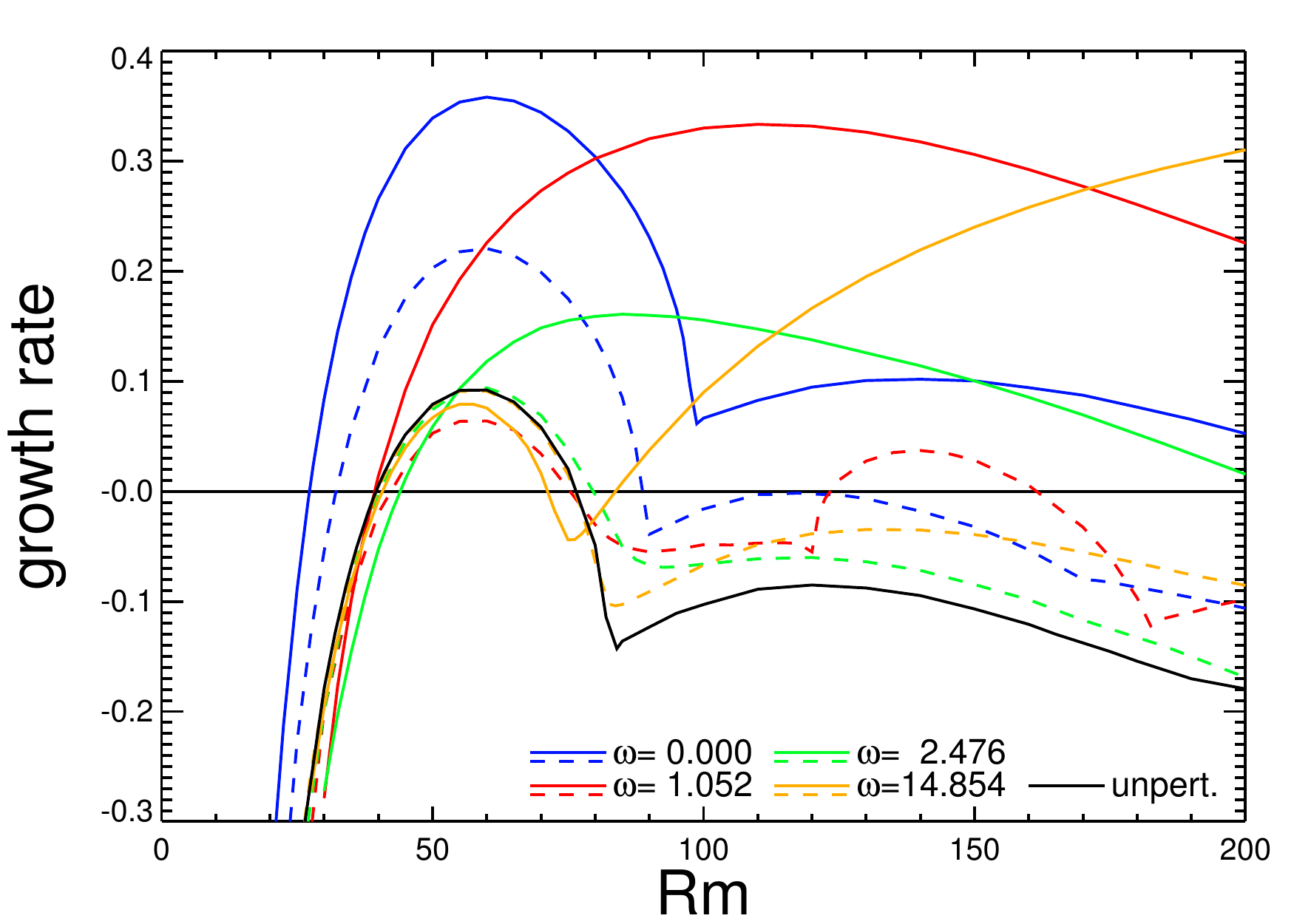}}
\caption{\label{fig::rmcrit_vs_omga}Growth rate versus ${\rm{Rm}}$
for $\epsilon=0.1$ (dashed curves) and $\epsilon=0.3$ (solid curves).  
The solid black curve denotes the case $\epsilon=0$.}
\end{figure}

\begin{figure}[h!]
\subfloat[\label{fig::rmcrit_vs_omgb}]  
{\includegraphics[width=0.49\textwidth]{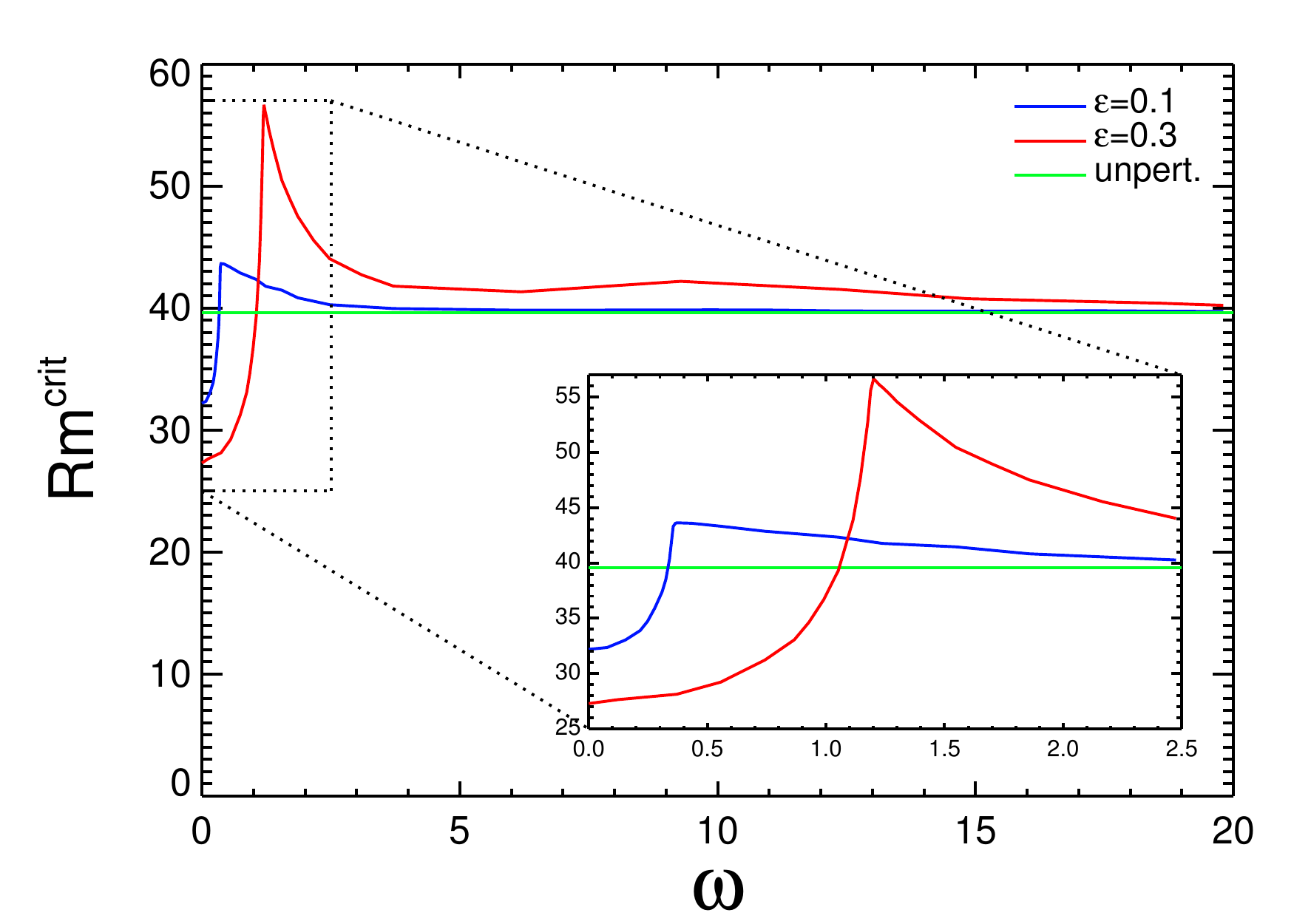}}
\subfloat[\label{fig::rmcrit_vs_omgc}]
{\includegraphics[width=0.49\textwidth]{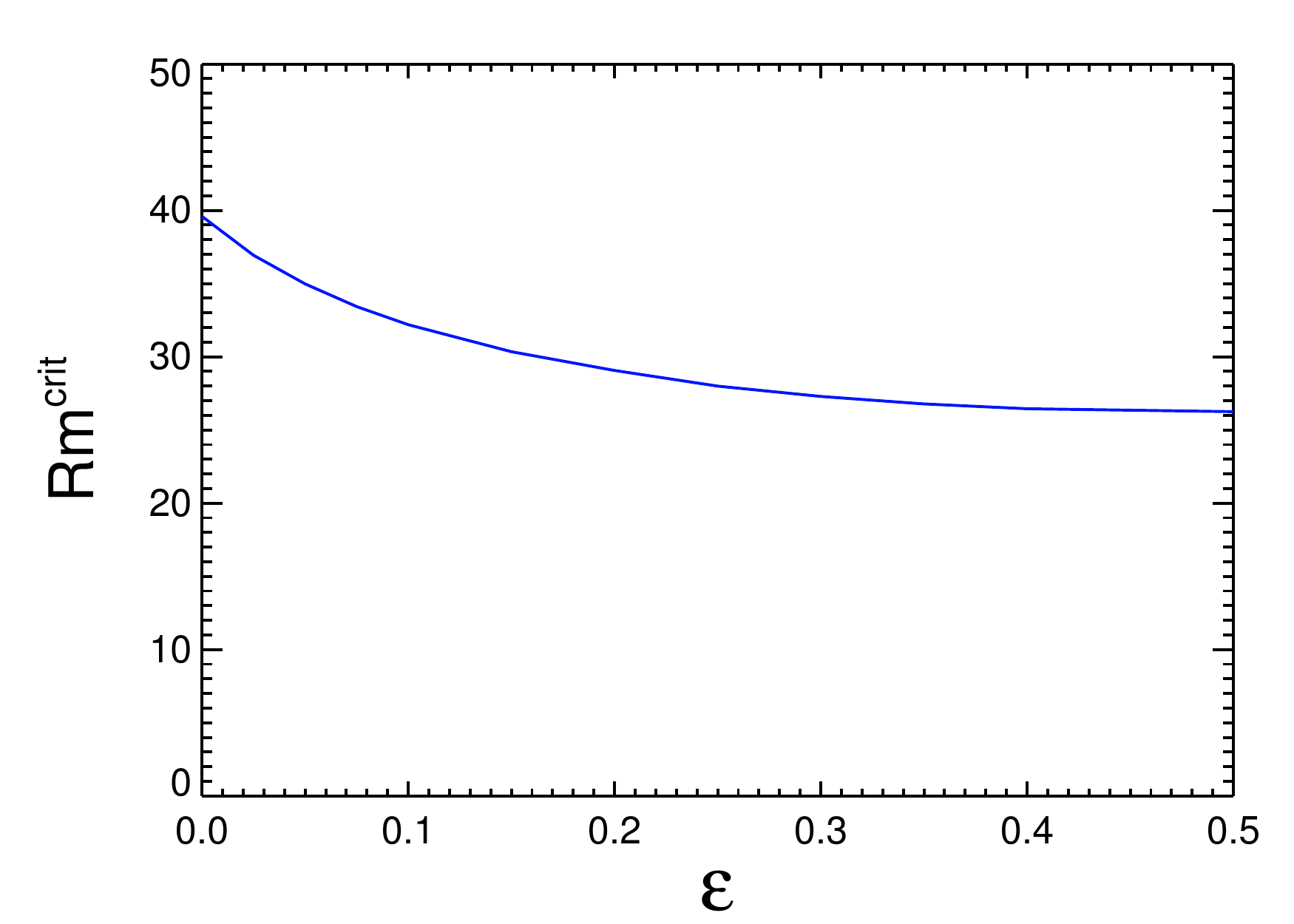}} 
\caption{
\label{fig::rmcrit_vs_omg} 
(a) Critical magnetic Reynolds numbers against
perturbation frequency $\omega$ for $\epsilon=0.1$ and
$\epsilon=0.3$. The horizontal green line denotes the case
$\epsilon=0$.  
(b) Critical magnetic Reynolds number versus perturbation
amplitude $\epsilon$ for $\omega=0$ where the maximum reduction of 
${\rm{Rm}}^{\rm{crit}}$ takes place.
}
\end{figure}

The boost of growth rates seen in Fig.~\ref{fig::gr_vs_vortex1} is
also reflected in the behavior of the critical magnetic Reynolds
number ${\rm{Rm}}^{\rm{crit}}$ that is required for the onset of
dynamo action. There are considerable variations of the growth rates
with ${\rm{Rm}}$ and with $\omega$ (see
Fig.~\ref{fig::rmcrit_vs_omga}). However, when restricting
${\rm{Rm}}$ to the crucial regime in the vicinity of the dynamo
threshold, the growth rates do not change much with the perturbation
frequency, and nearly all curves around the onset of dynamo action
take the same evolution (Fig.~\ref{fig::rmcrit_vs_omga}).  A
beneficial impact on dynamo action is found only for small $\omega$
(blue curves in Fig.~\ref{fig::rmcrit_vs_omga}). Accordingly, the
most significant decrease in ${\rm{Rm}}^{\rm{crit}}$ is found for
$\omega=0$ [Fig.~\ref{fig::rmcrit_vs_omgb}], and the reduction can
reach considerable values of up to $30\%$ [from
${\rm{Rm}}^{\rm{crit}}=39.6$ at $\epsilon=0$ to
${\rm{Rm}}^{\rm{crit}}=26.3$ at $\epsilon=0.5$,
Fig.~\ref{fig::rmcrit_vs_omgc}].
\renewcommand{\siz}{0.3cm}
\captionsetup[subfigure]{farskip=-0.0cm,nearskip=-0.0cm,margin=1.0cm,
                         singlelinecheck=false,format=plain,
                         indention=0.0cm,hangindent=0.0cm,
                         justification=justified,captionskip=0.1cm,
                         position=bottom} 
The behavior of ${\rm{Rm}}^{\rm{crit}}$ with the sharp reduction
around $\omega\approx 0$ essentially reflects the structure of the
growth rates at ${\rm{Rm}}=30$ with one sharp peak around $\omega=0$
as is shown in Figs.~\ref{sf::6a} and~\ref{sf::6b}. The secondary
regimes with enhanced growth rates that emerge at larger Reynolds
number and larger perturbation frequencies hardly impact the
primary onset of dynamo action around ${\rm{Rm}}\approx 30 \dots 40$
but can be related to a second dynamo regime with a threshold around
${\rm{Rm}}\approx 120$ (see, e.g., solid red curve in
Fig.~\ref{fig::rmcrit_vs_omga}) where no dynamo action is obtained
at all without perturbation.  The existence of a second regime with
dynamo action is less important in the context of experimental dynamos
but can be associated with subcritical phenomena in fully nonlinear
simulations and may be causative for possible deviations from ideal
scaling laws by providing additional energy for the dynamo process
\cite{2015AnRFM..47..163L}.


\section{Low-dimensional model}\label{sec::ldm}

\subsection{Amplitude equation}

In order to find a qualitative explanation for the behavior observed
in the simulations presented above, we derive a simple low-dimensional
analytical model for the amplitude of distinct dynamo modes.  The
framework presented in this section must be seen as a toy model, where
the parameters are chosen to reproduce qualitatively the observed
features in the former section.  Our model is based on an azimuthal
decomposition of the global magnetic field. In doing so, the dynamics
of each azimuthal mode is determined by one eigenmode with the largest
growth rate. The decomposition yields a linear system of equations
that couples all odd azimuthal wave numbers. From this model, we find
different scenarios leading to an enhancement of the growth rate,
which are not described by the classical parametric instability.
Although we assume a cylindrical domain, most of the following
considerations require only the existence of a well-defined symmetry
axis in order to allow a reasonable decomposition into azimuthal field
modes and are also valid, e.g., in spherical geometry.

We start again with the magnetic induction
equation~(\ref{eq::induction}). We assume a prescribed velocity field
$\vec{U}=\vec{U}(\vec{r},t)$ that is composed of an axisymmetric
component $\vec{U}_0$ and a periodic perturbation with the azimuthal
wave number $\widetilde{m}$ propagating in the azimuthal direction
with the frequency $\omega$:
\begin{flalign}
\!\!\!\vec{U}(\vec{r},t)\!=\!\vec{U}_{0}(r,z)
\!+\!\epsilon\left[\vec{u}_{\widetilde{m}}(r,z)
e^{i({\widetilde{m}}\varphi+\omega t)}
\!+\vec{u}_{-{\widetilde{m}}}(r,z)
e^{-i({\widetilde{m}}\varphi+\omega t)}\right], &
\label{eq::velocity}
\end{flalign}
where the parameter $\epsilon$ characterizes the amplitude of the
perturbation.  We now reduce the induction equation to a system of
equations for the amplitudes of azimuthal magnetic field modes
characterized by a wave number $m$. We assume that the magnetic field
$\vec{B}$ can be decomposed according to
\begin{equation}
\vec{B}=\sum\limits_{-M}^{M}\hat{b}_m^{\mbox{}}(t)\vec{b}_m(r,z)e^{im\varphi} 
\label{eq::modeldecomp}  
\end{equation}
with the complex amplitude $\hat{b}_m(t)\in \mathbb{C}$ that fulfills
$\hat{b}_m=\hat{b}_{-m}^*$ in order to ensure a real valued magnetic
field. We further suppose a normalization for the function $\vec{b}_m$
given by  
\begin{equation}
\iint\vec{b}_m^{\mbox{ }}\cdot\vec{b}_m^*rdrdz=1.
\end{equation}

The decomposition~(\ref{eq::modeldecomp}) assumes that the individual
modes are modulated only by a simple temporal varying amplitude which
in general is not correct (if the linear operator is nonnormal as in
case of the induction equation).  Furthermore, we consider only the
leading eigenmode for each azimuthal wave number, and we additionally
suppose that only one single mode is close to be
unstable. Consequently, the decomposition~(\ref{eq::modeldecomp}) into 
azimuthal modes can at least qualitatively be justified.

The coefficient $\hat{b}_m(t)$ can be computed by forming the scalar
product
\begin{equation}
\hat{b}_m(t)=\big<\vec{B},\vec{b}_m\big>
=\iiint\vec{B}\cdot\vec{b}^*_me^{-im\varphi}dV
\end{equation}
so that the temporal evolution of $\hat{b}_m(t)$ is governed by 
\begin{equation}
\frac{d}{dt}\hat{b}_m(t)=\frac{d}{dt}\bigg<\vec{B},\vec{b}_m\bigg>=
\bigg<\nabla\times\big(\vec{U}\times\vec{B}\big)+\eta\Delta\vec{B},\vec{b}_m\bigg>.
\end{equation}
Using the velocity field~(\ref{eq::velocity}) with an explicit
$\widetilde{m}=2$ distortion yields the evolution equation for the
amplitude $\hat{b}_m(t)$ with an explicit coupling to the magnetic modes
$\hat{b}_{m\pm 2}(t)$:
\begin{equation}
\frac{d}{dt}\hat{b}_m(t)=\alpha_{{m},{m}}\hat{b}_m
+\epsilon\left(e^{-i\omega t}\alpha_{{m},{m}+2}\hat{b}_{m+2}+
e^{i\omega t} \alpha_{{m},{m}-2}\hat{b}_{m-2}\right)\label{eq::linsystem}
\end{equation}
with
\begin{eqnarray}
\alpha_{{m},{m}}     & \! = \! & \!
\iiint\!\!\bigg(\nabla\!\!\times\!\big(\vec{U}_{0}\!\times\!\vec{b}_me^{im\varphi}\big) 
\!+\!\eta\Delta\vec{b}_me^{im\varphi}\!\bigg)\!\cdot\!\vec{b}_m^*e^{-im\varphi}dV,\nonumber
\\
&&\\[-0.2cm]
\alpha_{{m},{m}\pm2} & \! = \! & 
\! \iiint\!\!\bigg(\nabla\!\times\!\big(\vec{u}_{\mp 2}\times 
\vec{b}_{m\pm 2}e^{im\varphi}\big)\bigg)\cdot\vec{b}^*_me^{-im\varphi}dV.\nonumber
\end{eqnarray}
The magnetic field $\vec{B}$ must be a real valued function which
requires $\hat{b}_m=\hat{b}_{-m}^*$ and hence entails
\begin{equation}
\alpha_{{m},{m}}=\alpha^*_{-{m},-{m}} 
\mbox{  and  } 
\alpha_{{m},{m}\pm 2}=\alpha^*_{-{m},-{m}\mp 2}. 
\end{equation}
We now write system~(\ref{eq::linsystem}) in an explicit form
\begin{flalign}
\begin{array}{lclclcl}
& \vdots & & & & \\
\displaystyle\frac{d}{dt} \hat{b}_{-5}  \!\! & \!\! = \!\! &  \!
\epsilon e^{i\omega t}\alpha_{-5,-7}\hat{b}_{-7} \!\! & \! + \! 
& \!\! \alpha_{-5,-5}\hat{b}_{-5} \!\! & \! + \! & \!\!\epsilon e^{-i\omega t}\alpha_{-5,-3}\hat{b}_{-3}, 
\\[0.25cm]
\displaystyle\frac{d}{dt} \hat{b}_{-3} \!\! & \!\! = \!\! &  \!
\epsilon e^{i\omega t}\alpha_{-3,-5}\hat{b}_{-5} \!\! & \! + \! 
& \!\! \alpha_{-3,-3}\hat{b}_{-3} \!\! & \! + \! &  \!\!\epsilon e^{-i\omega t}\alpha_{-3,-1}\hat{b}_{-1},
\\[0.25cm]
\displaystyle\frac{d}{dt} \hat{b}_{-1} \!\! & \!\! = \!\! &  \!
\epsilon e^{i\omega t}\alpha_{-1,-3}\hat{b}_{-3} \!\! & \! + \! 
& \!\! \alpha_{-1,-1}\hat{b}_{-1} \!\! & \! + \! & \!\! \epsilon e^{-i\omega t}\alpha_{-1,1}\hat{b}_{1},
\\[0.25cm]
\displaystyle\frac{d}{dt} \hat{b}_{1} \!\! & \!\! = \!\! &  \!
\epsilon e^{i\omega t}\alpha_{1,-1}\hat{b}_{-1} \!\! & \! + \! 
& \!\! \alpha_{1,1}\hat{b}_{1} \!\! & \! + \! &  \!\!\epsilon e^{-i\omega t}\alpha_{1,3}\hat{b}_{3},
\\[0.25cm]
\displaystyle\frac{d}{dt} \hat{b}_{3} \!\! & \!\! = \!\! & \! 
\epsilon e^{i\omega t}\alpha_{3,1}\hat{b}_{1} \!\! & \! + \! 
& \!\! \alpha_{3,3}\hat{b}_{3} \!\! & \! + \! & \!\! \epsilon e^{-i\omega t}\alpha_{3,5}\hat{b}_{5},
\\[0.25cm]
\displaystyle\frac{d}{dt} \hat{b}_{5} \!\! & \!\! = \!\! &  \!
\epsilon e^{i\omega t}\alpha_{5,3}\hat{b}_{3 \!\!} & \! + \! 
& \!\! \alpha_{5,5}\hat{b}_{5} \!\! & \! + \! &  \!\!\epsilon e^{-i\omega t}\alpha_{5,7}\hat{b}_{7},
\\[0.25cm]
& \vdots & & & & \\
\end{array}\label{eq::linsystem_exp}
\end{flalign}
which can be written in compact form as a matrix equation%
\begin{equation}
\frac{d}{dt}{B}(t) = A(t)B(t)\label{eq::matrix}
\end{equation}
with
$B(t)=(\cdots,\hat{b}_{-m},\hat{b}_{-m+2},
\dots,\hat{b}_{-1},\hat{b}_{1},\dots,\hat{b}_{m-2},
\hat{b}_{m}\cdots)^{T}$ 
and a tridiagonal matrix $A(t)$ given by
\begin{equation}
A=
\left(
\begin{array}{cccccccccc}
&  &  \ddots &  &  &  &  &  & &  \\
\cdots&0 & \delta_3^* f^*_t  & \alpha_3^* & \delta_2^* f_t & 0 & \cdots &  &  & \\
&\cdots &  0   & \delta_1^*f^*_t & \alpha_1^* & \gamma^*f_t & 0  & \cdots & &\\
& & \cdots &  0   & \gamma f^*_t  & \alpha_1   & \delta_1f_t & 0 & \cdots & \\
&  &    &   \cdots    & 0 & \delta_2f^*_t & \alpha_3  & \delta_3 f_t & 0 & \cdots \\
&  &    &       &  &  &   &\ddots  & &  \\
\end{array}
\right),
\end{equation}
where we used the abbreviations $\alpha_j=\alpha_{j,j}$,
$\gamma=\alpha_{1,-1}$, $\delta_1=\alpha_{1,3}$,
$\delta_2=\alpha_{3,1}$, 
$\delta_3=\alpha_{3,5}$ and $f_t=\epsilon e^{-i\omega t}$.  
The nondiagonal elements of $A$ represent the
coupling between the azimuthal modes introduced by the
nonaxisymmetric flow perturbation.  In the unperturbed case these
nondiagonal elements vanish and all azimuthal modes $\hat{b}_m(t)$
decouple.  In that case the diagonal elements are the eigenvalues
\begin{equation}
\alpha_{j,j}=\alpha_j= \alpha_j^r+i\alpha_j^i,\nonumber
\end{equation}
where $\alpha^r_j$, the real part of $\alpha_j$,  denotes the growth rate
of the unperturbed case and implicitly incorporates the magnetic
Reynolds number, and $\alpha^i_j$, the imaginary part of
$\alpha_j$, denotes the frequency of the unperturbed case.
The off-diagonal parameters, originally labeled with
$\alpha_{m,m\pm 2}$, prescribe the interaction of the modes with
themselves and/or with adjacent modes. 

In case of a perturbation with an azimuthal wave number
$\widetilde{m}=2$ we achieve two classes of magnetic modes which
incorporate even azimuthal wave numbers and odd azimuthal wave
numbers. Here only the second class with odd wave numbers is relevant
because the even modes typically decay on a faster time scale, and we
could not find any growing solutions with even azimuthal symmetry in
our simulations presented above.

Note that the approach outlined above does not constitute a
perturbation theory in the strict mathematical sense. The
nonaxisymmetric spatio-temporally periodic perturbation changes the
structure of the solution (the geometry of the eigenvector) by
coupling different azimuthal modes, and the limiting case
$\epsilon\rightarrow 0$ is different from the case $\epsilon=0$.  This
means that the addition of a propagating wave with $\epsilon\neq 0$
changes the shape of the modes $\vec{b}_m$ in comparison with the
simple axisymmetric case so that the coefficients $\alpha_{m,m}$
depend implicitly on $\epsilon$ and $\omega$.  However, in the limit
$\epsilon\rightarrow 0$ we assume that $\alpha_{m,m}(\epsilon\neq 0)
\approx \alpha_{m,m}(\epsilon=0)$ and that the addition of a
nonaxisymmetric perturbation does not change the eigenvector of the
unperturbed problem.

\subsection{Direct solution for truncation at $m=\pm 1$}\label{sec::m=1}

In order to solve the set of ordinary differential equations for the
eigenvalues, i.e., growth rate and frequency, the system must be
truncated at a fixed azimuthal wave number $M$.  We start with the
most severe approximation and cut the system~(\ref{eq::linsystem_exp})
at $m=\pm1$. We obtain two coupled differential equations
\begin{eqnarray}
\frac{d\hat{b}_{-1}}{dt} & = &
\alpha^*\hat{b}_{-1}
+\epsilon\gamma^*e^{-i\omega t}\hat{b}_{1}, \label{eq::simple_m1_1}
\\
\frac{d\hat{b}_{1}}{dt} & = &
\epsilon\gamma e^{i\omega t}\hat{b}_{-1}+\alpha\hat{b}_{1},
\label{eq::simple_m1_2}
\end{eqnarray}
where we used the abbreviations
\begin{equation}
\begin{array}{rcccl}
\alpha & = & \alpha_{1,1} & = & \alpha^*_{-1,-1},
\\
\gamma & = & \alpha_{1,-1} & = & \alpha^*_{-1,1}.
\end{array}
\end{equation}
Taking the derivative of Eq.~(\ref{eq::simple_m1_2}) yields
\begin{equation}
\displaystyle\frac{d^2\hat{b}_1}{dt^2}=\epsilon\gamma e^{i\omega t} 
\left(i\omega \hat{b}_{-1}
+\frac{d\hat{b}_{-1}}{dt}\right)
+\alpha\frac{d\hat{b}_1}{dt}. 
\end{equation} 
We use~(\ref{eq::simple_m1_1}) to replace
$\displaystyle\frac{d\hat{b}_{-1}}{dt}$ and
then~(\ref{eq::simple_m1_2}) to ultimately get rid off 
$\hat{b}_{-1}$ which gives 
\begin{eqnarray}
\frac{d^2\hat{b}_1}{dt^2} & = & \left[-i\omega\alpha-|\alpha|^2
+\epsilon^2|\gamma|^2\right]\hat{b}_1
+\left(i\omega+2{\alpha_r}\right)\frac{d\hat{b}_1}{dt}.
\end{eqnarray}
We search for solutions $\hat{b}_1(t) \propto e^{\sigma t}$ which yields
the relation
\begin{equation}
\sigma^2- \left(i\omega+2{\alpha_r}\right) \sigma -
\left[-i\omega\alpha-|\alpha|^2+\epsilon^2|\gamma|^2\right] =0 
\end{equation}
with the solutions
\begin{equation}
\sigma_{1,2} =
{\alpha^r}+i\frac{\omega}{2}\pm\frac{1}{2}
\sqrt{4\epsilon^2|\gamma|^2-\left({2\alpha^i-\omega}\right)^2}. 
\label{eq::sol_with_coupling}
\end{equation}

 We distinguish two cases in dependence of the sign of
$4\epsilon^2|\gamma|^2-(2\alpha^i-\omega)^2$.

\noindent
(1) For $\displaystyle|2\alpha^i-{\omega}| < 2\epsilon|\gamma|$ we
obtain one frequency
\begin{equation}
\sigma^i=\displaystyle\frac{\omega}{2}
\end{equation}
and two different growth rates
\begin{equation}
\sigma^r_{1,2}=\alpha^r\pm\frac{1}{2}\sqrt{4\epsilon^2|\gamma|^2
-\left(2\alpha^i-\omega\right)^2}.
\end{equation}
For small deviations of the forcing frequency
${\omega}$ from twice the unperturbed frequency $2\alpha^i$ we
can write for the growth rate:
\begin{equation}
\sigma^r_{1,2}\approx\alpha^r\pm\epsilon|\gamma|
\left(1-\frac{(2\alpha^i-\omega)^2}{4\epsilon^2|\gamma|^2}\right),  
\end{equation}
and we find two extrema for the growth rate
$\sigma^{r,{\rm{max}}}_{1,2}=\alpha^r\pm\epsilon|\gamma|$ when the
perturbation frequency is equal to twice the intrinsic frequency
$\omega=2\alpha^i$. Around this maximum the growth rate is locally
parabolic.

\noindent
(2) For $\displaystyle |2\alpha^i-\omega| > 2\epsilon|\gamma|$ we get
only one growth rate
\begin{equation}
\sigma^r=\alpha^r
\end{equation}
and two frequencies
\begin{equation}
\sigma^i_{1,2}=\frac{\omega}{2}
\pm\frac{1}{2}\sqrt{\left(2\alpha^i-\omega\right)^2-4\epsilon^2|\gamma|^2}.  
\end{equation}
For large $\omega$ these frequencies tend to $\sigma^i_{1}=\omega$ and
$\sigma^i_{2}=\alpha^i$. 

\begin{figure}
\centering
\includegraphics[width=0.49\textwidth]{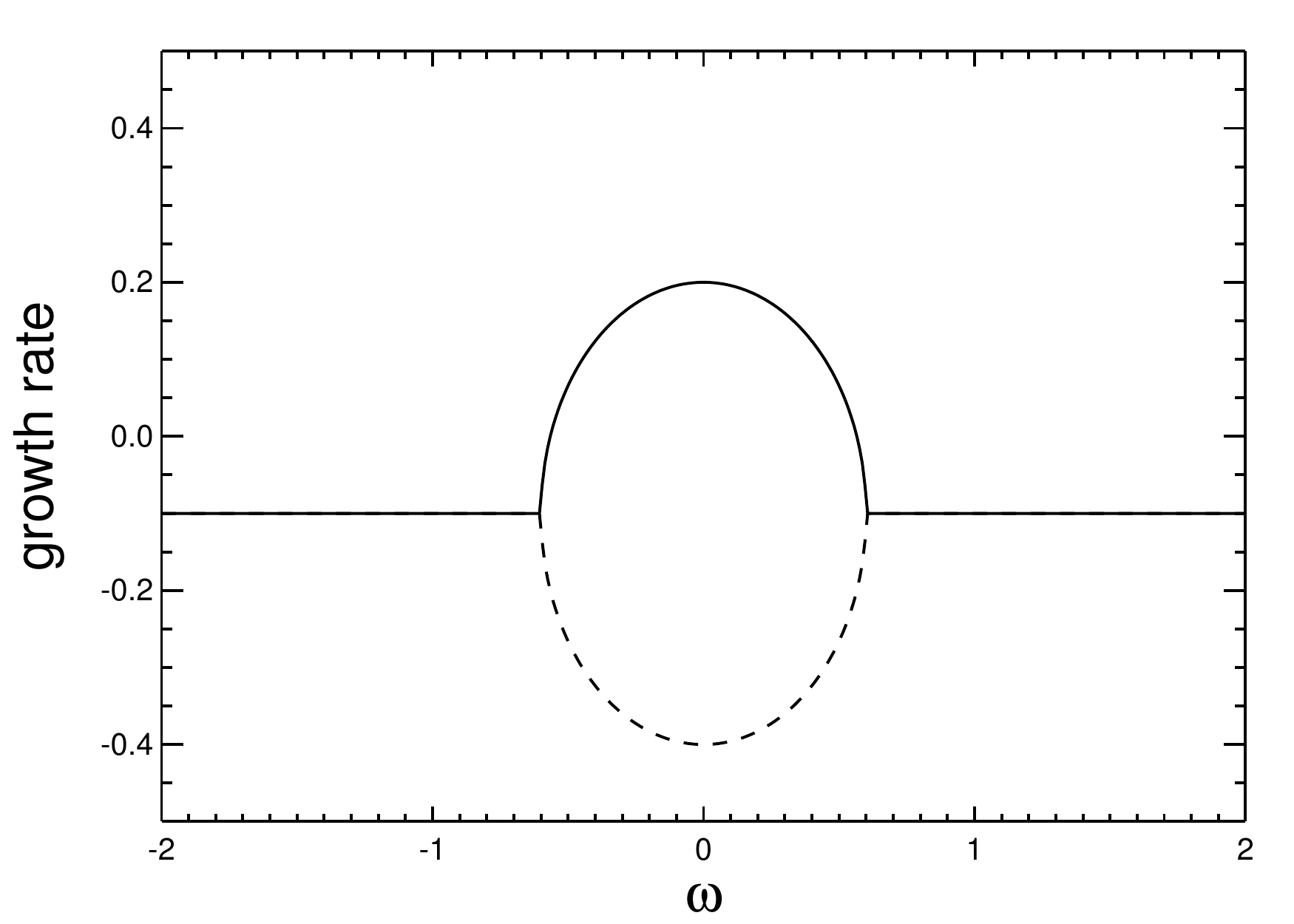}
\\
\includegraphics[width=0.49\textwidth]{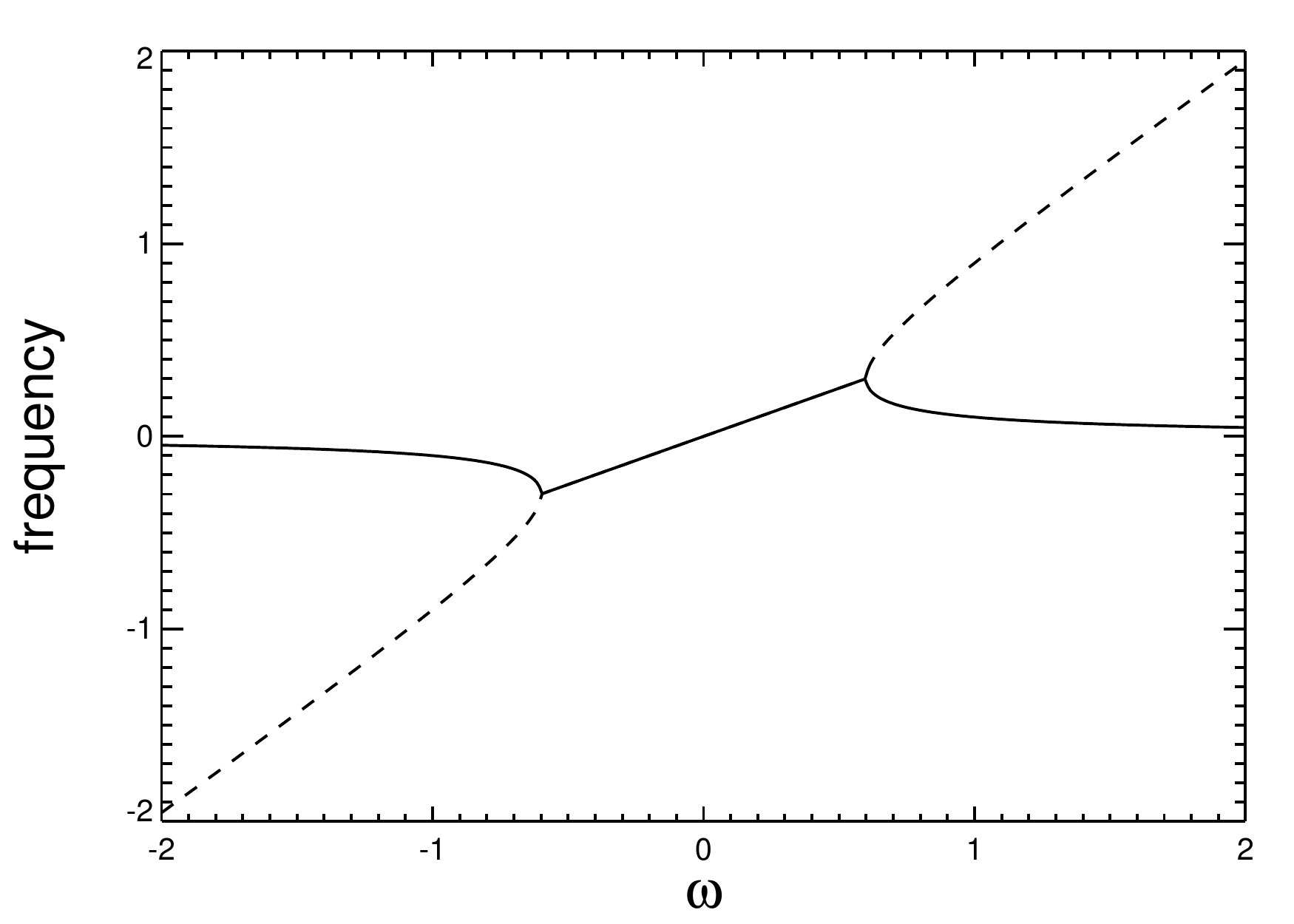}
\caption{
Growth rate (top) and frequency (bottom) versus ${\omega}$ for
$\alpha^r=0$, $\alpha^i=0$, $|\gamma|^2=1$, and
$\epsilon=0.3$. The solid (dashed) curve shows the
    solution corresponding to ${\sigma}_1
    ({\sigma}_2)$.   
\label{fig::ev_vs_omg_m1}}
\end{figure}

Figure~\ref{fig::ev_vs_omg_m1} shows the behavior of 
$\sigma$ (top panel: growth rate, bottom panel: frequency)
versus the perturbation frequency.   
In the resonant regime, i.e., for
$\displaystyle\left|2\alpha^i-{\omega}\right| < 2\epsilon|\gamma|$,
the growth rates of both interacting modes separate and form a
bubble-like pattern, and the location of the maximum is given by twice
the frequency of the unperturbed problem.  Width and height of the
resonance are determined by the interaction parameter $\gamma$ and by
the amplitude of the perturbation $\epsilon$ (height $\sim
\epsilon|\gamma|$ and width $\sim 4\epsilon|\gamma|$).

In the resonant regime, we find the frequency exactly proportional to
${\omega}/{2}$, which implies that the phase of the $m=1$ eigenmode
becomes locked to the $m=2$ velocity perturbation and the pattern of
the nonaxisymmetric magnetic field is enslaved to the azimuthal phase
of the flow perturbation.  Outside of the resonant regime, i.e., for
$\displaystyle\left|2\alpha^i-{\omega}\right| > 2\epsilon|\gamma|$ the
growth rates collapse, and we find two frequencies with different
asymptotic behavior for $|\omega|\gg1$.  One solution converges
against the unperturbed value, $\alpha^i$
(Fig.~\ref{fig::ev_vs_omg_m1}, solid curve in bottom panel), and the
second solution tends to the frequency of the perturbation $\omega$
(Fig.~\ref{fig::ev_vs_omg_m1}, dashed curve in bottom panel).

Two distinguished locations can be found exactly at the transition
from the nonresonant to the resonant regime. At these
{\it{exceptional points}} \cite{katobook,2004CzJPh..54.1039B}, where
$|\omega-2\alpha^i|=\pm 2\epsilon|\gamma|$, we see an abrupt change of
growth rates and frequencies in combination with a degeneration of the
eigenvalues and a collapse of the corresponding eigenfunctions.

\subsection{Coupling to larger azimuthal wave numbers}

\subsubsection{Application of Floquet theory}

We now treat the general case and consider
system~(\ref{eq::linsystem_exp}) for in principle arbitrary $m$
anticipating that later we will have to truncate our model at a fixed
value for $m_{\rm{max}}= \pm M$ in order to perform explicit
calculations of the eigenvalues.  We revert to Floquet theory (see,
e.g., Rewf.~\cite{LODE}), which implies that the time dependence of the
solution of a differential equation of the form
\begin{equation}
\frac{d}{dt}B(t)=A(t)B(t)\label{eq::diffeq}
\end{equation}
with a $T$-periodic matrix $A(t)=A(t+T)$ and a $n-$dimensional vector
$B$ is given by
\begin{equation}
B(t)=P(t) e^{Rt}\label{eq::floquetsolution}
\end{equation}
with a $T$-periodic invertible matrix $P(t)=P(t+T)$ and a constant
matrix $R$.  It is always possible to find a transformation, the
Lyapunov-Floquet transformation,
\begin{equation}
B(t)=P(t)X(t)\label{eq::transform},
\end{equation} 
such that the previously time dependent linear system~(\ref{eq::diffeq})
becomes a linear system  
\begin{equation}
\frac{d}{dt}X(t)=R X(t)\label{eq::transfsys}
\end{equation}
with the time-independent coefficient matrix $R$.
From~(\ref{eq::diffeq}) and~(\ref{eq::floquetsolution}) we get
\begin{equation}
\frac{d}{dt}B=\dot{P}e^{Rt}+PRe^{Rt}=APe^{Rt}
\end{equation}
so that
\begin{equation}
R=P^{-1}AP-P^{-1}\dot{P}.
\end{equation}
We now write 
\begin{equation}
P=e^{iDt}
\end{equation}
from which immediately follows 
\begin{equation}
P^{-1}\dot{P}=iD
\end{equation}
so that the constant coefficient matrix $R$ in the transformed system 
(\ref{eq::transfsys}) can be written as
\begin{equation}
R=e^{-iDt}Ae^{iDt}-iD.
\end{equation}

In our particular case we can write 
\begin{equation}
A(t)=\displaystyle e^{iD_{\omega}t} \hat{A} e^{-iD_{\omega}t}
\end{equation}
with
\begin{equation}
D_{\omega}=\left(
\begin{array}{ccccc}
-\frac{M}{2}\omega &  0                    &         &                      &                    \\
0                   & -\frac{M-2}{2}\omega & 0       &                      &                    \\
                    &  0                    &  \ddots & 0                    &                    \\
                    &                       & 0       & \frac{M-2}{2}\omega & 0                  \\
                    &                       &         & 0                    & \frac{M}{2}\omega 
\end{array}
\right)
\end{equation}
and $\hat{A}$ the matrix with the components of $A$ but without the 
time modulation $e^{\pm i\omega t}$. This gives
\begin{equation}
R=\hat{A}-iD_{\omega}
\end{equation}
so that from~(\ref{eq::transfsys}) we end up with the system
\begin{equation}
\frac{dX}{dt}=\left(\hat{A}-iD_{\omega}\right)X.
\end{equation}

The solutions are given by $X=X_0e^{\widetilde{\sigma} t}$ with
$\widetilde{\sigma}\in \mathbb{C}$ the eigenvalues in the transformed
system which are the roots of the characteristic equation
\begin{equation}
\left|\hat{A}-iD_{\omega}-\widetilde{\sigma}\mathbb{I}\right|=0, 
\end{equation}
where $|\cdots|$ denotes the determinant and $\mathbb{I}$ is the
identity matrix. 

\subsubsection{Truncation at $M = 1$}

In order to check the validity of the computations from the previous
section we solve the case $M=1$ and compare with the solution
obtained in Sec.~\ref{sec::m=1}.  We have
\begin{equation}
\left(
\begin{array}{l}
\displaystyle\frac{d}{dt}\hat{b}_{-1}\\[0.35cm]
\displaystyle\frac{d}{dt}\hat{b}_{1}
\end{array}
\right)
= 
\left(
\begin{array}{cc}
\vphantom{\displaystyle\frac{d}{dt}}
\alpha^* & \epsilon e^{-i\omega t}\gamma^* \\[0.35cm]
\vphantom{\displaystyle\frac{d}{dt}}
\epsilon e^{i\omega t} \gamma & \alpha
\end{array}
\right)
\left(
\begin{array}{c}
\vphantom{\displaystyle\frac{d}{dt}}
\hat{b}_{-1}\\[0.35cm]
\vphantom{\displaystyle\frac{d}{dt}}
\hat{b}_{1}
\end{array}
\right)
=
\left(
\begin{array}{cc}
\vphantom{\displaystyle\frac{d}{dt}}
e^{-i\frac{\omega}{2} t} & 0 \\[0.35cm]
\vphantom{\displaystyle\frac{d}{dt}}
0 & e^{i\frac{\omega}{2} t}
\end{array}
\right)
\left(
\begin{array}{cc}
\vphantom{\displaystyle\frac{d}{dt}}
\alpha^* & \epsilon\gamma^* \\[0.35cm]
\vphantom{\displaystyle\frac{d}{dt}}
\epsilon \gamma & \alpha
\end{array}
\right)
\left(
\begin{array}{cc}
\vphantom{\displaystyle\frac{d}{dt}}
e^{i\frac{\omega}{2} t} & 0 \\[0.35cm]
\vphantom{\displaystyle\frac{d}{dt}}
0 & e^{-i\frac{\omega}{2} t}
\end{array}
\right)
\left(
\begin{array}{l}
\vphantom{\displaystyle\frac{d}{dt}}
\hat{b}_{-1}\\[0.35cm]
\vphantom{\displaystyle\frac{d}{dt}}
\hat{b}_{1}
\end{array}
\right)
\end{equation}
so that 
\begin{eqnarray}
\hat{A}=
\left(
\begin{array}{cc}
\alpha^* & \epsilon\gamma^* \\
\epsilon\gamma & \alpha
\end{array}
\right)
\mbox{ and }
D_{\omega}=
\left(
\begin{array}{cc}
\nicefrac{-\omega}{2} & 0\\
0 & \nicefrac{\omega}{2}
\end{array}
\right).
\end{eqnarray}
The characteristic equation for the eigenvalues becomes
\begin{equation}
\left(\alpha^*+i\frac{\omega}{2}-\widetilde{\sigma}\right)
\left(\alpha-i\frac{\omega}{2}-\widetilde{\sigma}\right)
-\epsilon^2\left|\gamma\right|^2=0
\end{equation}
with the solutions
\begin{equation}
\widetilde{\sigma}_{1,2}=\alpha^r
\pm\frac{1}{2}\sqrt{4\epsilon^2|\gamma|^2-\left(2\alpha^i-\omega\right)^2},\label{eq::50}
\end{equation}
which, when taking into account the transformation
$B=P(t)X(t)=e^{iD_{\omega}t}X$ 
according to Eq.~(\ref{eq::transform}), is identical to the
solution~(\ref{eq::sol_with_coupling}) obtained previously in
Sec.~\ref{sec::m=1}.

\subsubsection{Truncation at $M = 3$}\label{subsec::m3}
 
In the next step we include higher order modes with $m=\pm 3$ so that
we have to compute the characteristic equation
\begin{equation}
\left|
\begin{array}{cccc}
\alpha^*_3\!+\!i\frac{3\omega}{2}\!
-\!\widetilde{\sigma}\! & \epsilon\delta^*_2 & 0 & 0 \\
\epsilon\delta^*_1 & \alpha_1^*\!
+\!i\frac{\omega}{2}\!
-\!\widetilde{\sigma}\! & \epsilon\gamma^* & 0 \\
0 & \epsilon\gamma & \alpha_1\!
-\!i\frac{\omega}{2}\!
-\!\widetilde{\sigma}\! & \epsilon\delta_1 \\
0 & 0 & \epsilon\delta_2 & \alpha_3\!
-\!i\frac{3\omega}{2}\!-\!\widetilde{\sigma}\!
\end{array}
\right|=0,\label{eq::determinant}
\end{equation}
where we used the abbreviations $\alpha_1=\alpha_{1,1},
\alpha_3=\alpha_{3,3}, \gamma=\alpha_{1,-1}, \delta_{1}=\alpha_{1,3}$
and $\delta_2=\alpha_{3,1}$.  Before we specify the general solution
of~(\ref{eq::determinant}) we briefly discuss three limit cases in
order to prove the validity of the solution and illustrate distinct
properties of the solutions resulting from different types of
coupling.

\paragraph{Weak coupling}

We first assume no coupling between $m=1$ and $m=3$ as well as
between $m=+1$ and $m=-1$ (i.e., $|\gamma|^2=|\delta_1\delta_2|=0$) so
that the matrix $A(t)$ is a diagonal matrix. We obtain the unperturbed
solutions for the $m=1$ mode and a second set for the $m=3$ mode,
\begin{equation}
\widetilde{\sigma}_{1,2}=\alpha_1^r
\pm i\left(\alpha_1^i-\frac{\omega}{2}\right) \mbox{ and }
\widetilde{\sigma}_{3,4}=\alpha_3^r
\pm i\left(\alpha_3^i-\frac{3\omega}{2}\right).
\end{equation}

If only the coupling between the $m=\pm 1$ mode and $m=\pm 3$ is weak
(i.e., $|\gamma|^2 \gg |\delta_1\delta_2|$),  
we obtain the characteristic equation
\begin{equation}
\Bigg[\bigg(\alpha_3^*+i\frac{3\omega}{2}
-\widetilde{\sigma}\bigg)
\bigg(\alpha_3-i\frac{3\omega}{2}
-\widetilde{\sigma}\bigg)\Bigg]  
\Bigg[\bigg(\alpha_1^*+i\frac{\omega}{2}
-\widetilde{\sigma}\bigg)
\bigg(\alpha_1-i\frac{\omega}{2}
-\widetilde{\sigma}\bigg)-\epsilon^2|\gamma|^2\Bigg] =  0,
\end{equation}
and we recover the previous solution~(\ref{eq::sol_with_coupling})
from section~\ref{sec::m=1} with the resonance from the interaction of
$m=1$ and $m=-1$ at $\omega=2\alpha_1^i$ and two separate solutions for
the $m=3$ mode:
\begin{equation}
\widetilde{\sigma}_{1,2} =
\alpha_{1}^r+i\frac{\omega}{2}
\pm\frac{1}{2}\sqrt{4\epsilon^2|\gamma|^2
-\left(\omega-2\alpha_1^i\right)^2}\mbox{  and  } 
\widetilde{\sigma}_{3,4} = 
\alpha_3^r\pm i\left(\alpha_3^i-\frac{3\omega}{2}\right).\label{eq::weakcoupling}
\end{equation}
Note that the first part of Eq.~(\ref{eq::weakcoupling}) also
corresponds to the solution denoted in Eq.~(\ref{eq::50}) for a
truncation at $M=1$.

\paragraph{Strong coupling between $m=1$ and $m=3$}

Now we assume that the coupling between $m=1$ and $m=-1$ can be
neglected but the coupling between $m=1$ and $m=3$ remains strong,
i.e., $|\delta_1\delta_2|\gg |\gamma|^2$. 
\captionsetup[subfigure]{margin=0.3cm,position=bottom,
                         captionskip=0.2cm,justification=centering}
\begin{figure}[t!]
\hspace*{-0.3cm}
\subfloat[Growth rate]
{\includegraphics[width=0.35\textwidth]{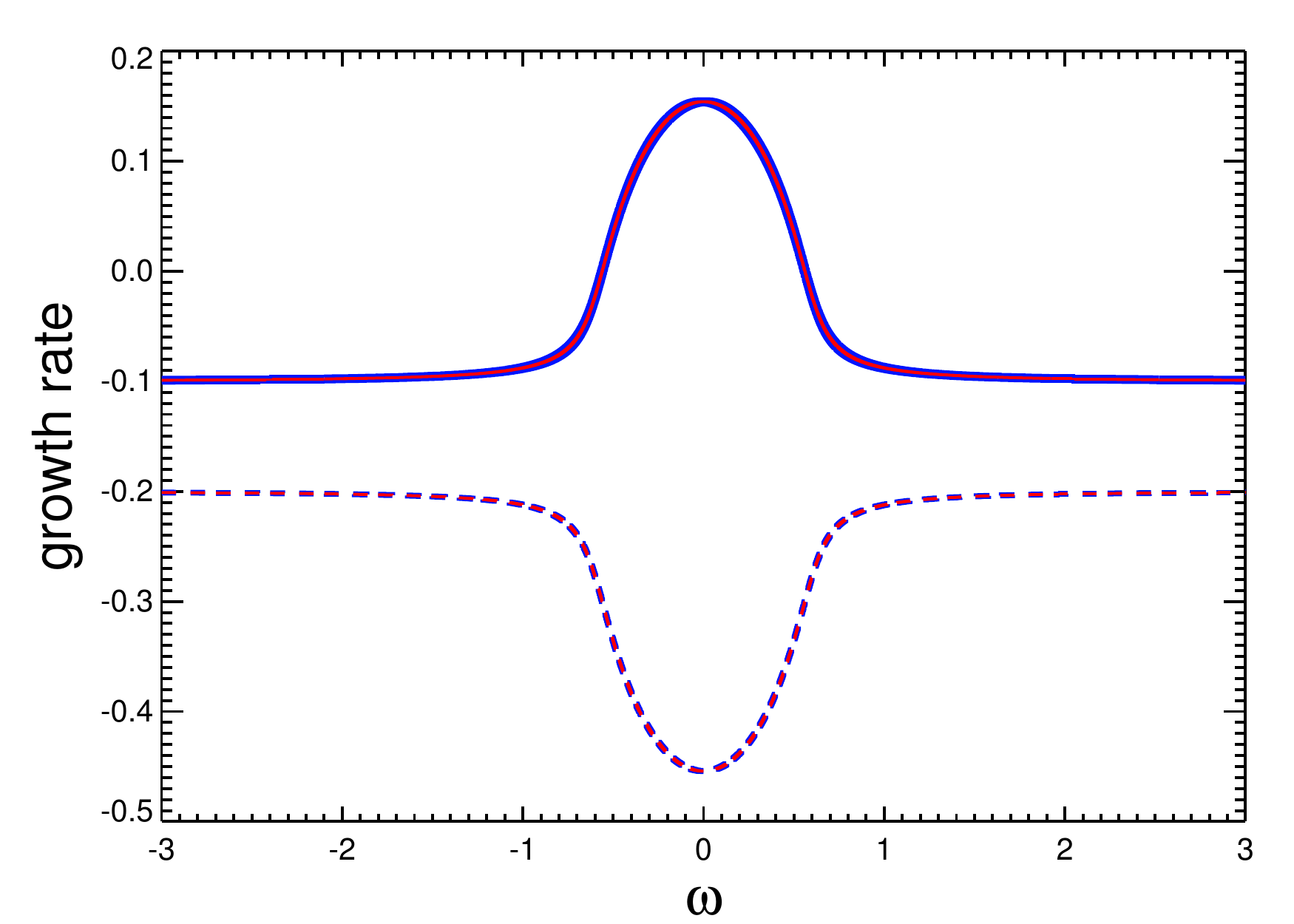}}
\hspace*{-0.2cm}
\subfloat[Frequency]
{\includegraphics[width=0.35\textwidth]{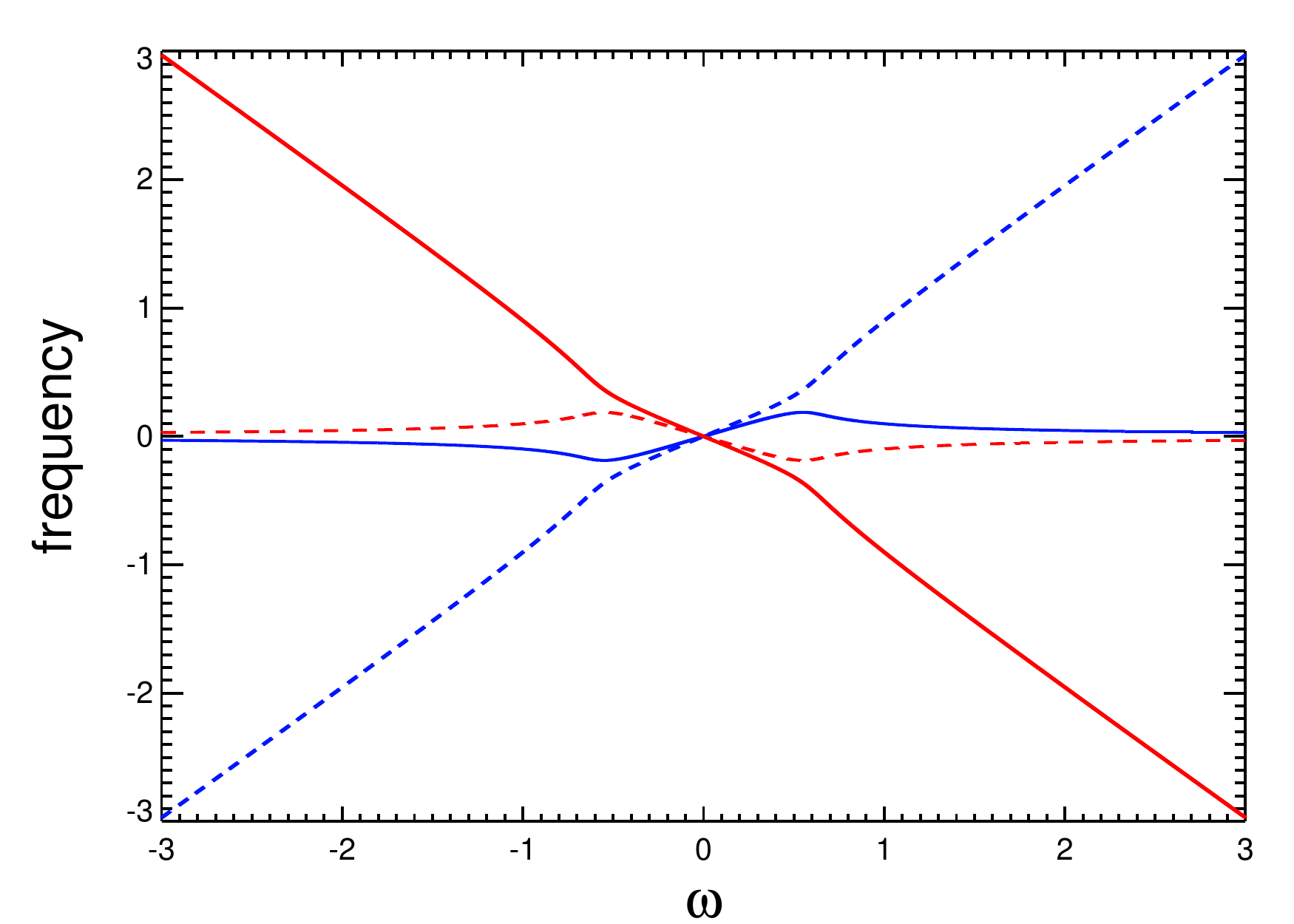}}
\hspace*{-0.2cm}
\subfloat[Frequency (close up)]
{\includegraphics[width=0.35\textwidth]{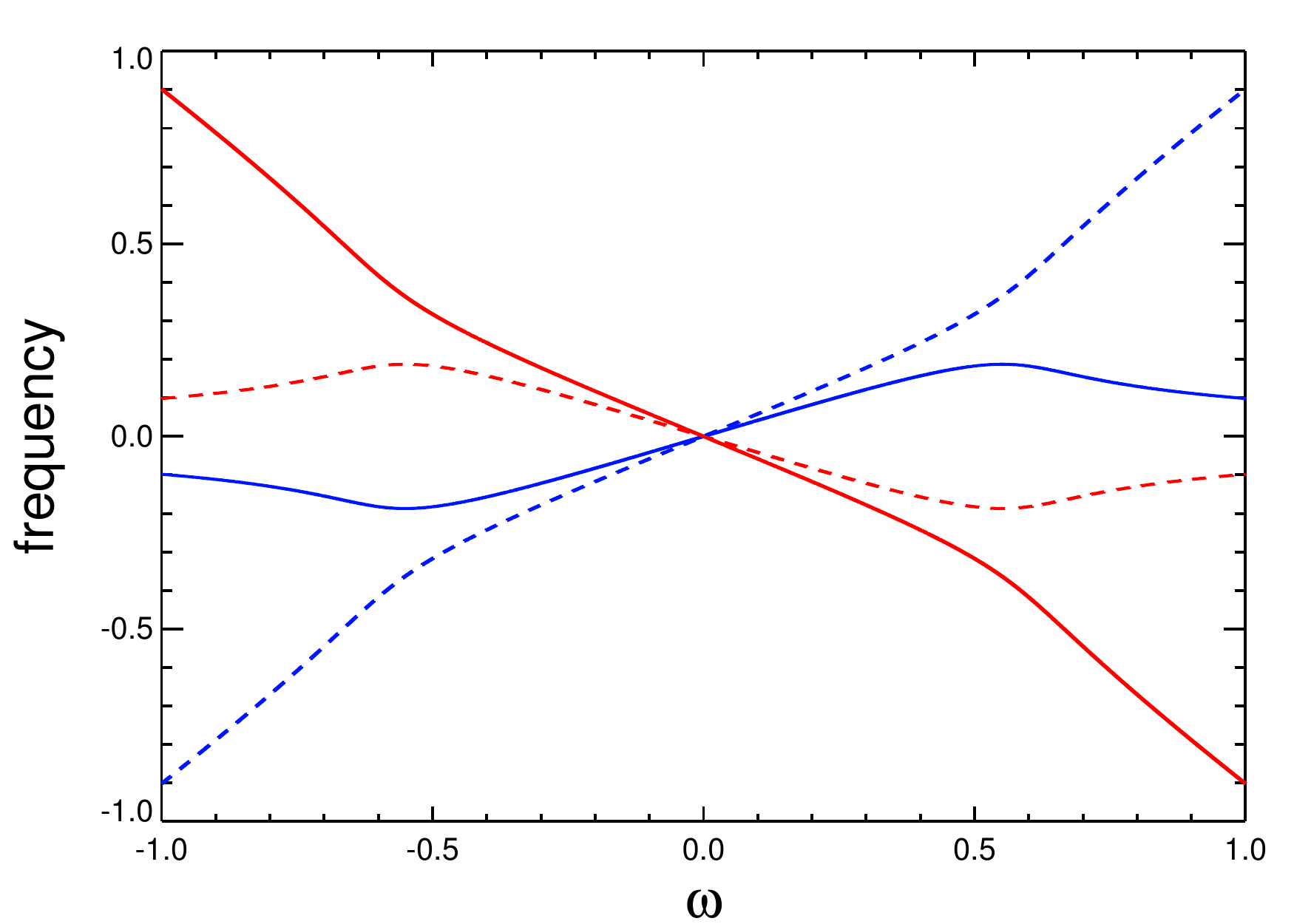}}
\caption{
Growth rate (real part of $\widetilde{\sigma}_1\dots\\widetilde{sigma}_4$) and frequencies
(imaginary part of $\widetilde{\sigma}_1\dots\widetilde{\sigma}_4$) versus $\omega$ for strong
coupling between $m=1$ and $m=3$ ($|\delta_1\delta_2|\gg|\gamma|^2$).
$\alpha^r_1=-0.1, \alpha_3^r=-0.2, \alpha_1^i=\alpha_3^i=0,
|\gamma|^2=0, \delta_1\delta_2=1, \epsilon=0.3$.  The blue curves show
the solutions $\widetilde{\sigma}_1$ (solid curves) and
$\widetilde{\sigma}_2$ (dashed curves), and the red curves show
$\widetilde{\sigma}_3$ (solid curves) and $\widetilde{\sigma}_4$
(dashed curves).  Note that the real parts of $\widetilde{\sigma}_1$ and
$\widetilde{\sigma}_3$ (solid curves in a) and of $\widetilde{\sigma}_2$ and $\widetilde{\sigma}_4$
(dashed curves in a) collapse exactly.
}\label{fig::couplingm1m3} 
\end{figure}
Then the characteristic equation for the calculation of the eigenvalues
reads 
\begin{equation}
\Bigg[\left(\alpha_3^*+i\frac{3\omega}{2}-\widetilde{\sigma}\right)
\Bigg(\alpha_1^*+i\frac{\omega}{2}-\widetilde{\sigma}\Bigg)
-\epsilon^2\delta_1^*\delta_2^*\Bigg] 
\Bigg[\left(\alpha_3-i\frac{3\omega}{2}-\widetilde{\sigma}\right)
\Bigg(\alpha_1-i\frac{\omega}{2}
-\widetilde{\sigma}\Bigg)-\epsilon^2\delta_1\delta_2\Bigg]=0.
\end{equation}
and we have two kinds of solutions, one for the coupled system of $m=-1$
and $m=-3$ and a second one for the coupled system with $m=1$ and
$m=3$ mode:
\begin{equation}
\widetilde{\sigma}_{1,2} = \quad\frac{1}{2}
\bigg[\alpha_1^r+\alpha_3^r+i\Big(\alpha_1^i
+\alpha_3^i-2\omega\Big)\bigg]
\pm\frac{1}{2}\sqrt{4\epsilon^2\delta_1\delta_2+\bigg[\alpha_1^r
-\alpha_3^r+i\Big(\alpha_1^i-\alpha_3^i
+\omega\Big)\bigg]^2},\label{eq::57a}  
\end{equation}
\begin{equation}
\widetilde{\sigma}_{3,4} = \quad\frac{1}{2}
\bigg[\alpha_1^r+\alpha_3^r-i\Big(\alpha_1^i
+\alpha_3^i-2\omega\Big)\bigg]
\pm\frac{1}{2}\sqrt{4\epsilon^2\delta_1\delta_2+\bigg[\alpha_1^r
-\alpha_3^r-i\Big(\alpha_1^i-\alpha_3^i
-\omega\Big)\bigg]^2}.\label{eq::57b}  
\end{equation}
The expressions~(\ref{eq::57a}) and (\ref{eq::57b}) generalize the
resonance between two independent eigenmodes.  Whereas the real parts
of $\widetilde{\sigma}_{1,2}$ and $\widetilde{\sigma}_{3,4}$ collapse
(blue and red curves in Fig.~\ref{fig::couplingm1m3}a), we obtain
four different solutions for the frequencies that describe two sets of
counter-rotating eigenmodes (Figs.~\ref{fig::couplingm1m3}b and c).
As in the previous case, for large $|\omega|$ we see an asymptotic
behavior according to $\pm\omega$ respectively $\alpha^i_1$ and/or
$\alpha^i_3$ after back transformation to the original system.
Note that the product $\delta_1\delta_2$ is always real and positive,
and the second term under the square root in~(\ref{eq::57a}) and
in~(\ref{eq::57b}) remains complex for all $\omega$ (provided
$\alpha^r_1-\alpha^r_3$ is non zero) so that frequency locking cannot be
expected unlike the case of a parametric instability described in
Eq.~(\ref{eq::sol_with_coupling}).
Nevertheless, the growth rates again exhibit a regime with
amplification but with the maximum at $\omega=\alpha^i_3-\alpha^i_1$,
and, in contrast to the previous case with $m=1$ and $m=-1$ coupling,
we do not see any intersection or merging of growth rates
(Fig.~\ref{fig::couplingm1m3}a).  The transition between the regime
with nearly unchanged growth rate and the regime with amplification is
smooth and goes along with a change of the behavior of the frequencies
(Figs.~\ref{fig::couplingm1m3}b and c).

\paragraph{General solution}

In general the computation of the determinant~(\ref{eq::determinant})
yields a characteristic equation for the eigenvalues $\sigma$ given by
a polynomial of order four that reads
\begin{equation}
\sigma^4+p\sigma^3+q\sigma^2+r\sigma+s=0\label{eq::poly_order4}
\end{equation}
with 
\begin{eqnarray}
p  &  =  &  -2(\alpha_1^r+\alpha_3^r),  \nonumber\\
q  &  =  & 
\left|\alpha_1'\right|^2
+\left|\alpha_3'\right|^2+4\alpha_1^r\alpha_3^r
-2\epsilon^2{\delta^r}
-\epsilon^2|\gamma|^2,
\nonumber\\
&&\\[-0.3cm]
r  &  =  & 
-2(\alpha_3^r|\alpha_1'|^2+\alpha_1^r|\alpha_3'|^2)
+2\epsilon^2|\gamma|^2\alpha_3^r
+2\epsilon^2\Big(\big(\alpha_1^r
+\alpha_3^r\big)\delta^r
+(\alpha_1^i+\alpha_3^i-2\omega)\delta^i\Big),
\nonumber\\
s  &  =  & 
-2\epsilon^2\bigg[\delta^i\bigg(\alpha_1^i\alpha_3^r
+\alpha_1^r\alpha_3^i
-\frac{\omega}{2}(3\alpha_1^r+\alpha_3^r)\bigg)
+\delta^r\bigg(\frac{\omega}{2}(3\alpha_1^i+\alpha_3^i)
+\alpha_1^r\alpha_3^r-\alpha_1^i\alpha_3^i
-\frac{3}{4}\omega^2\bigg)\bigg]
\nonumber\\
&&-\epsilon^2\left|\alpha_3'\right|^2|\gamma|^2
+\left|\alpha_1'\right|^2\left|\alpha_3'\right|^2+\epsilon^4|\delta|^2,
\nonumber
\end{eqnarray}
and the abbreviations $\alpha_j'=\alpha_j-i\frac{j\omega}{2}$ and
$\delta=\delta_1\delta_2$.  All coefficients of the
polynomial~(\ref{eq::poly_order4}) are real valued, and we obtain four
complex solutions $\sigma_1,\dots,\sigma_4$.

\begin{figure}[b!]
\subfloat[$\quad\delta_1\delta_2=-3, 0 \le |\gamma|^2 \le
5$]{\includegraphics[width=0.49\textwidth]{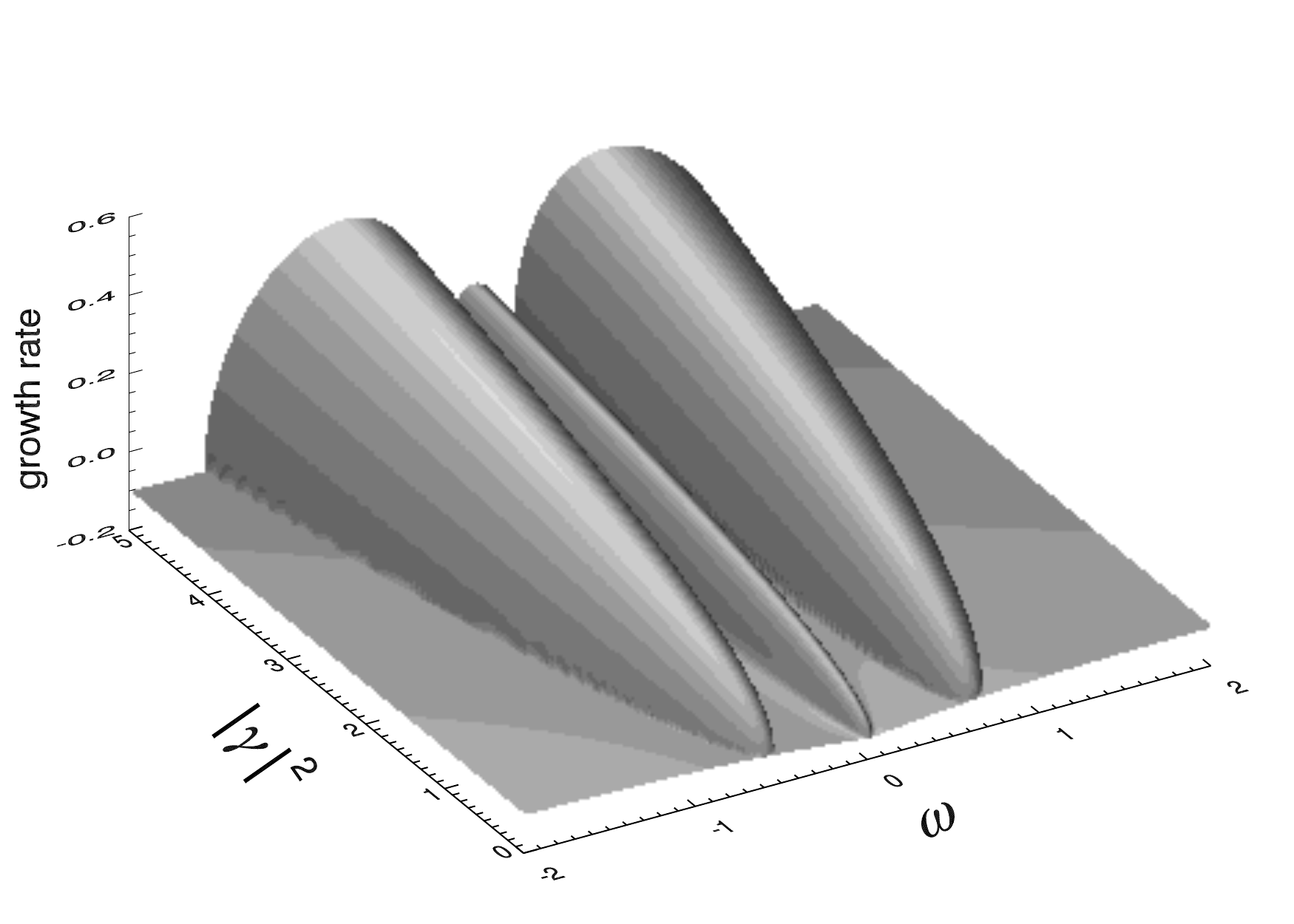}}
\subfloat[$\quad|\gamma|^2=1, -5 \le \delta_1\delta_2 \le
5$]{\includegraphics[width=0.49\textwidth]{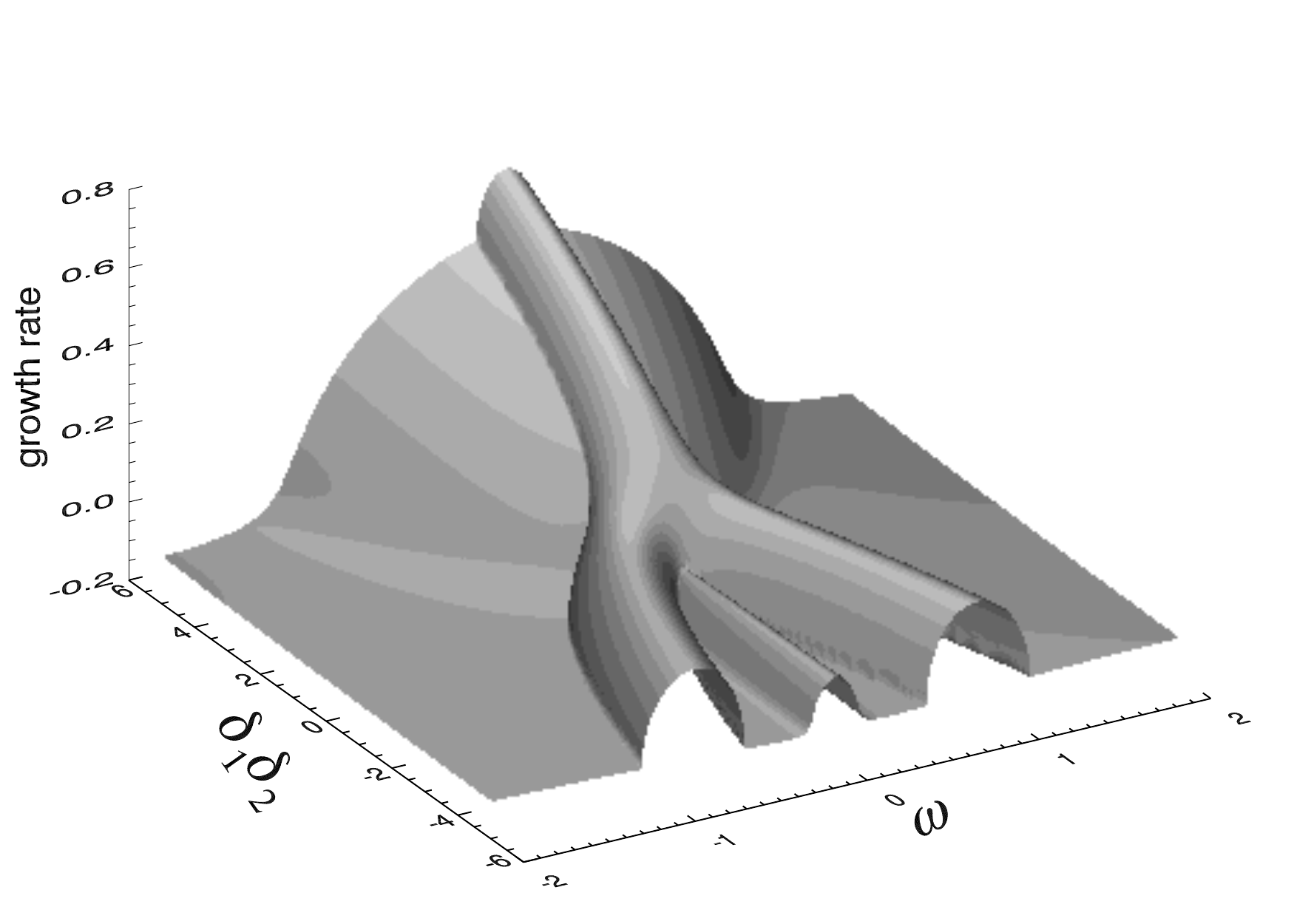}}

\caption{
(a) Growth rates versus $|\gamma|^2$ and $\omega$. (b) Growth rates versus
$\delta_1\delta_2$ and $\omega$ (b). 
Only the leading mode is shown. 
Details on the behavior at $|\gamma|^2=1$ and $\delta_1\delta_2=\pm 3$
can be extracted from Fig.~\ref{fig::mpm3}.
In both cases $\alpha^r_1=-0.1,
\alpha^r_3=-0.2, \alpha^i_1=\alpha^i_3=0$, and $\epsilon=0.3$.  
\label{fig::3d_gr_vs_gam_delt_omg}
}
\end{figure}

In the following, we always apply negative values for the unperturbed
growth rates, i.e., we consider models with a magnetic Reynolds number
below the dynamo threshold without the perturbation.  The interaction
parameters $|\gamma|^2$ and $\delta_1\delta_2$ are chosen such that
characteristic and representative solutions are obtained with
properties reminiscent of the simulations.
Figure~\ref{fig::3d_gr_vs_gam_delt_omg} shows the impact of the
interaction parameters $|\gamma|^2$ and $\delta_1\delta_2$ for the
case $\alpha_1^r=-0.1, \alpha_3^r=-0.2$ and $\alpha_1^i=\alpha_3^i=0$,
with the perturbation amplitude fixed at $\epsilon=0.3$.  For fixed
$\delta_1\delta_2=-3$ the growth rates increase monotonically with
$|\gamma|^2$ which parameterizes the interaction of $m=1$ and $m=-1$
modes (Fig.~\ref{fig::3d_gr_vs_gam_delt_omg}a). In contrast, the
variation of $\delta_1\delta_2$, which parameterizes the interaction of
$m=1$ and $m=3$ (respectively $m=-1$ and $m=-3$), may significantly
change the structure of the growth rates
(Fig.~\ref{fig::3d_gr_vs_gam_delt_omg}b).
\captionsetup[subfigure]{margin=0.0cm,singlelinecheck=false,format=plain,
                         indention=0.3cm,justification=justified,
                         captionskip=0.2cm,position=bottom,
                         nearskip=0.0cm,farskip=0.0cm}
\begin{figure}[b!]
\subfloat[Growth rates for $\delta_1\delta_2=-3$]
{\includegraphics[width=0.485\textwidth]{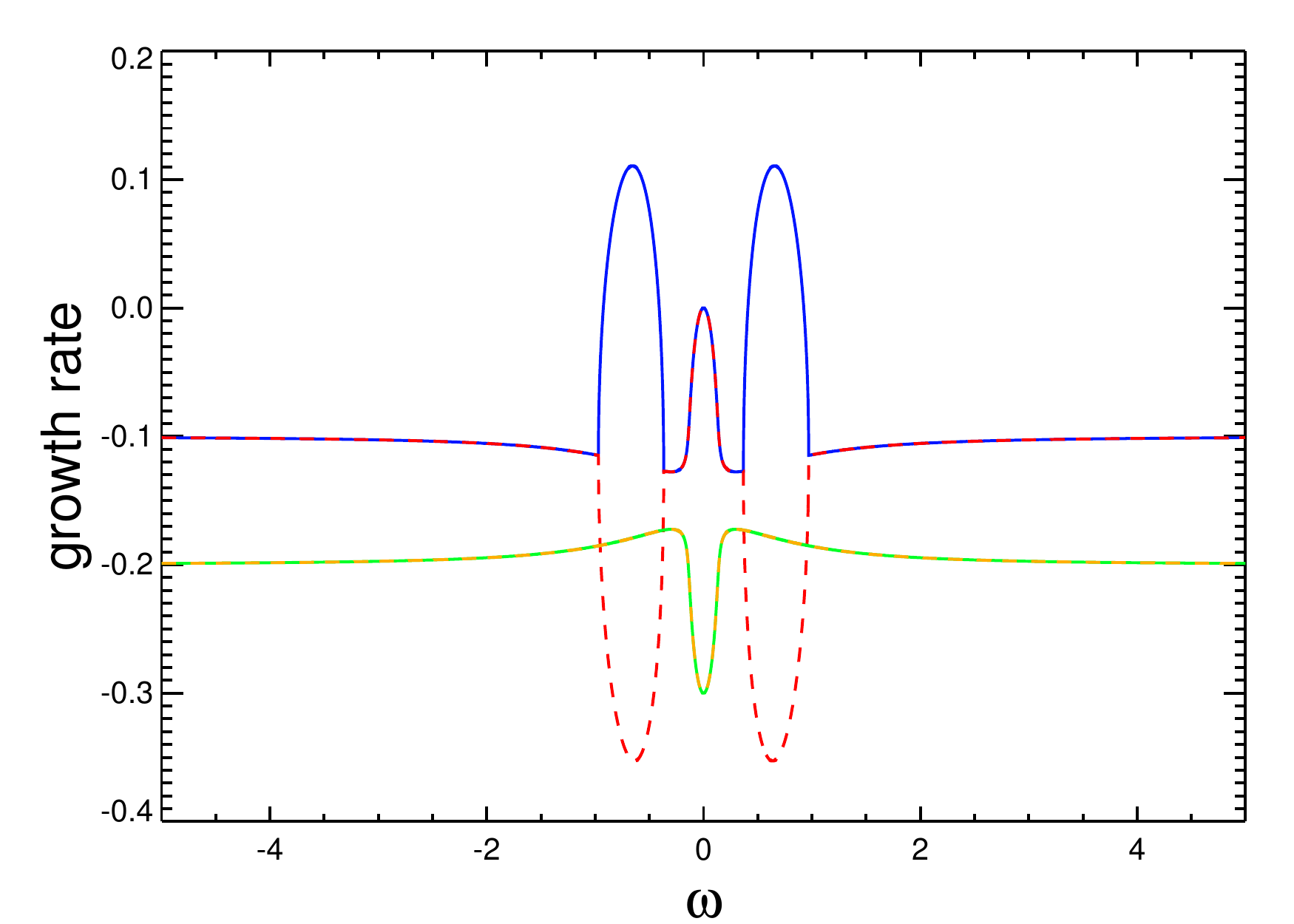}}
\subfloat[Growth rates for $\delta_1\delta_2=+3$]
{\includegraphics[width=0.485\textwidth]{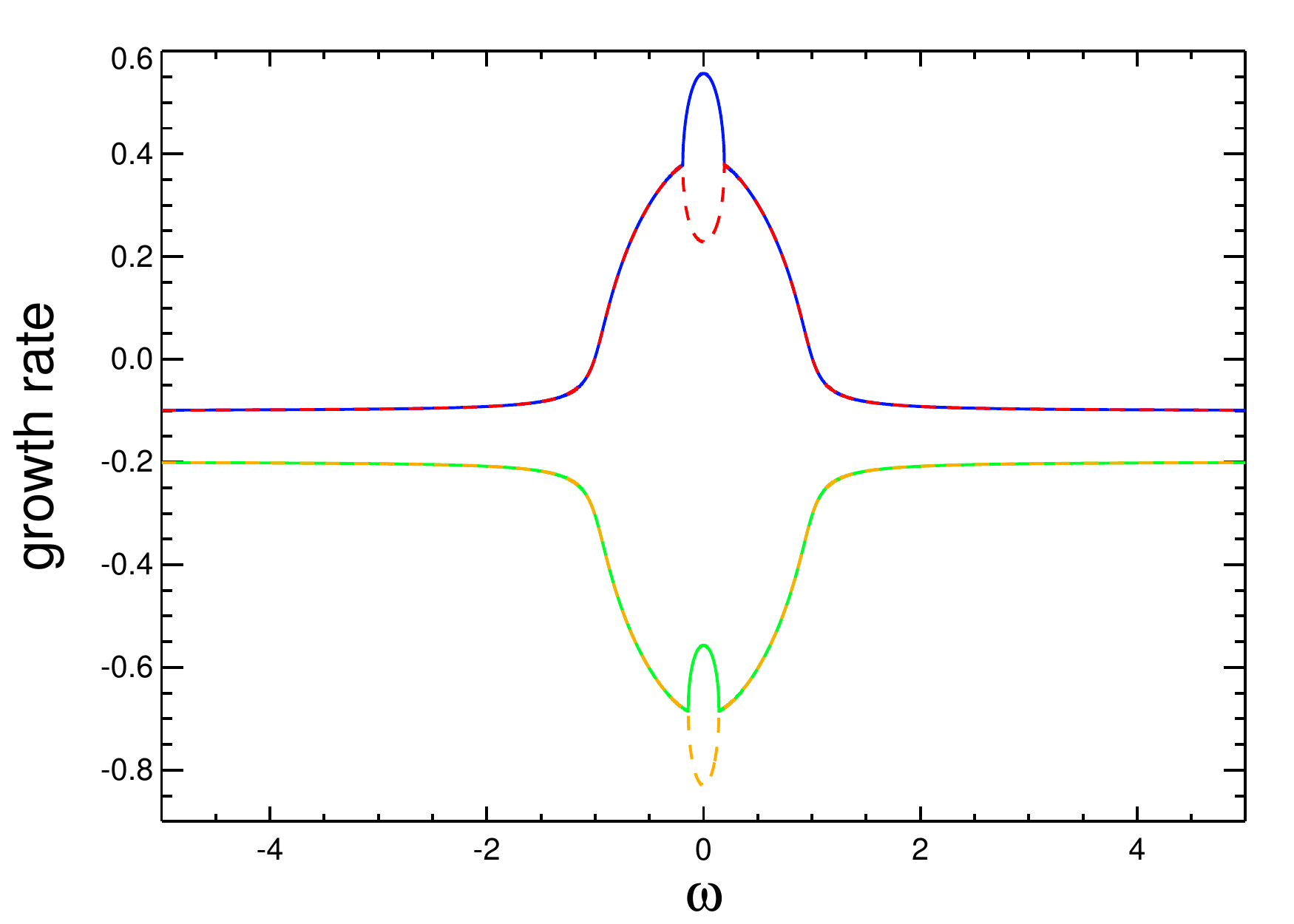}}
\\
\subfloat[Frequencies for $\delta_1\delta_2=-3$]
{\includegraphics[width=0.485\textwidth]{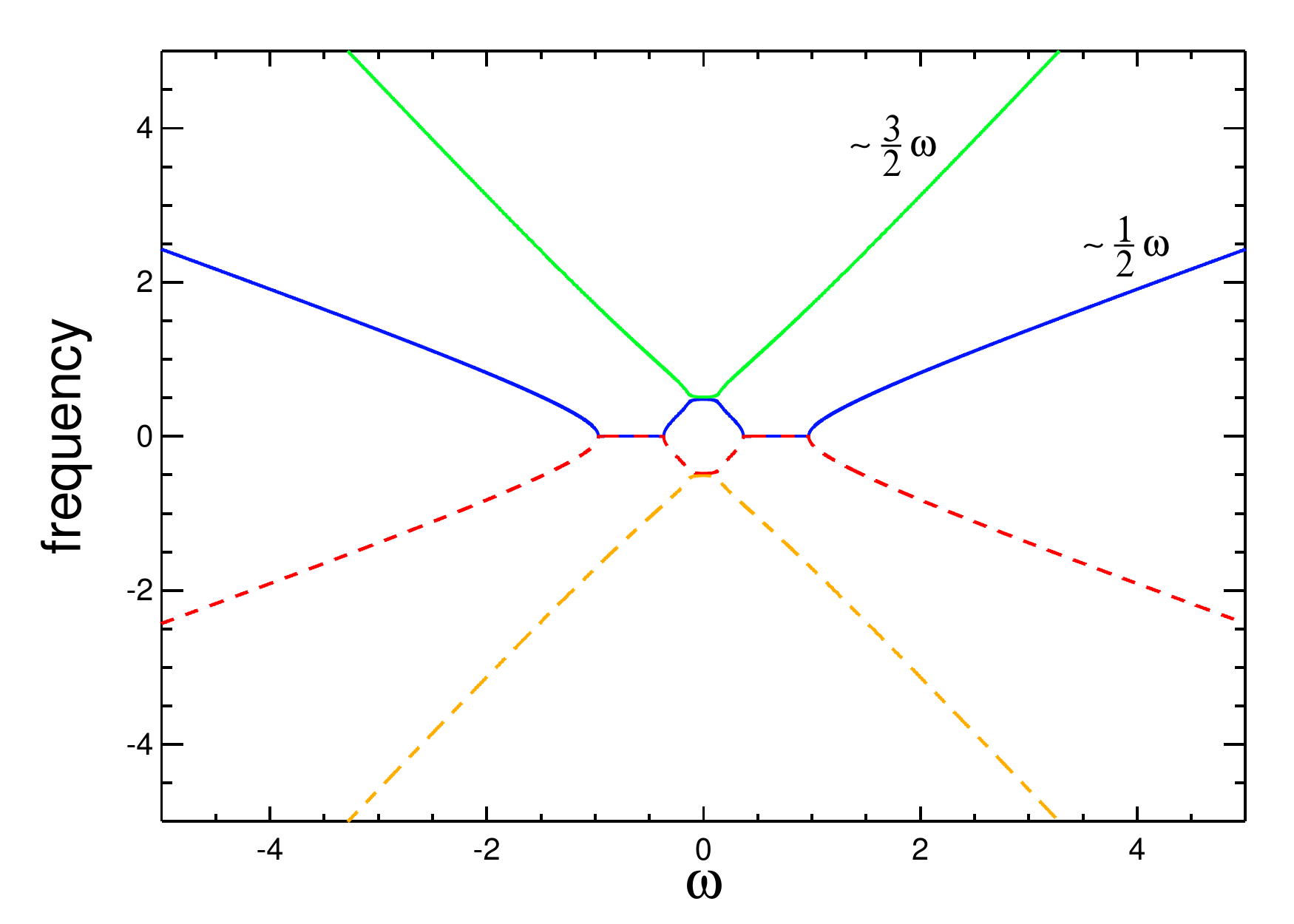}}
\subfloat[Frequencies for $\delta_1\delta_2=+3$]
{\includegraphics[width=0.485\textwidth]{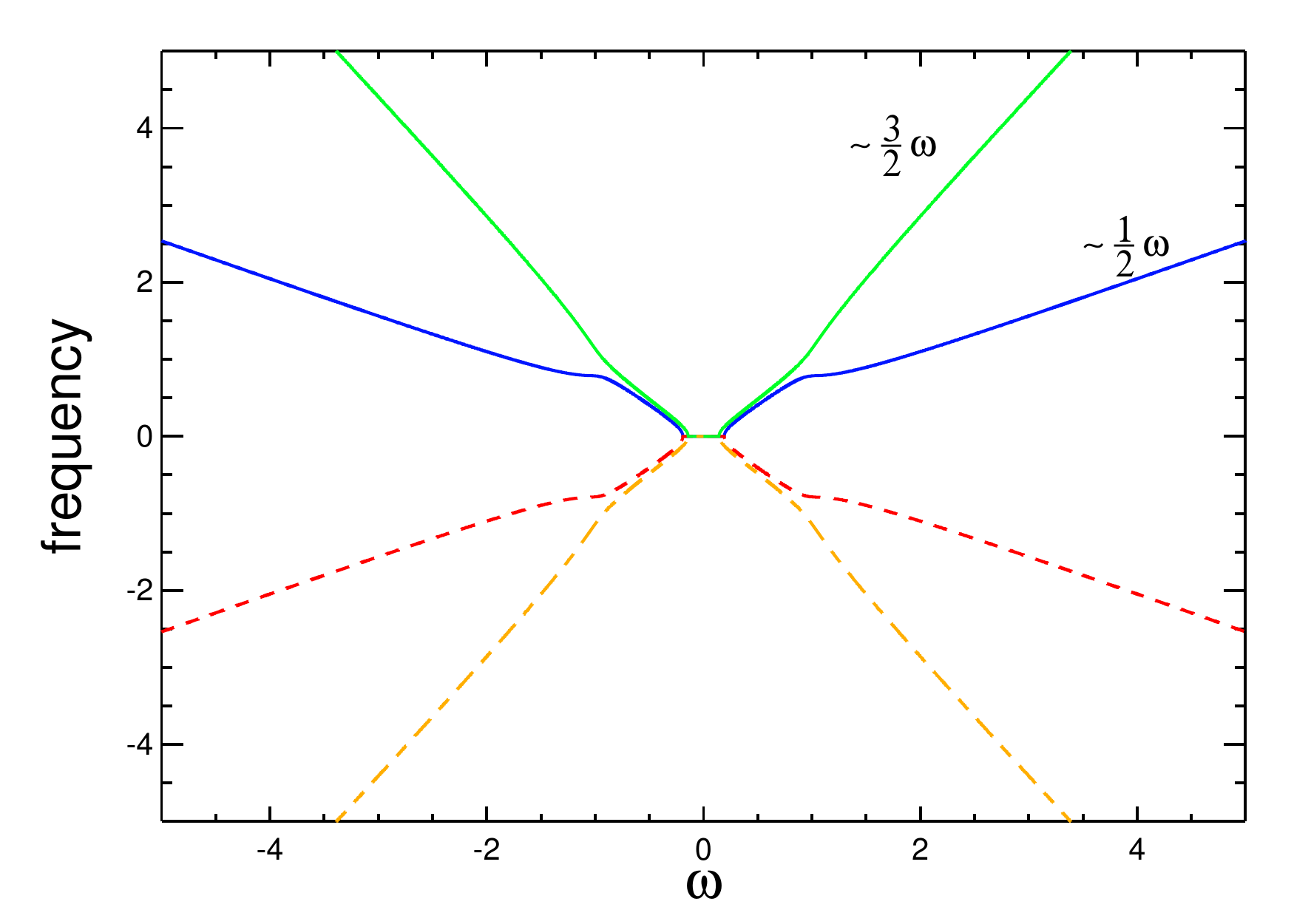}}
\caption{
Characteristic pattern of the growth rates (a,b) and frequencies
(c,d) in dependence of $\omega$ for $\delta_1\delta_2=-3$ (a, c) and
$\delta_1\delta_2=+3$ (b, d).  For both cases $\alpha_1^r=-0.1,
\alpha_3^r=-0.2, |\gamma|^2=1, \epsilon=0.3$.  The blue and red curves
denote the solutions belonging to $\widetilde{\sigma}_{1,2}$, and the
green and orange curves denote the solutions belonging to
$\widetilde{\sigma}_{3,4}$.  Note that the frequency plots are not
transformed into the original system.
\label{fig::mpm3}} 
\end{figure}
Basically, we obtain two different types of solutions, which are shown
in detail in Fig.~\ref{fig::mpm3}.  For a $\delta_1\delta_2 < 0$ the
growth rates have three maxima, and we have two regimes with a
parametric resonance symmetrically around the origin. The third
maximum corresponds to a parametric amplification and is embedded
between the parametric resonances. The pattern is rather similar to
the behavior for small perturbation frequencies found in the
simulations at ${\rm{Rm}}=120$ (see Figs.~\ref{sf::6c}
and~\ref{sf::6d}).  For a positive value of $\delta_1\delta_2$ we
obtain one absolute maximum at $\omega=0$. However, in this case the
single maximum emerges on top of a regime with amplification so that
the overall pattern is different from the case shown in
Figs.~\ref{sf::6a} and~\ref{sf::6b}.

\begin{figure}[t!]
\hspace*{-0.3cm}
\subfloat[Growth rate  versus perturbation frequency $\omega$ and 
perturbation amplitude $\epsilon$]
{\includegraphics[width=0.34\textwidth]{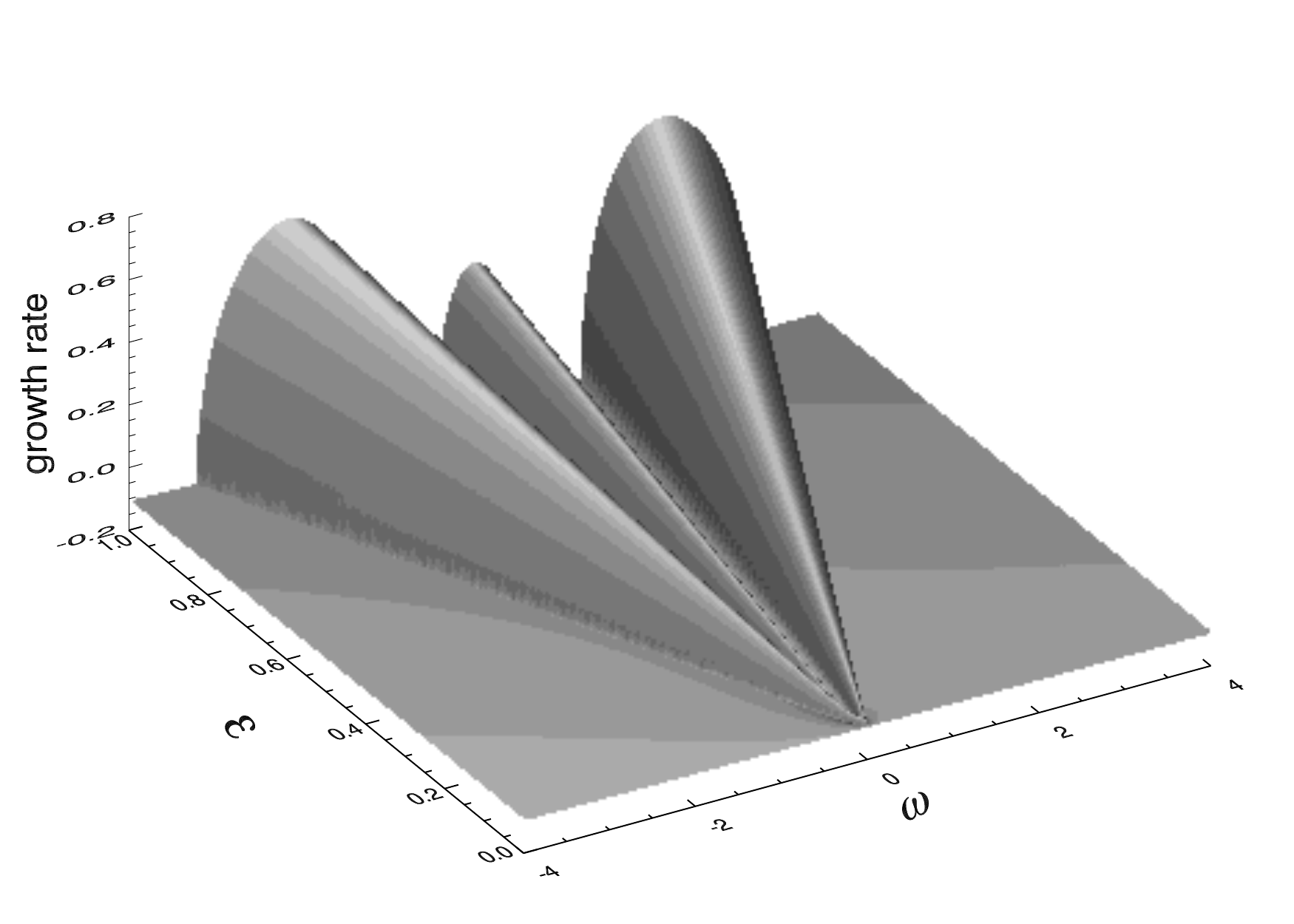}}        
\hspace*{-0.2cm}
\subfloat[Variation of the resonance frequency $\omega_{\rm{max}}$
with the perturbation amplitude]
{\includegraphics[width=0.34\textwidth]{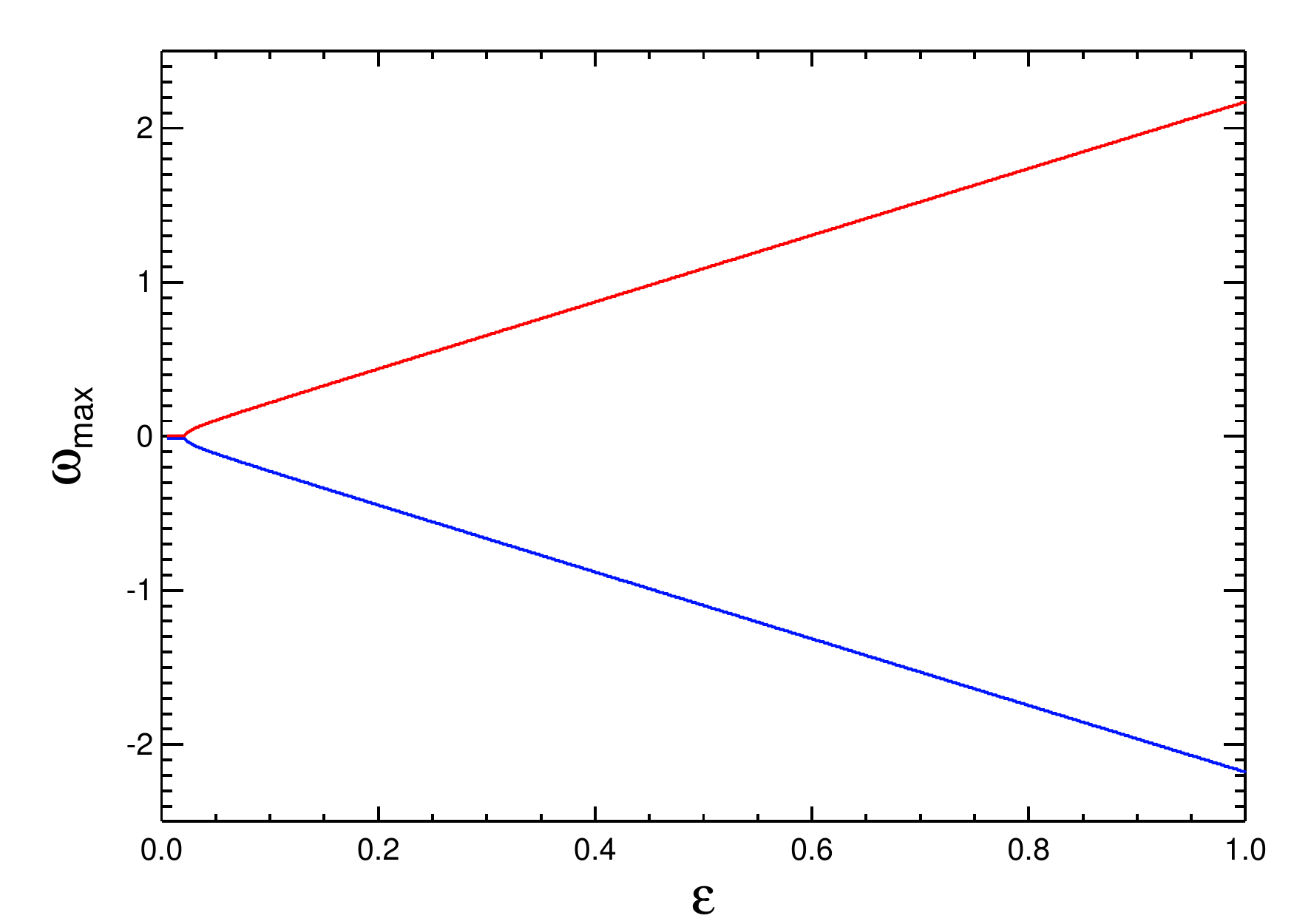}} 
\hspace*{-0.2cm}
\subfloat[Maximum of the growth rate versus perturbation amplitude] 
{\includegraphics[width=0.34\textwidth]{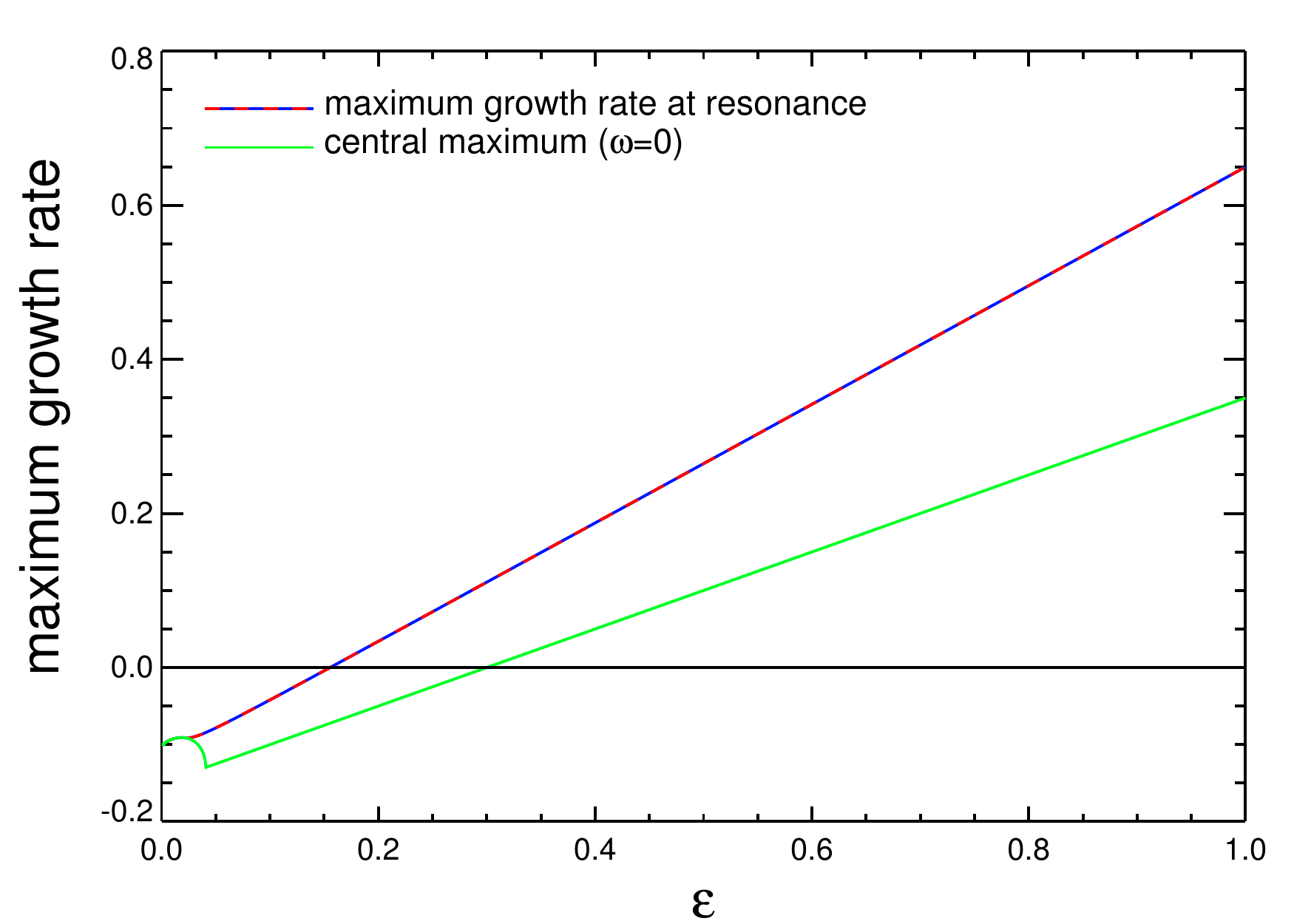}}        
\caption{Characteristics of the eigenvalues for 
  $\alpha_1^r=-0.1, \alpha_3^r=-0.2, \alpha_1^i=\alpha_3^i=0, |\gamma|^2=1$,
  and $\delta_1\delta_2=-3$.} 
\label{fig3d::gr_vs_omg_vs_eps} 
\end{figure}

The exemplary solutions show a combination of
the phenomena that arose in the various limit cases discussed in the
previous section.  
We find two resonance
maxima symmetric with respect to the origin (for $\epsilon=0.3$ at
$\omega_{\rm{max}}=\pm 0.65528$) resulting from the 
coupling of $m=1$ and $m=-1$ (see red and blue curves in
Fig.~\ref{fig::mpm3}a), 
but the resonance condition is no longer determined by a single and
simple relation that involves the 
unperturbed frequencies $\alpha_1^i$ and/or $\alpha_3^i$.  
The location of the resonance maximum slightly depends on the amplitude of
the perturbation. This is shown in
Fig.~\ref{fig3d::gr_vs_omg_vs_eps}, which presents the
growth rates against $\omega$ and $\epsilon$
[Fig.~\ref{fig3d::gr_vs_omg_vs_eps}(a)], the respective maximum
against $\epsilon$ [Fig.~\ref{fig3d::gr_vs_omg_vs_eps}(b)], and the
increase of the maximum of the growth rate versus the perturbation
amplitude $\epsilon$ [Fig.~\ref{fig3d::gr_vs_omg_vs_eps}(c)].
\captionsetup[subfigure]{margin=0.3cm,singlelinecheck=false,format=plain,
                         indention=0.45cm,justification=raggedright,
                         captionskip=0.2cm,position=bottom,
                         nearskip=0.0cm,farskip=0.0cm}
Beside the linear dependence of the location of the regimes with
parametric resonance on the perturbation frequency, a peculiar feature
is the existence of the third maximum around the origin.  This smaller
third local maximum results from an indirect interaction between
$m=\pm 1$ and $m=\pm 3$ without merging or splitting of the growth
rates and without phase locking [see Fig.~\ref{fig::mpm3}(c)].  A
similar phenomenon has been found for small $\epsilon \la 0.1$ in the
simulations at ${\rm{Rm}}=120$ (Fig.~\ref{fig::gr_vs_vortex1}).

\begin{figure}[b!]
\subfloat[Growth rates]{\includegraphics[width=0.49\textwidth]{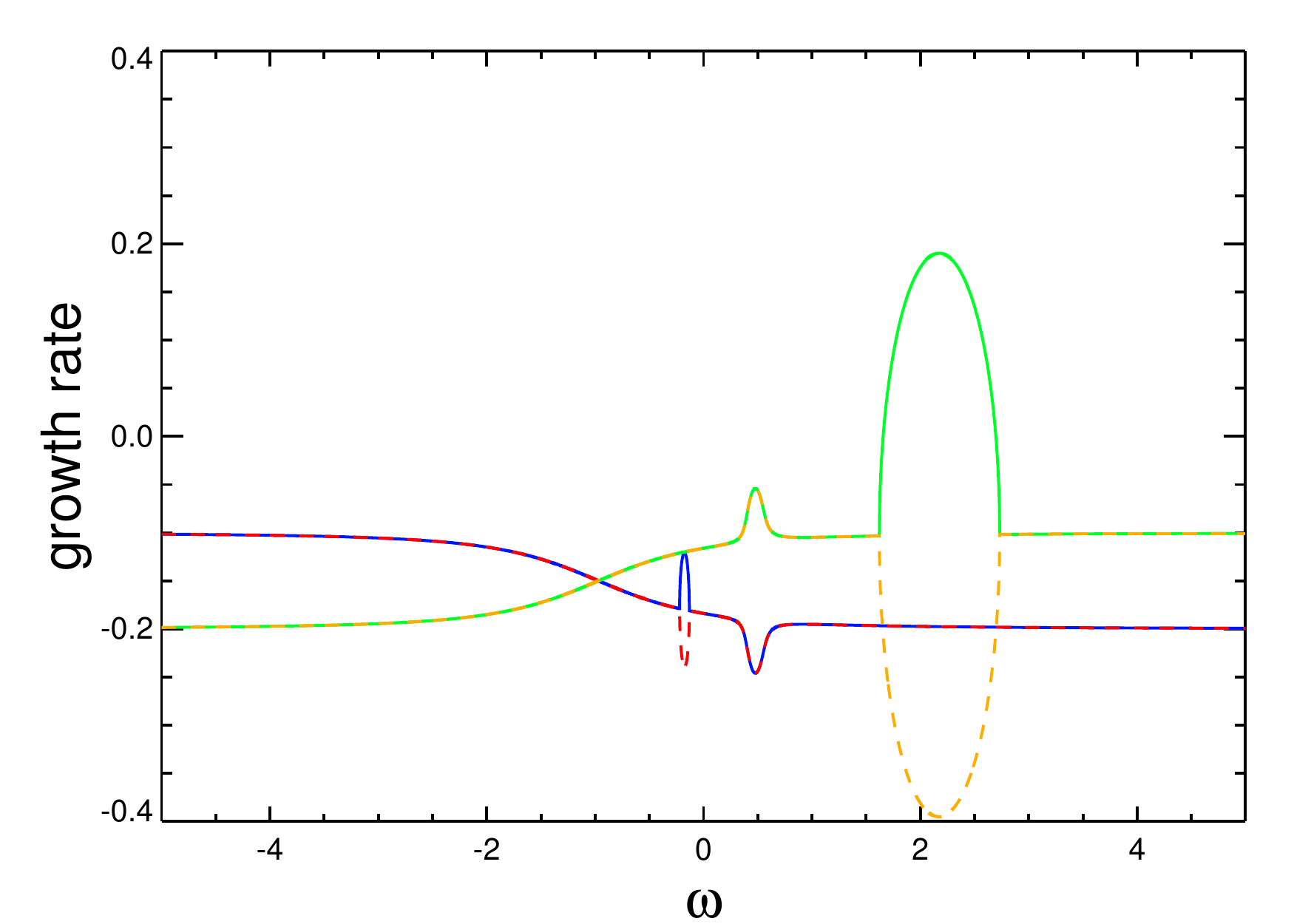}}
\subfloat[Frequencies]{\includegraphics[width=0.49\textwidth]{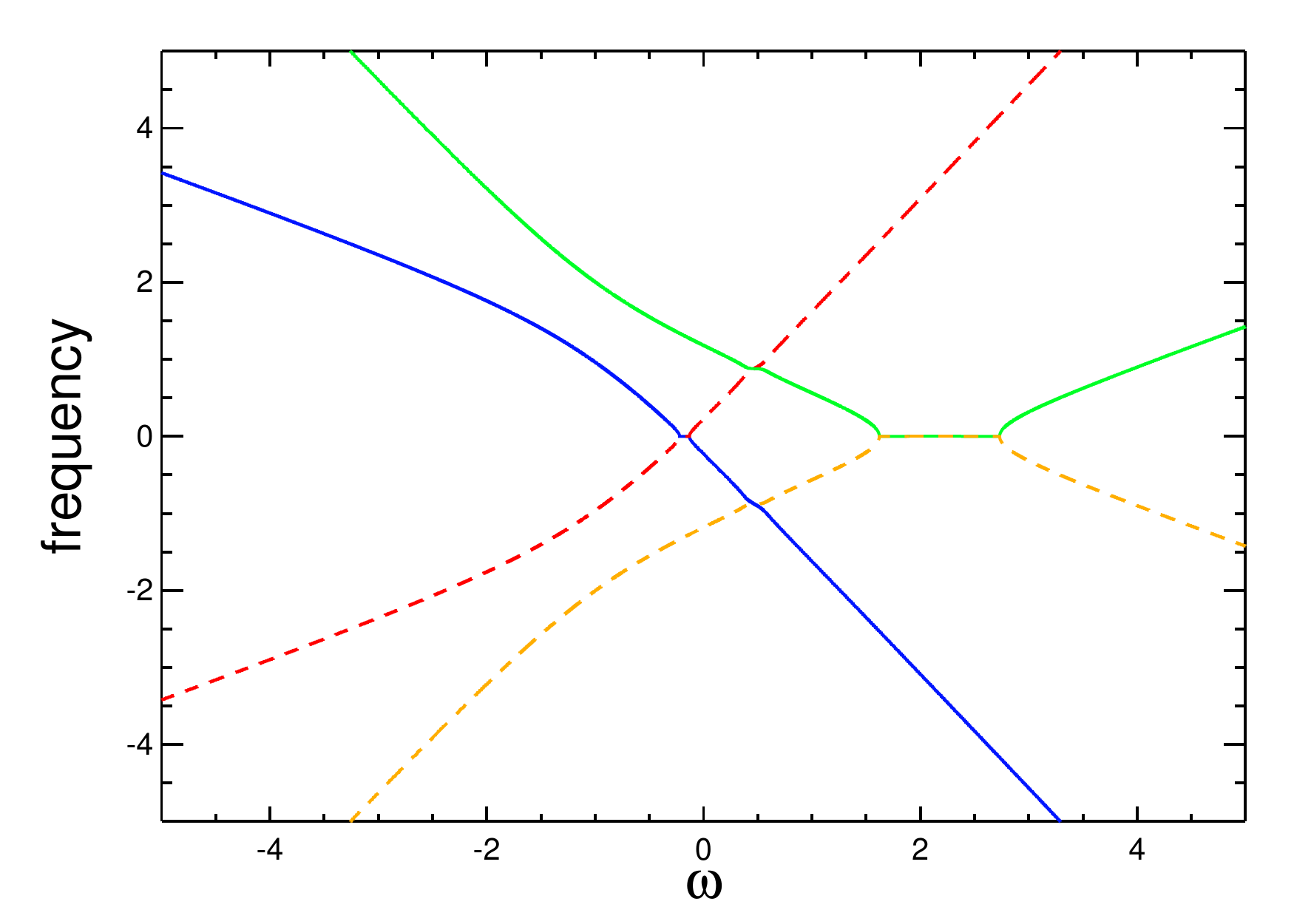}}
\caption{Asymmetric behavior of growth rates (a) and frequencies (b)
when the natural frequency of the system is $\neq
0$. $\alpha_1^r=-0.1, \alpha_1^i=1.0, \alpha_3^r=-0.2, \alpha_3^i=0,
|\gamma|^2=1, \delta_1\delta_2=-3,\epsilon=0.3$.  The blue and red
curves denote the solutions belonging to $\widetilde{\sigma}_{1,2}$,
and the green and orange curves denote the solutions belonging to
$\widetilde{\sigma}_{3,4}$.
}\label{fig::m1m3_symbreak}  
\end{figure}

Regarding the frequencies, we abandon the presentation of the
back-transformed frequencies in Fig.~\ref{fig::mpm3}.  We see a quite 
complex interaction of frequencies around the origin. Note in
particular the merging and splitting of the frequencies that belong to
the $m=-1$ and the $m=1$ branches [red and blue curve in
Figs.~\ref{fig::mpm3}c and d) which indicate the regions with phase
locking.  For large $\omega$ we obtain linear scalings $\propto \pm
\nicefrac{1}{2}{\omega}$ and $\propto \pm \nicefrac{3}{2}{\omega}$
which after the back-transformation with $P(t)=e^{iD_{\omega}t}$
correspond to a behavior $\propto \pm \omega$ and $\propto \pm
2\omega$.  Further complicated patterns are possible in particular when
considering complex couplings and/or a nonvanishing frequency for the
base modes ($\alpha_{1}^i$ and $\alpha_{3}^i$).
In contrast to the previous cases, a nonvanishing imaginary part of
an eigenvalue of the unperturbed state, like it occurs, for example in
the case of a parity-breaking flow that yields azimuthally propagating 
eigenmodes, results in a system of equations that is no longer
symmetric with respect to a sign change of the perturbation frequency
$\omega$.   
This is shown, for example, in
Fig.~\ref{fig::m1m3_symbreak} with a clear asymmetric behavior with
respect to $\omega$ when $\alpha_1^i \neq 0$.
\captionsetup[subfigure]{margin=0.2cm,singlelinecheck=false,format=plain,
                         indention=0.45cm,justification=justified,
                         captionskip=0.1cm,position=bottom}
Nevertheless, as previously, we find that for sufficiently high
frequencies the 
influence of the perturbation on the growth rate is negligible and the
growth rates approach again their unperturbed values (but note the
exchange of the $m=\pm 1$ and $m=\pm 3$ branches in
Fig.~\ref{fig::m1m3_symbreak}).  This behavior may allow an
allocation of each curve to the original azimuthal modes and to
determine which azimuthal modes couple or interact in order to form
the coupled eigenfunction of the perturbed problem.

\subsubsection{Cut off at $M = 5$}

Not surprisingly the behavior becomes more complex when increasing the
truncation level to $M = 5$.  Here we abstain from specifying the
characteristic equation which by virtue of its length and its
complicated structure does not offer any additional insights. We limit
ourselves to three examples that present typical solution patterns for
growth rate and frequency (Fig.~\ref{fig::m1m3m5}).  The extension
of the truncation level goes along with new parameters that describe
the eigenvalues of the unperturbed new mode, denoted with $\alpha_5^r$
and $\alpha_5^i$, and the interaction of the new mode with itself
and/or the $m=3$ mode, denoted by $\delta_3\delta_4$.  Again we see a
combination of resonances with phase locking and amplification, and we
have a symmetric pattern with respect to the sign of the perturbation
frequency as long as the imaginary parts of the involved dynamo modes
and the coupling coefficients remain zero.

\newcommand{\wid}{0.33}
\captionsetup[subfigure]{margin=0.4cm,singlelinecheck=false,format=plain,
                         indention=0.39cm,justification=justified,
                         captionskip=0.1cm,position=bottom}
\begin{figure}[h!]
\subfloat[$\alpha_1^r\!=\!-0.1, \alpha_3^r\!=\!-0.2,\alpha_5^r\!=\!-0.3,
           \alpha_{1,3,5}^i\!=\!0,\newline |\gamma|^2=1, \delta_1\delta_2=\delta_3\delta_4=-10$]
{\label{sf::17a}\includegraphics[width=\wid\textwidth]{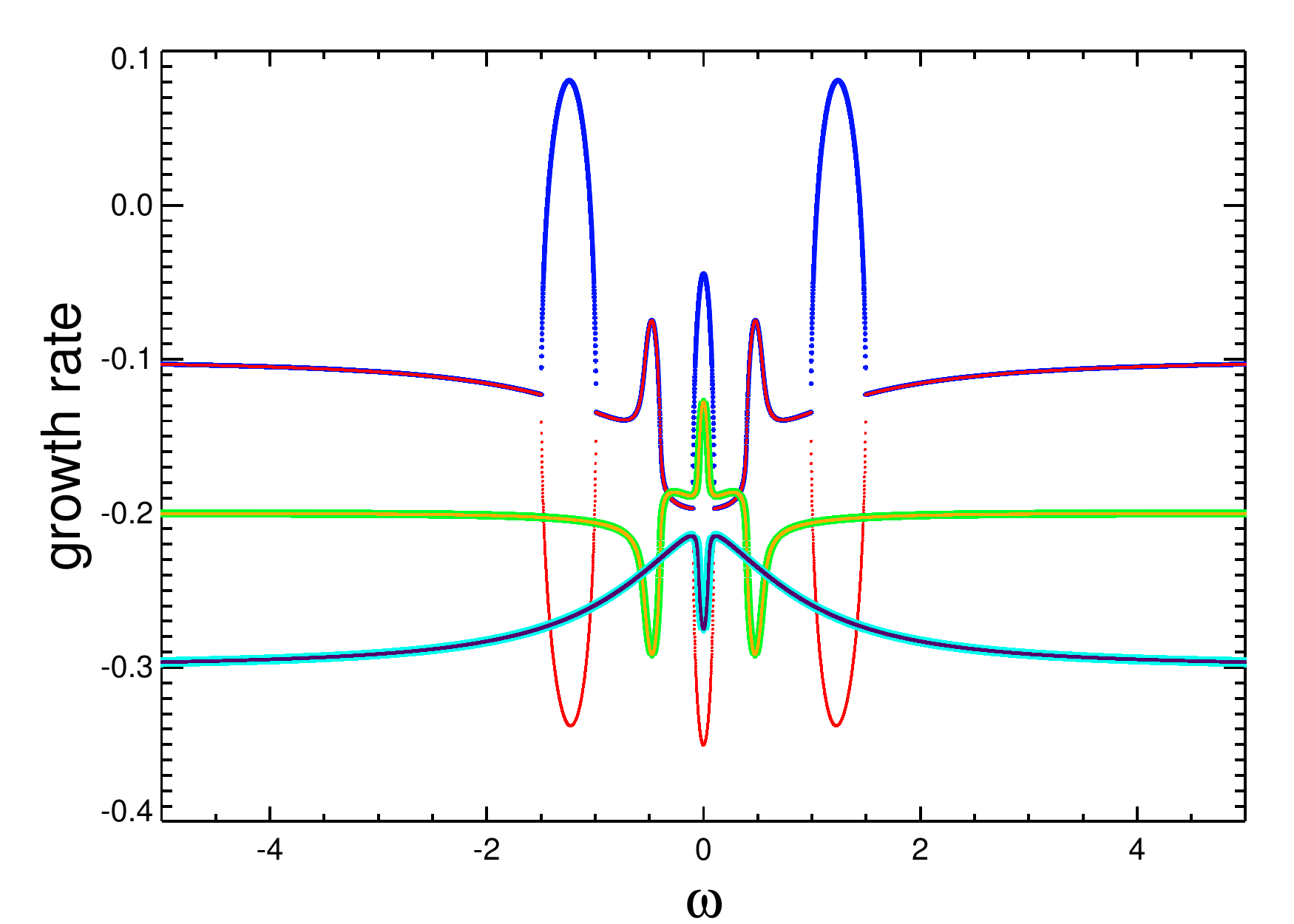}}
\subfloat[\label{sf::17b}$\alpha_1^r\!=\!-0.1, \alpha_3^r\!=\!-2.0,
\alpha_3^r\!=\!-3.0, \alpha_{1,3,5}^i\!=\!0, \newline |\gamma|^2=1, \delta_1\delta_2=\delta_3\delta_4=-10$]
{\includegraphics[width=\wid\textwidth]{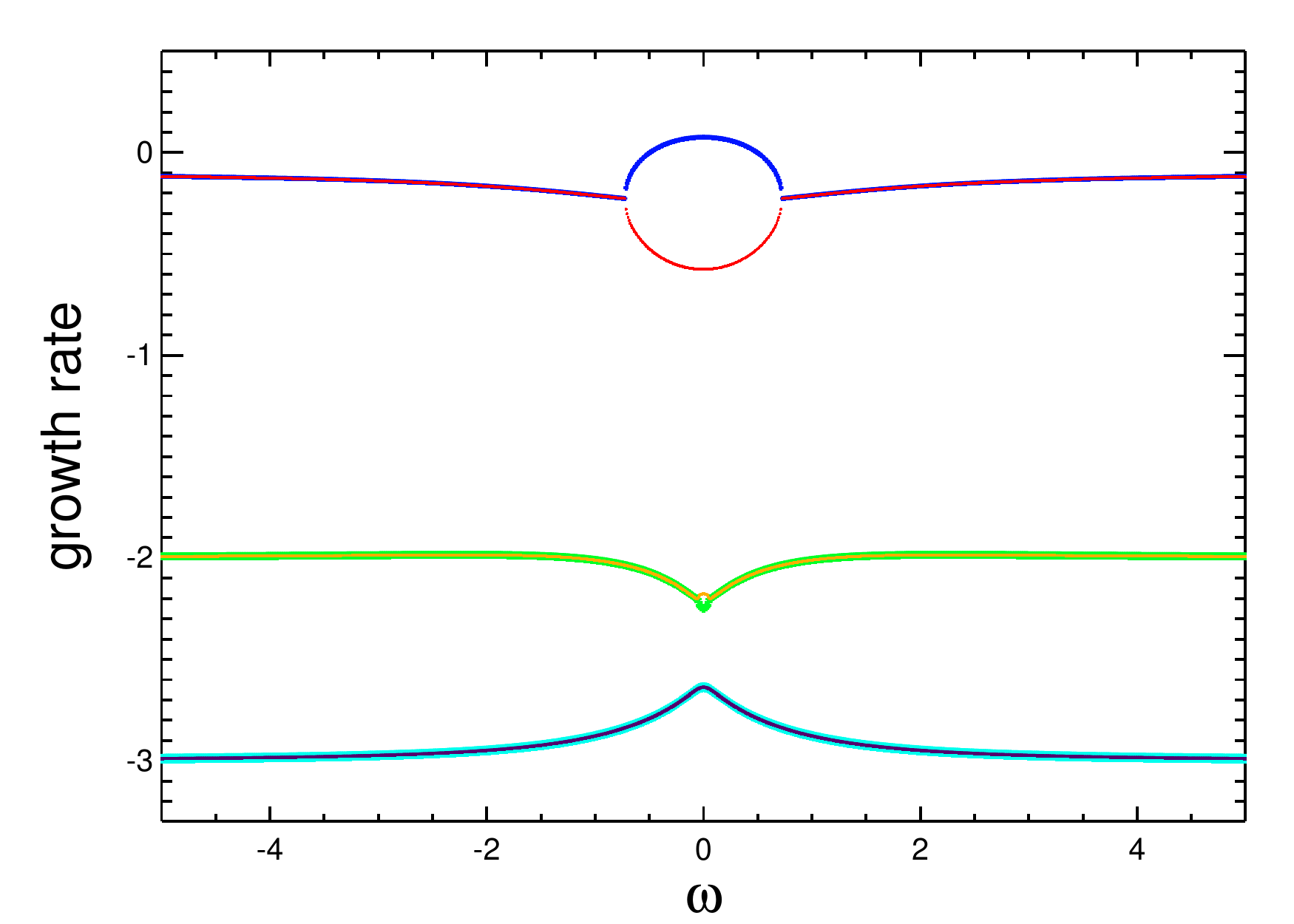}}
\subfloat[$\alpha_1^r\!=\!-0.1,\alpha_3^r\!=\!-0.2,\alpha_5^r\!=\!-0.3,
\alpha_{1,3,5}^i\!=\!0,\newline |\gamma|^2=5,\delta_1\delta_2=-3, \delta_3\delta_4=+3$]
{\label{sf::17c}\includegraphics[width=\wid\textwidth]{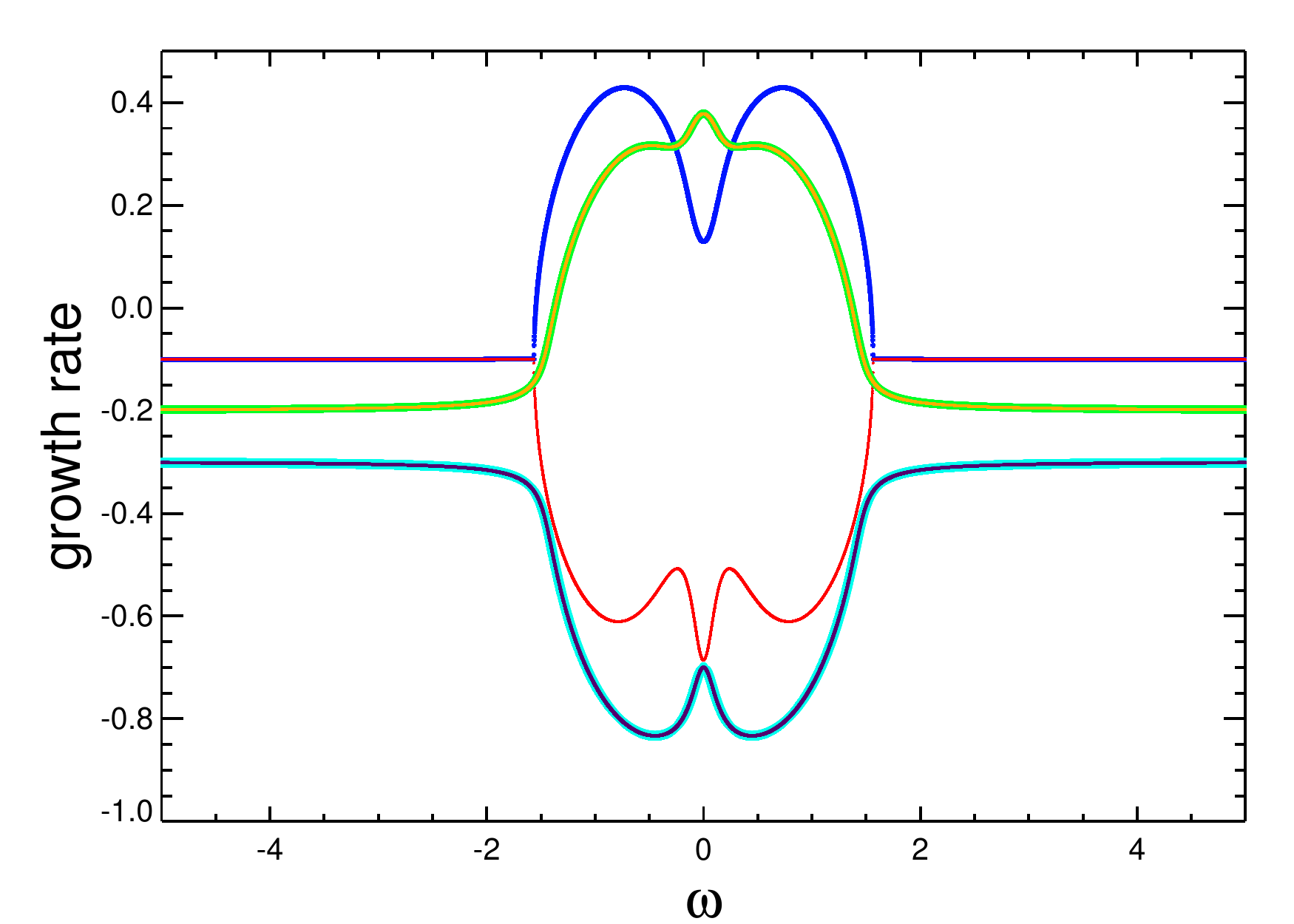}}
\\
\subfloat[$\alpha_1^r\!=\!-0.1, \alpha_3^r\!=\!-0.2,\alpha_5^r\!=\!-0.3,
\alpha_{1,3,5}^i\!=\!0,\newline |\gamma|^2=1, \delta_1\delta_2=\delta_3\delta_4=-10$]
{\label{sf::17d}\includegraphics[width=\wid\textwidth]{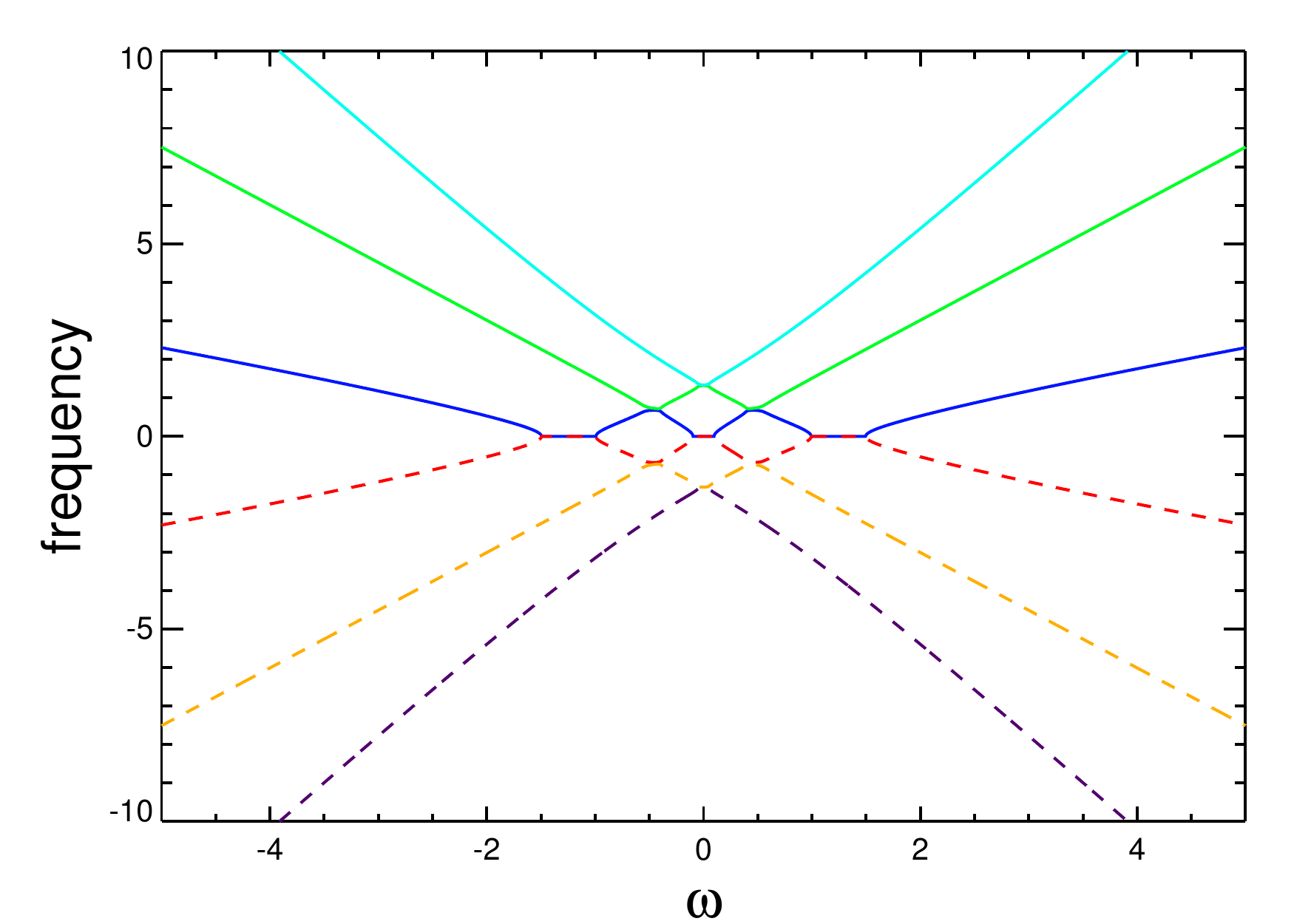}}
\subfloat[\label{sf::17e}$\alpha_1^r\!=\!-0.1, \alpha_3^r\!=\!-2.0,
\alpha_3^r\!=\!-3.0, \alpha_{1,2,3}^i\!=\!0, \newline |\gamma|^2=1,
\delta_1\delta_2=\delta_3\delta_4=-10$]
{\includegraphics[width=\wid\textwidth]{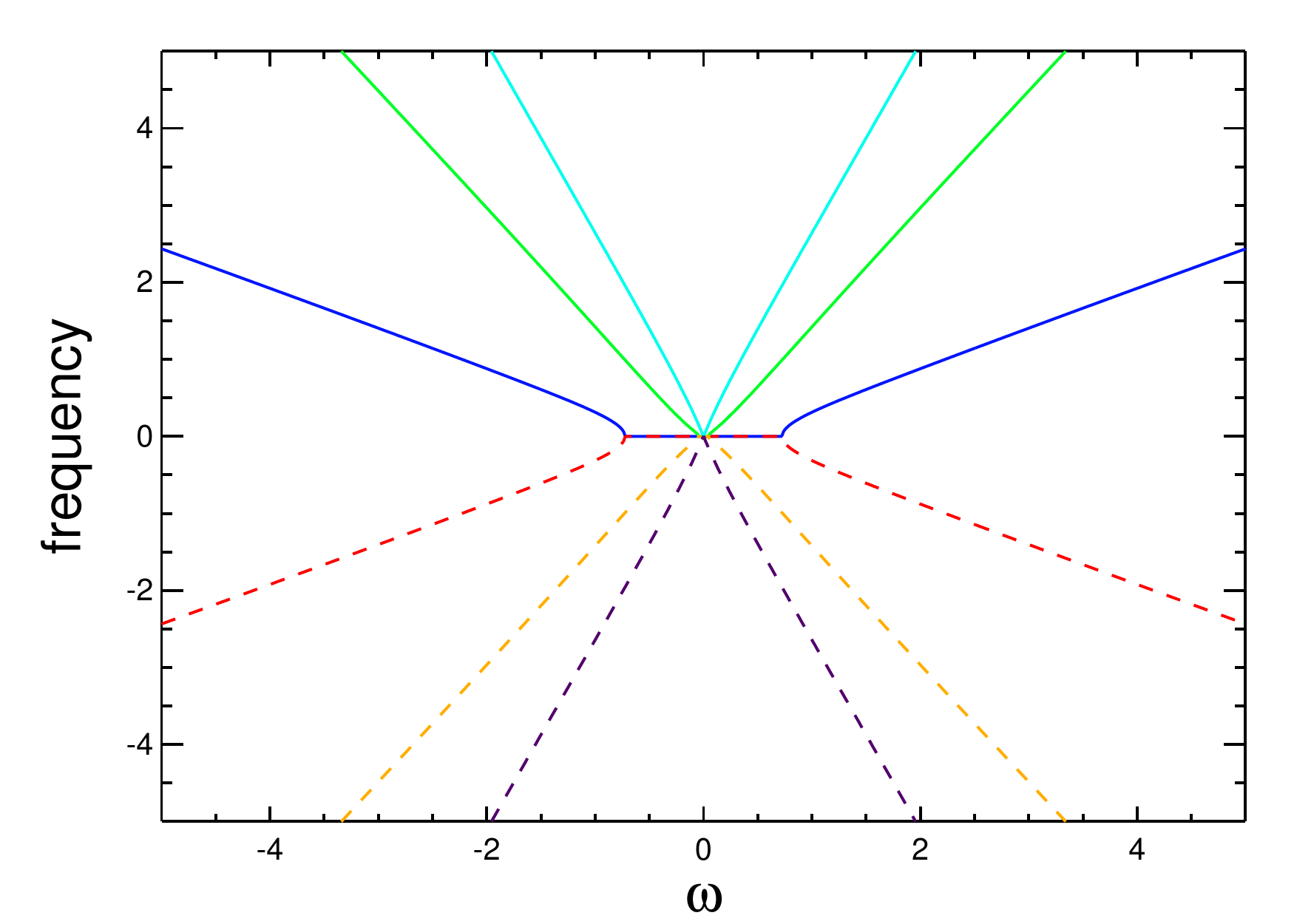}}
\subfloat[$\alpha_1^r\!=\!-0.1,\alpha_3^r\!=\!-0.2,\alpha_5^r\!=\!-0.3,
\alpha_{1,3,5}^i\!=\!0,\newline |\gamma|^2=5,\delta_1\delta_2=-3, \delta_3\delta_4=+3$]
{\label{sf::17f}\includegraphics[width=\wid\textwidth]{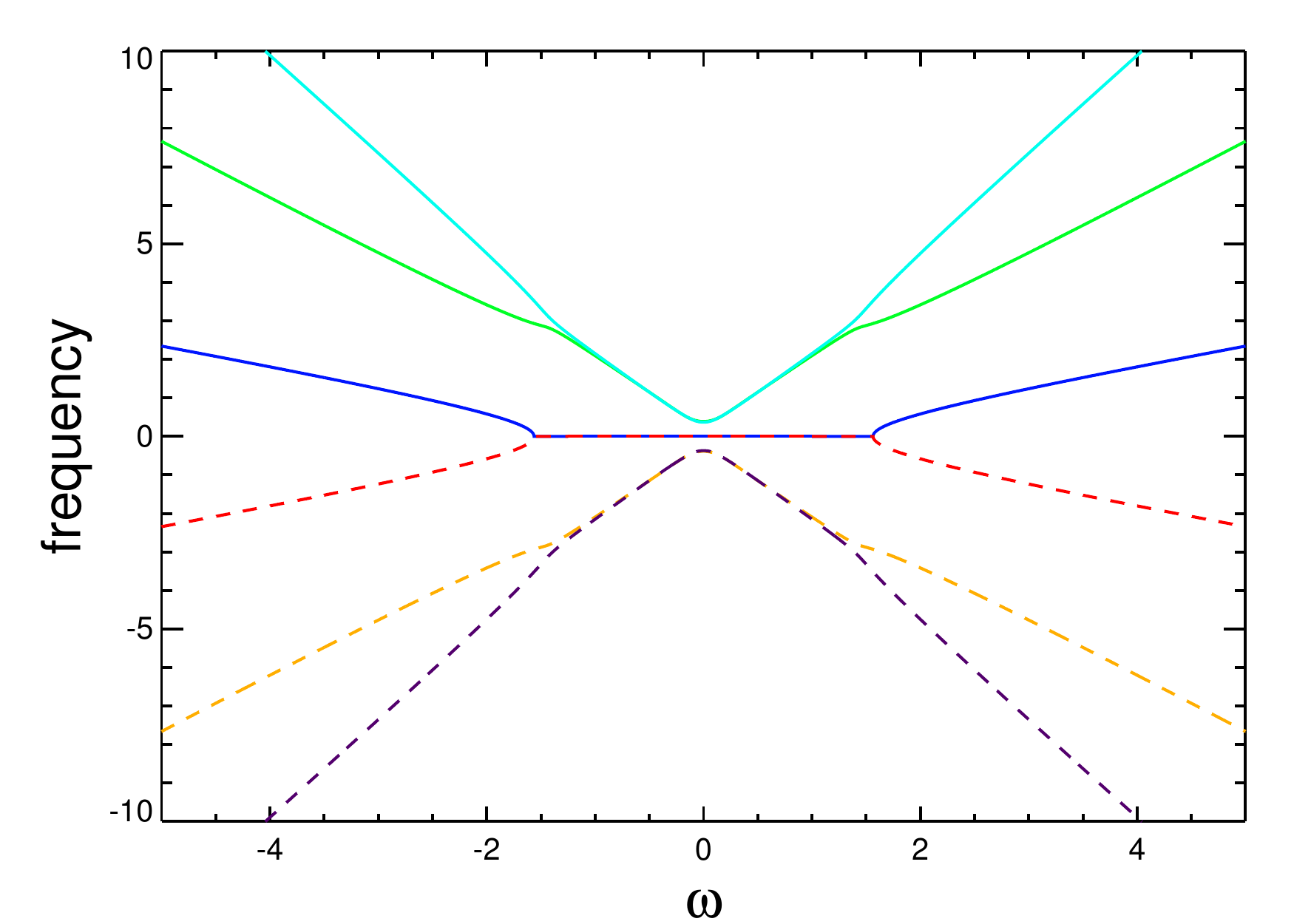}}
\caption{
Growth rates (a,b) and frequencies (c,d) versus perturbation frequency
for $\epsilon=0.3$ Note the large difference of the unperturbed growth
rates on the case shown in (b) and (e).  The red and blue (green and
yellow, light-blue and violet) curves denote the solutions
$\widetilde{\sigma}_{1,2}$ ($\widetilde{\sigma}_{3,4}$,
$\widetilde{\sigma}_{5,6}$).
\label{fig::m1m3m5}}
\end{figure}

The behavior of the growth rates confirms the impression from the
previous paragraph. A resonant behavior characteristic for a
parametric instability with phase locking results only from the
coupling of $m=1$ and $m=-1$ (blue and red curves in
Fig.~\ref{fig::m1m3m5}) whereas an interaction of $m=\pm 1$ with
$m=\pm 3$ or $m=\pm 5$ happens only indirectly, producing a smoother
parametric amplification patterns.  Furthermore, we do not only see
parabolic shapes around individual resonance maxima but also broader
regimes composed of several local maxima [see, e.g., blue curves in
Fig.~\ref{fig::m1m3m5}(c)].  Particularly striking is the fact that
the regimes with an enhancement of the growth rate are significantly
broadened, allowing a transition from a stable solution to an unstable
solution for a wide range of perturbation frequencies.  The pattern
becomes simpler when the original modes have larger differences in
their unperturbed growth rates (Fig.~\ref{sf::17b}) so that the
interaction of modes with different $|m|$ would require larger values
for the corresponding interaction parameters.


\subsection{The impact of the truncation level $M$}

\captionsetup[subfigure]{margin=0.2cm,singlelinecheck=false,format=plain,
                         indention=0.45cm,justification=justified,
                         captionskip=-0.0cm,position=bottom}
\begin{figure}[h!]
\subfloat[$\delta_1\delta_2=\delta_3\delta_4=0$ corresponding to a
truncation at $M=1$]
{\includegraphics[width=0.32\textwidth]{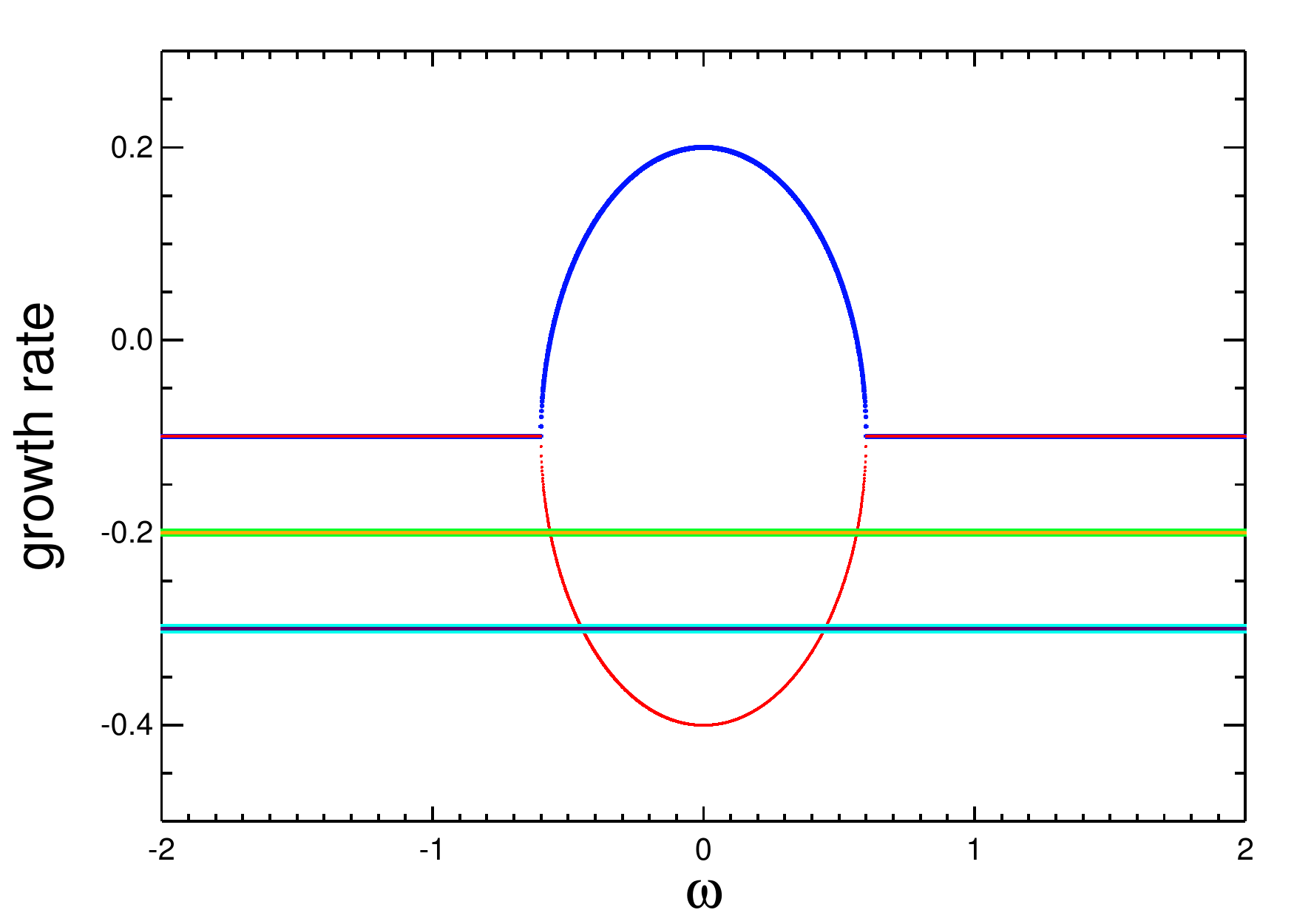}}
\subfloat[$\delta_1\delta_2=-3, \delta_3\delta_4=0$ corresponding to a
truncation at $M=3$]
{\includegraphics[width=0.32\textwidth]{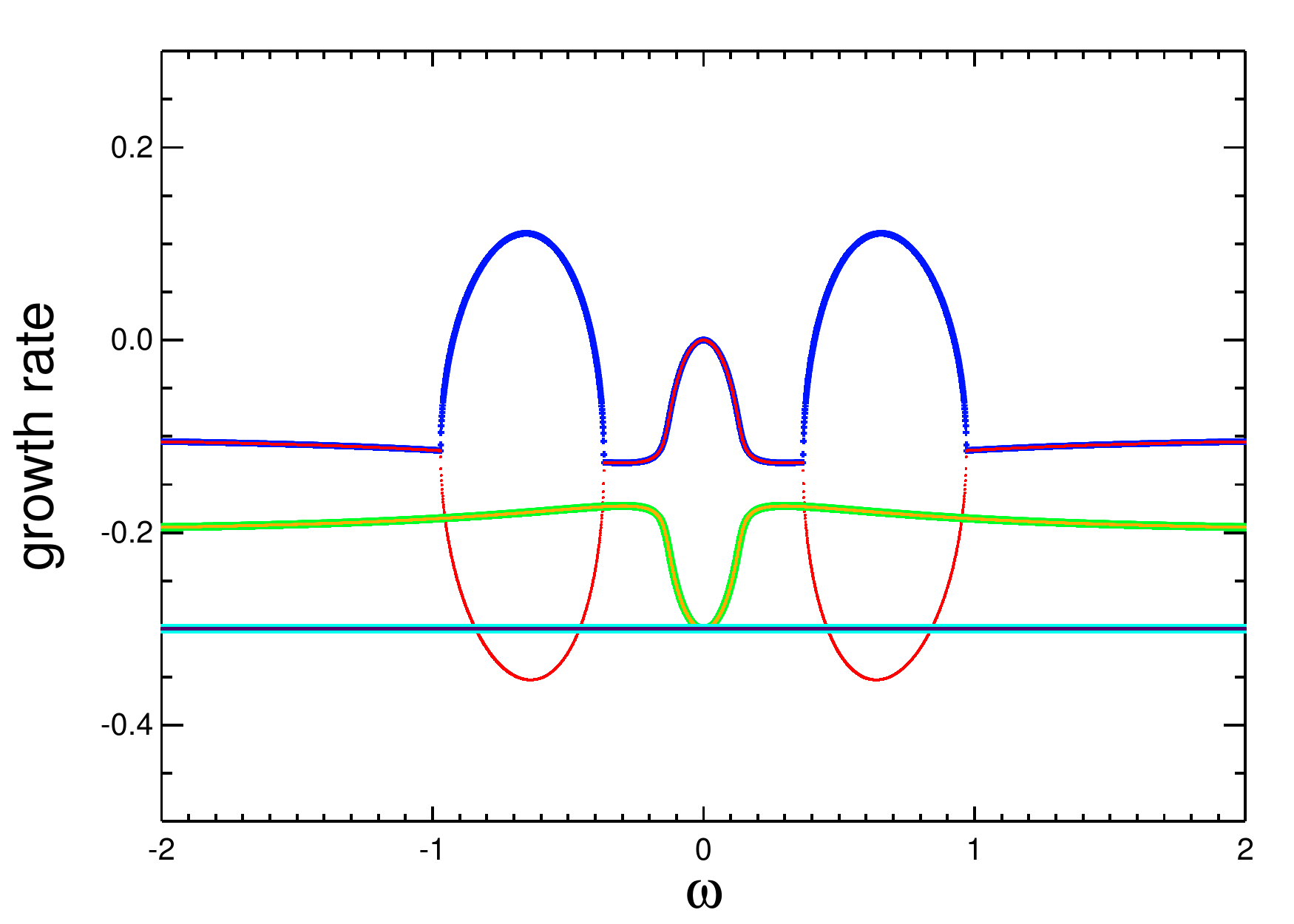}}
\subfloat[$\delta_1\delta_2=-3, \delta_3\delta_4=-3$ corresponding to a
truncation at $M=5$]
{\includegraphics[width=0.32\textwidth]{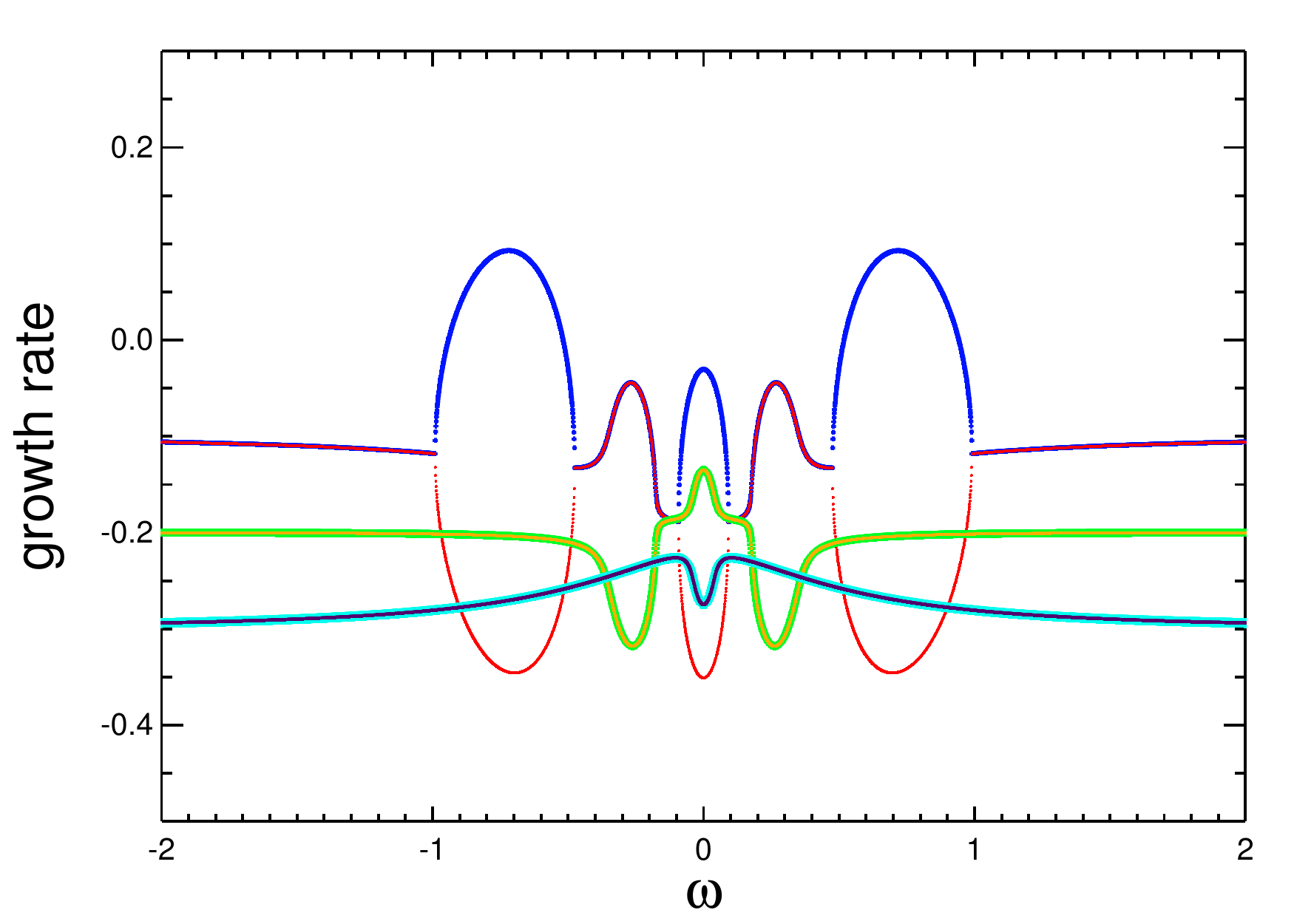}}
\caption{
Comparison of the solutions for the truncation level $M=1\cdots 5$
using different values for the interaction parameters
$\delta_1\delta_2$ and $\delta_3\delta_4$.  The plots correspond to
solutions for increasing order of truncation.  For all cases we use
$\alpha_1^r=-0.1, \alpha_3^r=-0.2, \alpha_5^r=-0.3$,
$\alpha_{1,3,5}^i=0$, $|\gamma|^2=1$, and $\epsilon=0.3$.  The red and
blue (green and yellow, light-blue and violet) curves denote the
solutions $\widetilde{\sigma}_{1,2}$ ($\widetilde{\sigma}_{3,4}$,
$\widetilde{\sigma}_{5,6}$).
\label{fig2d::gr_vs_omg_check_trunc}}
\end{figure}

\captionsetup[subfigure]{margin=0.2cm,singlelinecheck=false,format=plain,
                         indention=0.45cm,justification=justified,
                         captionskip=-0.0cm,position=bottom}
\begin{figure}[h!]
\subfloat[$\delta_3\delta_4=1$]{\includegraphics[width=0.32\textwidth]{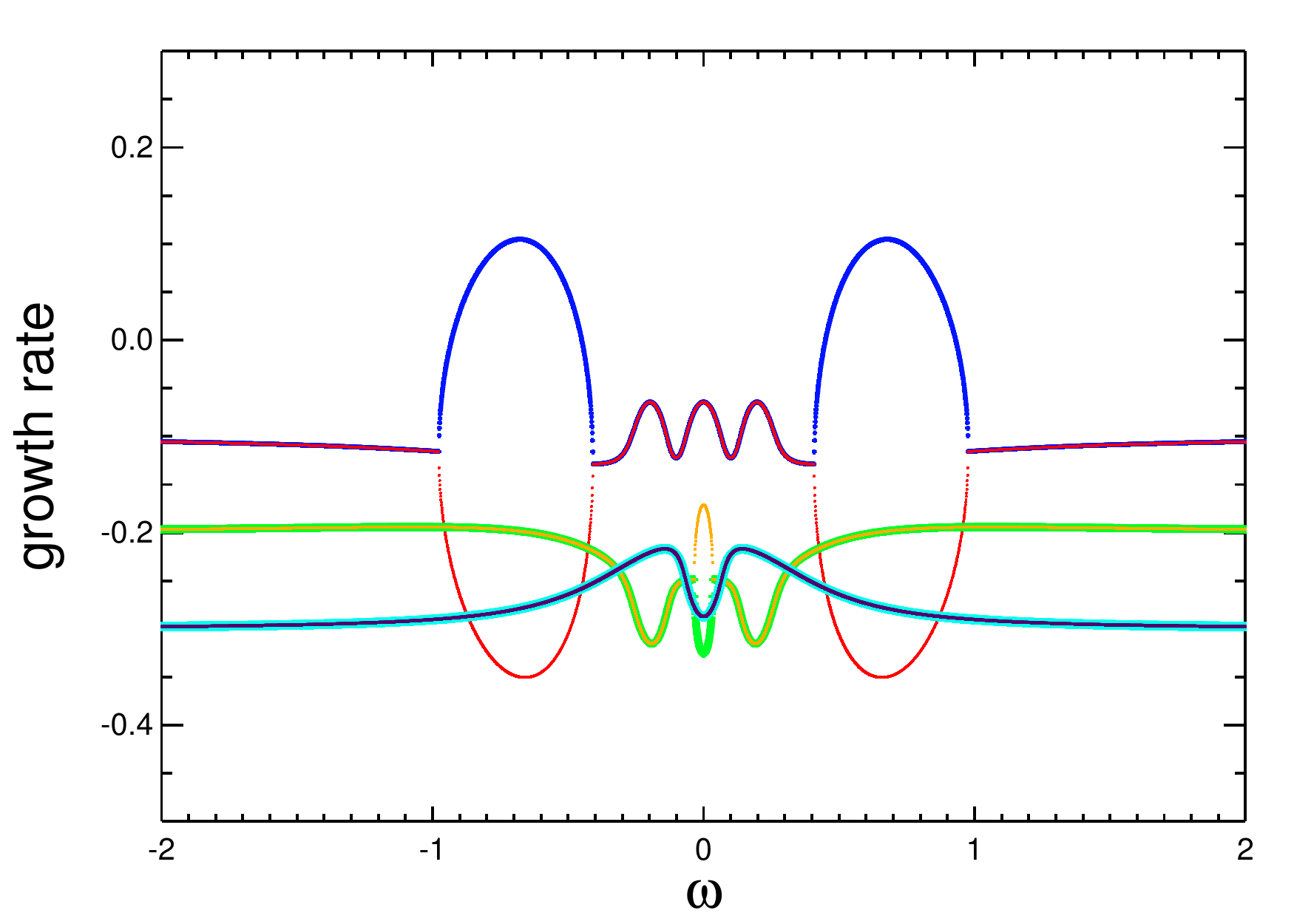}}
\subfloat[$\delta_3\delta_4=0.3$]{\includegraphics[width=0.32\textwidth]{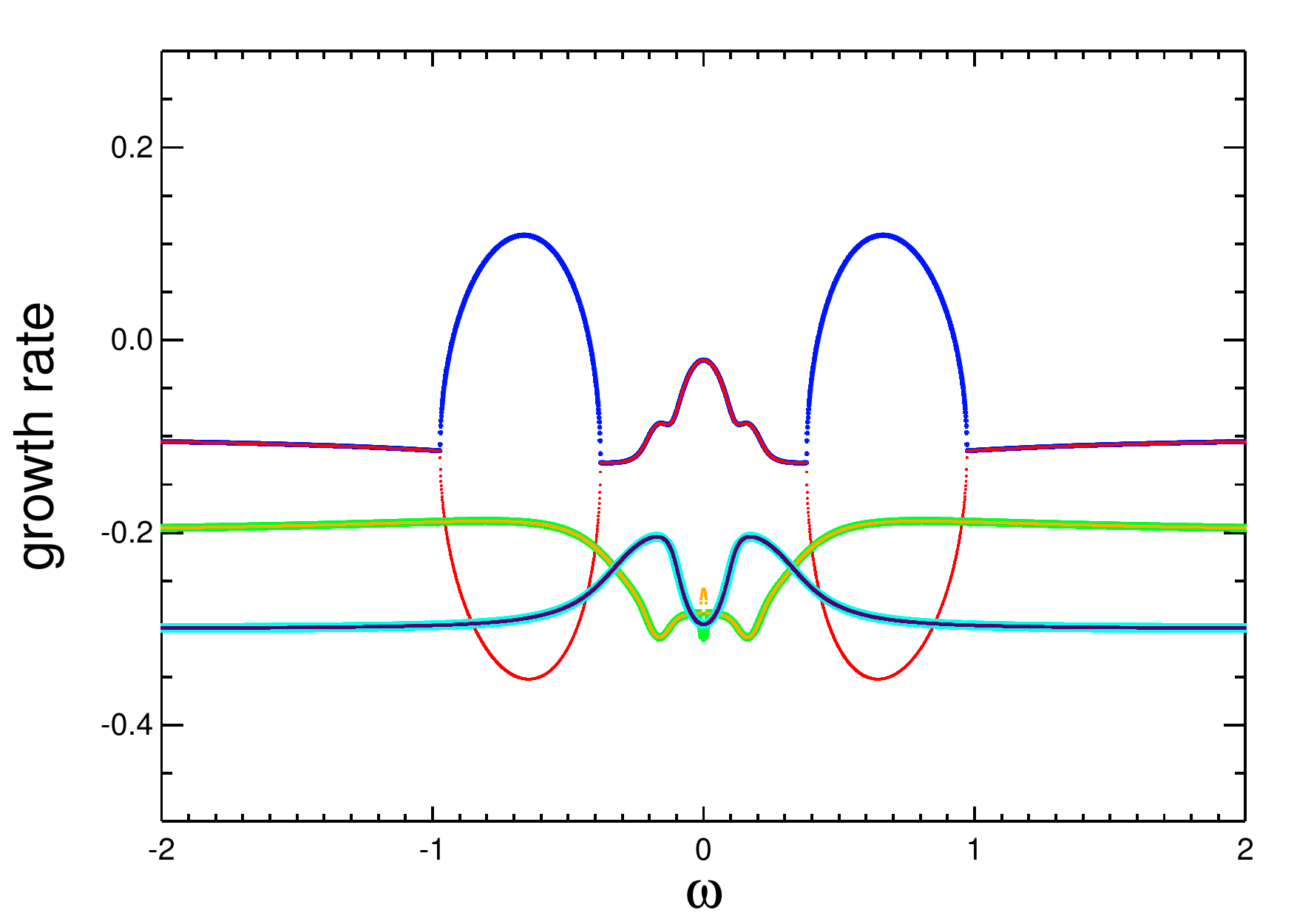}}
\subfloat[$\delta_3\delta_4=0.1$]{\includegraphics[width=0.32\textwidth]{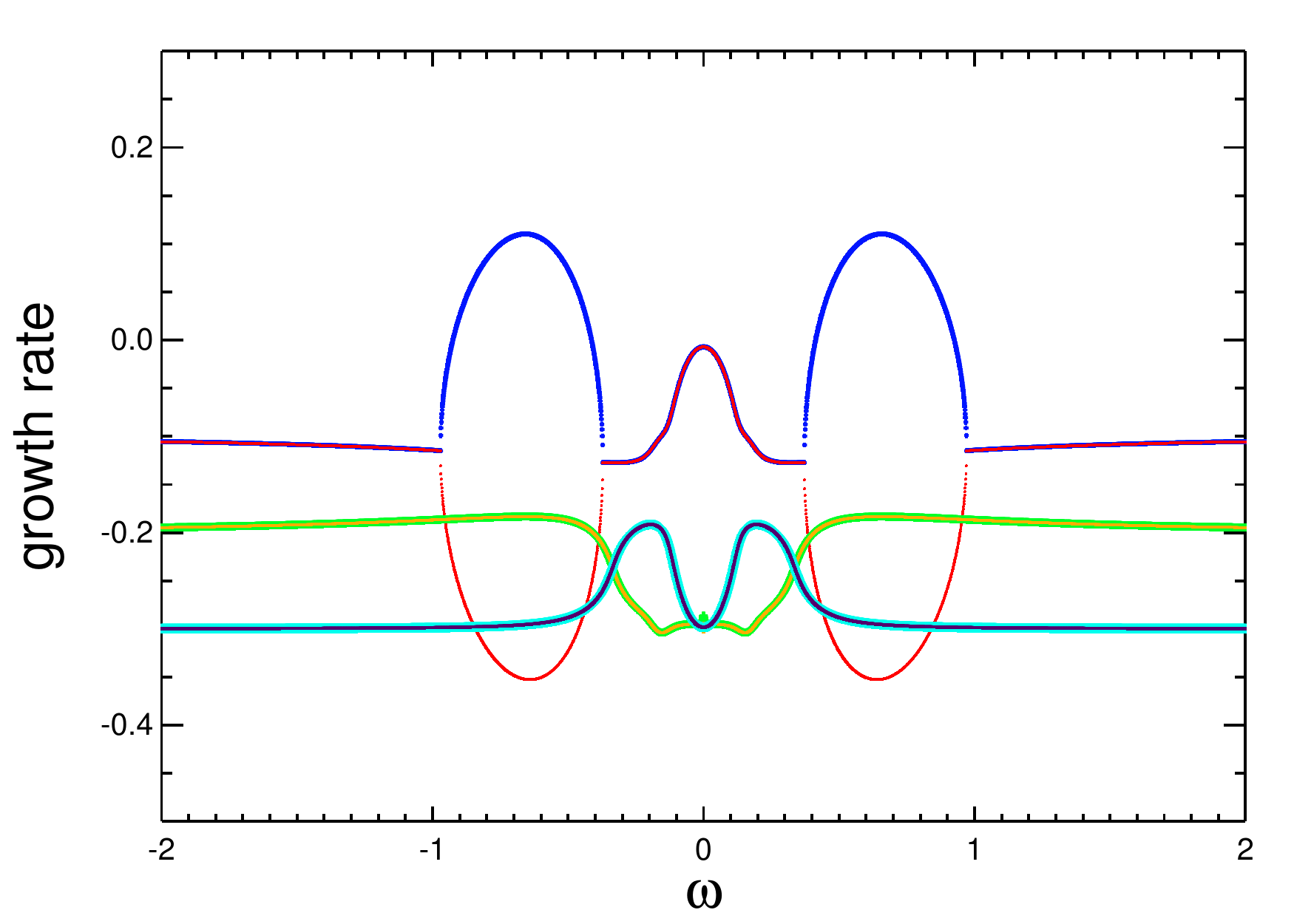}}
\caption{
Comparison of the solutions for the truncation level $M=5$ and for
decreasing values of $\delta_3\delta_4$.  For all cases we use
$\alpha_1^r=-0.1, \newline \alpha_3^r=-0.2, \alpha_5^r=-0.3$,
$\alpha_{1,3,5}^i=0$, $|\gamma|^2=1, \delta_1\delta_2=-3,$, and
$\epsilon=0.3$.  The red and blue (green and yellow, light-blue and
violet) curves denote the solutions $\widetilde{\sigma}_{1,2}$
($\widetilde{\sigma}_{3,4}$, $\widetilde{\sigma}_{5,6}$).
\label{fig2d::gr_vs_omg_check_trunc2}}
\end{figure}

The discussion of the impact of the truncation order is not
straightforward, because when increasing the order from $M$ to $M+2$,
new parameters get involved that parametrize the interaction of the
new modes with themselves and with the adjacent mode with $m=M-2$.  In
fact, putting the new parameters to zero always leads to the solutions
corresponding to the lower order truncation plus a solution of the
type denoted in Eq.~(\ref{eq::weakcoupling}) for a single mode that
emerges without interaction with the
system. Figure~\ref{fig2d::gr_vs_omg_check_trunc} demonstrates the
change in the solutions when increasing the truncation level.  Note
that for a truncation level $M=5$ the solutions for
$\delta_1\delta_2=\delta_3\delta_4=0$ correspond to the solutions at
order $M=1$ [Fig.~\ref{fig2d::gr_vs_omg_check_trunc}(a)]. Likewise,
for $\delta_1\delta_2\neq 0$ and $\delta_3\delta_4=0$ we recover the
solutions at order $M=3$ [Fig.~\ref{fig2d::gr_vs_omg_check_trunc}(b)].

Furthermore, the addition of new modes that come along with an
increase of the truncation level results in a change of the structure
of the solutions so that the behavior of the growth rates becomes more
complex.  The enlargement of the system by adding new modes always
results in the emergence of new features in the pattern of the growth
rates due to the interaction that become possible with the new modes.
The structural changes become less important when the parameters that
come along with increasing the truncation level decrease in amplitude.
Figure~\ref{fig2d::gr_vs_omg_check_trunc2} shows the results for a
truncation at $M=5$ with a fixed parameter $\delta_1\delta_2=-3$ and
for decreasing values of $\delta_3\delta_4$ (the parameter that
describes the interaction between $m=\pm 3$ and $m=\pm 5$ modes). The
resulting behavior of the growth rates corresponds to the smooth
transition of the state shown in
Fig.~\ref{fig2d::gr_vs_omg_check_trunc}(c) to the state with
$\delta_3\delta_4=0$ (corresponding to the truncation $M=3$) shown in
Fig.~\ref{fig2d::gr_vs_omg_check_trunc}(b).  We conclude, that the
analytical model only makes sense when for increasing truncation level
the supervened parameters become negligible at a certain $M$ so that
the consideration of further modes with large $M$ does not cause any
further significant change in the structure of the solutions.


\section{Discussion and Conclusions}

We have examined dynamo action driven by a mean axisymmetric flow
subject to spatio-temporal periodic perturbations. We have compared
the outcome of a three-dimensional numerical dynamo model in
cylindrical geometry with analytic results obtained from a simple low
dimensional model for the magnetic field amplitude.  Our computations
show that a periodic excitation by means of a nonaxisymmetric flow
perturbation can be beneficial for dynamo action when the perturbation
frequency is in the right range.  The essential properties shown by
growth rates and frequencies of the eigenmode in the simulations can
be qualitatively reproduced in the analytical model.  This concerns in
particular the occurrence of a localized strong increase of the growth
rate and broader regimes with weaker enhancement when different
azimuthal field modes are coupled.  We also find less obvious details
that occur in both approaches, like the symmetrical appearance of
parametric resonances and the weak linear dependence of the location
of the maximum on the amplitude of the perturbation.  However, a
quantitative agreement between the two models does not persist. This
could not be expected because of the simplicity of the analytical
model that does not take into account the particular geometry of the
flow and is based on a decomposition of the magnetic field into
azimuthal modes that are not exact eigenmodes of the (disturbed)
system.  In that sense our analytical approach is not equivalent to a
perturbation method in a systematic mathematical meaning.

The parametric resonances found in our simulations are characterized
by a strong gain in the growth rate in a narrow range of excitation
frequencies, a nonoscillating amplitude of the field and the linking
of the magnetic phase to the phase of the perturbation.  The
resonances have also been found in previous simulations with a different
spatial structure of the periodic disturbance (see also our previous study
\cite{2012PhRvE..86f6303G}) as well as in our analytical model in
which the shape of the disturbance is not specified at all. Hence we
suppose that the shape of the perturbations is not important for the
occurrence of the observed resonances.

Depending on the parameters in both approaches, simulations and
analytical model, we obtain either a single peak around $\omega=0$ or
two peaks symmetrical located around the origin.  Broader regimes with
an enhancement of the growth rates apparently emerge due to the
coupling of eigenmodes with a different azimuthal wavenumber.

Our analytic model suggests that the parametric resonances in the
perturbed system are based on the coupling of the azimuthal modes with
the same modulus of the azimuthal wave number but with different sign
($m=1$ and $m=-1$ or $m=3$ and $m=-3$ in the example shown in
Fig.~\ref{fig::m1m3_symbreak}).  This corrects the speculations from
our previous study where it was assumed that the parametric resonance
in the perturbed dynamo is based on the interaction of $m=1$ and $m=3$
\cite{2012PhRvE..86f6303G}.

The analytical low-dimensional model is not intended to perfectly
reproduce the simulations, which, indeed, is not even possible,
because neither the basic flow field nor the pattern of the
perturbation is considered in the low dimensional Ansatz.  Instead it
demonstrates how different eigenmodes of an unperturbed state become
coupled by a perturbation and thus impact the dynamo ability of the
system. In that sense the low dimensional model reflects some
essential properties of our three-dimensional simulations and provides
a plausible explanation for the complex behavior of the growth rates
in different regimes with amplification of the field generation
process like appear in our simulations at ${\rm{Rm}}=120$.
Individual features, such as the single peak around $\omega = 0$ for
${\rm{Rm}}=30$, the occurrence of the double maximum symmetrical about
the origin with a smaller maximum in-between, or the existence of
further broad secondary maxima without parametric resonance, are well
reproduced.  However, there is no real consistency between the
analytic model and the simulations and a direct comparison or the
reproduction of the particular pattern seen e.g. in
Figs.~\ref{sf::6c} and ~\ref{sf::6d} remains impossible.

The resonant behavior is similar to the behavior of periodically
perturbed mechanical systems but the resonance condition only
approximately fulfills the well-known relation
$\omega_{\rm{res}}=2\omega_0$ (with $\omega_0$ the natural frequency
of the unperturbed system) when a larger number of azimuthal field
modes is involved.  Furthermore, in accordance with the findings of
Ref.~\cite{1992A&A...264..319S}, the eigenvalues are no longer determined
by a Mathieu-like equation even though the solutions show a similar
behavior.

We find additional extended regimes with a significant enhancement of
the growth rate but less pronounced than in the resonant case and
without locking of the field to the phase of the perturbation.  Our
analytic model shows that the emergence of the amplification regimes
requires the interaction of dynamo modes with distinct azimuthal wave
numbers and a basic state which consists of dynamo modes with growth
rates that are sufficiently close to each other.  The vast regimes
with parametric amplification have not been found in previous models,
because either the higher magnetic field modes were not even
considered (e.g., in the galactic dynamo models by
Ref.~\cite{1990MNRAS.244..714C}) or they decayed on a very fast time scale
because the interaction triggered by the nonaxisymmetric perturbation
has been too weak to become effective \cite{2012PhRvE..86f6303G}.

The very parametric resonance is limited to a narrow regime of rather
small perturbation frequencies so that a realization in natural
dynamos might be rather unlikely but cannot be ruled out.
Nevertheless, the vast regimes with significant amplification caused
by the interaction of modes with different azimuthal wave number in
the analytical model and in the simulations may be still sufficient to
turn a subcritical system into a dynamo with exponentially growing
magnetic field even when the perturbation amplitude remains small.
This nonresonant parametric amplification can also be of importance
for long-term variations of magnetic activity as, for example, found in
the solar dynamo models of Ref.~\cite{2015AstL...41..374K}.

We have restricted our interest to linear models with a prescribed
flow. However, it has been found that nonaxisymmetric perturbations
also impact the nonlinear state of a dynamo
\cite{1999A&A...347..860R} and may even change the fundamental
character of the dynamo by triggering hemispheric asymmetries or
cyclic changes of the large-scale magnetic field orientation known as
flip-flop phenomenon or active longitudes \cite{2015ApJ...813..134P}.
We further restricted our examinations to a perturbation pattern with
azimuthal wave number $\widetilde{m}=2$ resulting, e.g., from tidal
forces in a two-body system.  A possible astrophysical application may
be the impact on flow driving in planetary dynamos in particular when
considering large Jupiter-like exo-planets that closely surround their
host stars or resonance effects for stellar dynamo models
\cite{Stefani2016}.  A possibility for a {\it{natural}} appearance of
a perturbation pattern with higher azimuthal wave number, say,
$\widetilde{m}=5$ and/or $\widetilde{m}=6$, are free inertial waves
that can be excited via triadic resonances, e.g., in a precession driven
flow \cite{2015NJPh...17k3044G,2015PhFl...27l4102H}, which in turn may
couple various magnetic field modes and thus improve the dynamo
capability of the system.

It is tempting to make use of the beneficial impact of a space-time
periodic regular perturbation on top of a given basic flow in order to
excite dynamo action in experiments like the French
von-K{\'a}rm{\'a}n-sodium (VKS) dynamo \cite{2007PhRvL..98d4502M} or
the Madison dynamo \cite{PhysRevLett.96.055002}.  Both experiments
utilize a flow of liquid sodium driven by two opposing and
counter-rotating impellers with the time-averaged flow similar to the
flow applied in our study\footnote{The Madison dynamo is running in a
sphere.}.  So far the VKS dynamo exhibited magnetic field
self-generation only in case of a flow driven by impellers made of a
ferromagnetic alloy \cite{2013PhRvE..88a3002M} whereas the Madison
dynamo did not show dynamo action at all.  For both experiments, a
reduction of the critical magnetic Reynolds number by $30\%$, as
found in our simulations, would be of great relevance.  Interestingly,
nonaxisymmetric vortex-like flow structures, which might represent
the role of a suitable disturbance, were discovered for both
configurations in water experiments
(\citep{2007PhRvL..99e4101D,2008JFM...601..339R,2009PhFl...21b5104C}
in a cylinder) as well as in nonlinear three-dimensional simulations
dedicated to the Madison dynamo \cite{2009PhRvE..80e6304R}.  Kinematic
dynamo simulations based on various manifestations of the flow field
obtained from these nonlinear hydrodynamic simulations yield a
beneficial impact of the nonaxisymmetric time-dependent flow
perturbations, whereas neither the time-averaged flow nor
time-snapshots of the velocity field were able to drive a dynamo
\cite{2009PhRvE..80e6304R}.  This effect was interpreted as dynamo
action based on nonnormal growth \cite{2008PhRvL.100l8501T}.
Nonnormal growth describes a perpetual increase of mode amplitudes by
virtue of the appropriate mixing of nonorthogonal eigenstates even if
the contributing eigenstates alone correspond to decaying
solutions. However, in contrast to the dynamo models from
\citet{2009PhRvE..80e6304R} and \citet{2008PhRvL.100l8501T} the boost
of the growth rates observed in our study already occurs when applying
stationary nonaxisymmetric perturbations and is not related to the
time scale given by transient growth during the initial phase of the
simulations which would depend on the initial conditions.  Hence, we
believe that the behavior found in our study is not based on
nonnormal growth.  In any case, our simulations show that, although
perturbations are able to considerably boost the growth rates in a
wide range of parameters (namely for a wide range of frequencies), a
beneficial impact for the onset of dynamo action by reducing the
critical magnetic Reynolds number can only be expected for stationary
and/or slowly drifting nonaxisymmetric perturbations (much slower
than the advective time scale based on the maximum axisymmetric
azimuthal flow).  Hence, the direct experimental realization of a
beneficial impact provoked by the inherent nonaxisymmetric vortices
is difficult because this would require a technical mechanism to
control the azimuthal frequency of the observed nonaxisymmetric flow
structures without significantly altering the basic flow field, which
is hardly conceivable. More promising might be the flow driving
mechanism at the Madison Plasma Dynamo Experiment (MPDX)
\cite{2014PhPl...21a3505C} where a conducting unmagnetized plasma is
exposed to a torque generated by external currents that interact with
a multi-cusp magnetic field at the boundary. The resulting forcing
drives a flow that is supposed to range deeply into the core plasma
and may provide a possibility to impose appropriate nonaxisymmetric
perturbations.

\bigskip

\acknowledgments{
The authors acknowledge support from the Helmholtz-Allianz LIMTECH. AG
is grateful for support provided by CUDA Center of Excellence hosted
by the Technical University Dresden (http://ccoe-dresden.de). This
research was supported in part by the National Science Foundation
under Grant No. NSF PHY-1125915. A.G. acknowledges participating in
the program {\it{Wave-Flow 
    Interaction in Geophysics, Climate, Astrophysics, and Plasmas}} at
the Kavli Institute for Theoretical Physics, UC Santa Barbara.
}

%
\end{document}